\documentclass[a4paper,11pt]{article}
\pdfoutput=1 
\usepackage[symbol]{footmisc} 
\usepackage{jheppub} 
	
\usepackage{color, colortbl}
\usepackage[T1]{fontenc}
\usepackage[utf8]{inputenc}
\usepackage[T1]{fontenc}
\usepackage{epsfig,latexsym}
\usepackage{amsmath}
\usepackage[dvipsnames]{xcolor}
\usepackage{adjustbox}
\usepackage{tensor}
\usepackage{graphicx}
\usepackage{verbatim}
\usepackage{mathrsfs}
\usepackage{amssymb}
\usepackage{multirow}
\usepackage{epsfig}
\usepackage{color,colordvi}
\usepackage{appendix}
\usepackage{slashed}
\usepackage{array}
\usepackage{cancel}
\usepackage{float}
\usepackage[font=footnotesize, justification=centering]{caption}
\usepackage{epsf}
\usepackage{amsmath}
\usepackage{tikz-feynhand}
\usepackage{physics}
\usepackage{MnSymbol}
\usepackage{rotating}
\usepackage{multirow}
\usepackage[normalem]{ulem}
\usepackage{tcolorbox}
\usepackage{bbold}
\DeclareMathAlphabet{\mathpzc}{OT1}{pzc}{m}{it}

\newcommand{\vtext}[1]{\begin{sideways}{#1}\end{sideways}}
 \csname
@addtoreset\endcsname{equation}{section}

\newcolumntype{x}[1]{>{\centering\arraybackslash\hspace{0pt}}p{#1}}

\newcommand{\beq}{\begin{equation}}
\newcommand{\eeq}{\end{equation}}
\renewcommand{\[}{\left[}
\renewcommand{\]}{\right]}
\renewcommand{\(}{\left(}
\renewcommand{\)}{\right)}

\newcommand{\be}{\begin{eqnarray}}
\newcommand{\ee}{\end{eqnarray}}
\newcommand{\bea}{\begin{eqnarray}}
\newcommand{\eea}{\end{eqnarray}}
\newcommand{\bi}{\begin{itemize}}
\newcommand{\ei}{\end{itemize}}
\newcommand{\ben}{\begin{enumerate}}
\newcommand{\een}{\end{enumerate}}
\def\bes{\begin{equation*}}
\def\ees{\end{equation*}}
\def\bead{\begin{aligned}}
\def\eead{\end{aligned}}
\def\bmat{\left(\begin{matrix}}
\def\emat{\end{matrix}\right)}

\def\Re{\text{Re}}
\def\Im{\text{Im}}
\def\diag{\text{diag}}

\def\cI{{\cal I}}
\def\cJ{{\cal J}}
\def\cL{{\cal L}}

\def\cO{{\cal O}}
\def\CKM{\text{CKM}}

\definecolor{Ecolor}{RGB}{106,157,235}
\definecolor{lightgray}{RGB}{220,220,220}

\title{Beyond Jarlskog: 699 invariants for CP violation in SMEFT}

\author[a]{Quentin Bonnefoy,}
\author[a,b]{Emanuele Gendy,}
\author[a,c]{Christophe Grojean,}
\author[d]{Joshua T. Ruderman}

\affiliation[a]{DESY, Notkestra{\ss}e 85, 22607 Hamburg, Germany}
\affiliation[b]{Institute of Theoretical Physics, Universit\"at Hamburg, 22761 Hamburg, Germany}
\affiliation[c]{Institut f\"ur Physik, Humboldt-Universit\"at zu Berlin, 12489 Berlin, Germany}
\affiliation[d]{Center for Cosmology and Particle Physics, Department of Physics, New York University, New York, NY 10003, USA}

\emailAdd{quentin.bonnefoy@desy.de}
\emailAdd{emanuele.gendy@desy.de}
\emailAdd{christophe.grojean@desy.de}
\emailAdd{ruderman@nyu.edu}

\abstract{As SMEFT is a framework of growing importance to analyze high-energy data, understanding its parameter space is crucial. The latter is commonly split into CP-even and CP-odd parts, but this classification is obscured by the fact that CP violation is actually a collective effect that is best captured by considering flavor-invariant combinations of Lagrangian parameters. First we show that  fermion rephasing invariance imposes that several coefficients associated to dimension-six operators can never interfere with operators of dimension $\leq4$ and thus cannot appear in any physical observable at $\order{1/\Lambda^2}$. For those that can, instead, we establish a one-to-one correspondence with CP-odd flavor invariants, all linear with respect to SMEFT coefficients.
We explicitly present complete lists of such linear CP-odd invariants, and carefully examine their relationship to CP breaking throughout the parameter space of coefficients of dimension $\leq 4$. Requiring that these invariants all vanish, together with the Jarlskog invariant, the strong-CP phase, and the 6 CP-violating dimension-6 bosonic operators, provides $699(+1+1+6)$ conditions for CP conservation to hold in any observable at leading order, $\cO\(1/\Lambda^2\)$.}

\begin{document} 
\begin{flushright}
DESY 21-205\\
HU-EP-21/50
\end{flushright}
\maketitle
\flushbottom

\section{Introduction}\label{section:intro}

The predictions of the $SU(3)_C \times SU(2)_L \times U(1)_Y$ gauge-invariant operators of dimension four or less built with the field content of the Standard Model, which we denote SM$_4$, have been remarkably confirmed by experiments. In particular, the observed pattern of CP-violation (CPV) \cite{ParticleDataGroup:2020ssz} remarkably confirms the Cabibbo--Kobayashi--Maskawa (CKM) description of three fermionic generations of matter \cite{Cabibbo:1963yz,Kobayashi:1973fv}. Especially, two key aspects of CPV as we know it, namely the facts that it is quite suppressed despite an $\order{1}$ phase and that it demands the existence of at least three generations, are explained in the CKM formalism by its collective origin: CPV arises in the SM$_4$ from the interplay between the gauge and Yukawa sectors \cite{Branco:1999fs,herzQuinn}. In other words, CP is broken by the simultaneous presence of non-vanishing physical parameters, which makes the breaking accidentally small in the SM$_4$\@. The notion of collective breaking has application beyond CPV, for instance in little Higgs models \cite{Arkani-Hamed:2001nha,Arkani-Hamed:2002sdy,Arkani-Hamed:2002ikv}.

The two notions of ``simultaneous presence'' and ``physical parameter'' in the previous paragraph suggest that a measure of CPV should be carefully defined, in particular when expressed in terms of Lagrangian parameters. An efficient way to do this is to use flavor-invariants, namely quantities which do not depend on the precise labeling of each fermion generation. As we will review below, a single quantity suffices to describe CPV in the fermionic sector of the SM$_4$,
\beq
J_4 \equiv \mathrm{Im} \, \mathrm{Tr} \left[ Y_{\smash{u}\vphantom{d}}^{\mathstrut} Y_{\smash{u}\vphantom{d}}^\dagger, Y_d^{\mathstrut} Y_d^\dagger \right]^3=3 \, \mathrm{Im} \, \mathrm{Det} \left[ Y_{\smash{u}\vphantom{d}}^{\mathstrut} Y_{\smash{u}\vphantom{d}}^\dagger, Y_d^{\mathstrut} Y_d^\dagger \right] \ .
\label{eq:Jarlskoginvariant}
\eeq
where $Y_{u,d}$ are the $3\times 3$ up and down quark Yukawa matrices in the SM$_4$ Lagrangian. The quantity $J_4$ goes under the name of Jarlskog invariant \cite{Jarlskog:1985ht,Jarlskog:1985cw,Bernabeu:1986fc}, and vanishes iff CP is conserved. The structure of $J_4$ is such that it is not modified by unitary reshuffling of the quark fields, which means that it corresponds to a physical quantity. In addition, its expression shows that CPV in the SM$_4$ is not a feature of $Y_u$ or $Y_d$ alone, but a feature of the whole model which can only be assessed with the knowledge of both matrices (and in particular, of the fact whether they can be simultaneously diagonalized). This ``collective'' property of CPV, namely the fact that it depends on several Lagrangian parameters at once, is a key property of the SM$_4$, as well as of its extensions. This also holds for strong CPV, whose order parameter is given by $\theta_{QCD}-\arg\det\(Y_uY_d\)$, and resides simultaneously within the $\theta$-term of QCD and within the quark Yukawa matrices (see appendix \ref{appendix:thetaQCD} for more details).

While $J_4$ is the only order parameter of CPV in the fermionic sector of the SM$_4$, additional ones need to be specified whenever the SM$_4$ is extended, see {\it e.g.}~Ref.\,\cite{Mendez:1991gp,Botella:1994cs,Lavoura:1994fv,Gunion:2005ja,Trautner:2018ipq,Bento:2021hyo} for multi-Higgs doublet models, Ref.\,\cite{Branco:1986kf,Lebedev:2002wq,Botella:2004ks} for the case of supersymmetric extensions of the SM$_4$, Ref.\,\cite{Botella:1985gb,Bernabeu:1986fc,Gronau:1986xb,Bering:2021hln} for the case of additional generations of matter, Ref.\,\cite{Branco:1986gr,Branco:1998bw,Dreiner:2007yz,Jenkins:2009dy,Yu:2019ihs,Yu:2020gre,Lu:2021aar,Wang:2021wdq,Yu:2021cco} for the inclusion of neutrino masses, Ref.\,\cite{Aguilar-Saavedra:1997sbo} for vector-like extensions, Ref.\,\cite{DiLuzio:2020oah} for CP-violating ALP EFTs or Ref.\,\cite{Valenti:2021rdu} for models of spontaneous CP breaking. In this paper, we assume that there are no new light degrees of freedom (d.o.f.) below, or close to, the weak scale, but we remain agnostic about the presence of heavy states. In that case, the SM$_4$ should be understood as the low-energy approximation of some fundamental UV dynamics, {\it i.e.}~it becomes necessary to extend it into an effective field theory (EFT). Under the assumption that a decoupling limit can be consistently taken, the adequate description is the so-called Standard Model Effective Field Theory (SMEFT)~\cite{Buchmuller:1985jz}. There, the dynamics of the SM$_4$ d.o.f.\ are derived from the SM$_4$ Lagrangian supplemented by a tower of higher-dimensional operators,
\beq
\cL=\cL_{\text{SM}_4}+\sum_i\frac{C_i}{\Lambda^{d_i-4}}\cO_i \ ,
\eeq
where $\cO_i$ is a local operator of dimension $d_i>4$, $C_i$ a complex coefficient (sometimes called Wilson coefficient) generically of order one (modulo possible selection rules and after taking $\hbar$ dimensions into account) and $\Lambda$ is a dimensionful scale associated to heavy new physics. If large enough, $\Lambda$ allows for a power expansion of any phenomenological prediction in inverse powers of itself. Complete bases of operators up to dimension 8 can be found in Refs.~\cite{Grzadkowski:2010es,Li:2020gnx,Murphy:2020rsh}.

The coefficients $C_i$ are generically complex and introduce a large number of new sources of CPV in SMEFT (see the counting in Ref.\,\cite{Alonso:2013hga} at dimension $d_i\leq 6$). In this paper, we are interested in CPV associated to flavorful Wilson coefficients, whose analysis requires the careful extraction of basis-independent physical parameters, which account for the collective properties of CPV\@. Therefore, the new CPV phases should be captured by CP-odd flavor-invariants, similar to $J_4$. We focus here on the CPV phases found in the fermionic sector, since the bosonic sources of CPV in the dimension-six SMEFT are trivially flavor-invariant.\footnote{For the 6 CP-odd bosonic operators appearing in the basis in Ref.\,\cite{Grzadkowski:2010es}, indeed, the condition for CP conservation is simply for their coefficient to vanish.} Moreover, we focus on the limit of vanishing neutrino masses throughout the whole work.

One could ask: why do we need invariant quantities? For instance, it is common to associate a complex top-quark Yukawa coupling with a new source of CPV, without referring to invariants. However a complex top Yukawa only signals CPV if one works in a given flavor basis where the top is a mass eigenstate of real mass.  
One may wonder how to describe CPV in SMEFT, without any specific reference to the IR physics, such as the masses or the electroweak vacuum. This picture is for instance justified if one cares about new sources of CPV which arise from the matching to a given UV model, which should be analyzed at an energy scale above the electroweak vacuum expectation value (vev) and should not depend on the details of the IR dynamics, such as specific flavor bases motivated by low energy considerations. Flavor-invariants are well-suited to answer such questions, as they allow one to capture physical and collective properties of the model, to parametrize CP-odd observables in a basis-independent way, and also to make the matching with UV models and their properties easier, by decorrelating the parametrization of CPV quantities from flavor bases connected to the IR properties of SM$_4$ particles.

Let us use the aforementioned example of the top-quark complex Yukawa coupling to offer a preview of the flavor-invariants considered in this paper. As said above, a top-Higgs Lagrangian with a complex Yukawa coupling
\beq
-\cL\supset m_t\bar t t + \frac{m_t}{v}\bar t(\kappa_t+i\tilde \kappa_t\gamma_5)th=m_t\bar t_L t_R + \frac{m_t}{v}\bar t_L(\kappa_t+i\tilde \kappa_t)t_Rh+\text{h.c. }\ ,
\label{topComplexYukawa}
\eeq
violates CP$ $. It can originate in SMEFT from the dimension-four and six Yukawa couplings,\footnote{At order $1/\Lambda^2$ (and for one generation only) the correspondence between the different coefficients reads \cite{Giudice:2007fh,Fuchs:2020uoc}
	\bes
	m_t=\frac{y_t v}{\sqrt{2}}\left(1+\frac{1}{2}\frac{\Re{\, C_{tH}}v^2}{y_t\Lambda^2}\right)\ , \quad \kappa_t+i\tilde{\kappa}_t=1+\frac{\Re{\,C_{tH}}}{y_t}\frac{v^2}{\Lambda^2}+i\frac{\Im{\,C_{tH}}}{y_t}\frac{v^2}{\Lambda^2}
	\ees
	provided we start in a basis where $y_t$ is already real.
}
\beq
-\cL\supset y_t\bar Q_L t_R\tilde H +\frac{C_{tH}}{\Lambda^2}\bar Q_L t_R\tilde H\abs{H}^2+\text{h.c. }\ .
\eeq
The above expression generalizes to three generations of matter in SMEFT upon replacing $Q_L,t_R,y_t,C_{tH}\to Q_{L,i},t_{R,j},Y_{u,ij},C_{uH,ij}$. Focusing on the diagonal entries of $C_{uH}$ in a basis where $Y_{u}$ is diagonal and real (with non-degenerate entries, as is experimentally relevant), their three imaginary parts $\text{Im}\, C_{uH,ii}$ violate CP\@. They can be captured by three independent flavor-invariants, whose expressions read
\beq
L_{\smash{k=1,2,3}\vphantom{d}}^{\vphantom{\dagger}}=\text{Im}\Tr(X_{\smash{u}\vphantom{d}}^{k-1\vphantom{\dagger}} C_{\smash{uH}\vphantom{d}}^{\vphantom{\dagger}}Y_{\smash{u}\vphantom{d}}^{\dagger})
\label{diagonalSetCuH}
\eeq
where $X_{\smash{u}\vphantom{d}}^{\vphantom{\dagger}}\equiv Y^{\vphantom{\dagger}}_{\smash{u}\vphantom{d}}Y^\dagger_{\smash{u}\vphantom{d}}$. In the basis where $Y_{u}^{\vphantom{\dagger}}=\text{diag }y_{u,i}$ is diagonal and real, one has $L_{k=1,2,3}=y_{u,i}^{2k-1}\text{Im }C_{uH,ii}^{\mathstrut}$, with an implicit sum over $i$. Therefore, at fixed, non-vanishing and non-degenerate $y_{u,i}$, the set of three $L_{k=1,2,3}$ maps to that of three $C_{uH,ii=1,2,3}$ in a bijective fashion, hence those three invariants capture the three new sources of CPV associated to up quark complex Yukawa couplings, as in Eq.\,\eqref{topComplexYukawa}. In the generic case however, $C_{uH}$ has off-diagonal entries even in the basis where $Y_{u}$ is diagonal, all of which can be complex, such that one needs nine flavor-invariants to capture the nine new sources of CPV in $C_{uH}$. 
Although there is no unique choice, one possible set of invariants reads
\beq
L_{k=1,...,9}=\left\{ \ \begin{matrix}
\Im\Tr\(C_{\smash{uH}}^{\phantom{\dagger}} Y_{\smash{u}}^\dagger\)\quad\quad&\quad&\Im\Tr\(X_{\smash{u}}^{\vphantom{\dagger}}
X_{\smash{d}}^{\vphantom{\dagger}}C_{\smash{uH}}^{\vphantom{\dagger}} Y_{\smash{u}}^{\smash{\dagger}}\)&\quad&\Im\Tr\(X_{\smash{d}}^{\smash{2}^{\vphantom{\dagger}}}X_{\smash{u}}^{\smash{2}{\vphantom{\dagger}}} C_{\smash{uH}}^{\vphantom{\dagger}}Y_{\smash{u}}^{\smash{\dagger}^{\vphantom{\dagger}}}\)\quad\quad\\
\Im\Tr\(X_{\smash{u}}^{\vphantom{\dagger}} C_{\smash{uH}}^{\vphantom{\dagger}}  Y_{\smash{u}}^\dagger\)&\quad&\Im\Tr\(X_{\smash{d}}^{\vphantom{\dagger}}X_{\smash{u}}^{\vphantom{\dagger}}  C_{\smash{uH}}^{\vphantom{\dagger}}  Y_{\smash{u}}^\dagger\)&\quad&\Im\Tr\(X_{\smash{u}}^{\vphantom{\dagger}} X_{\smash{d}}^{\smash{2}^{\vphantom{\dagger}} }X_{\smash{u}}^{\smash{2}\vphantom{\dagger} } C_{\smash{uH}}^{\vphantom{\dagger}} Y_{\smash{u}}^\dagger\)\\
\Im\Tr\(X_{\smash{d}}^{\vphantom{\dagger}}  C_{\smash{uH}}^{\vphantom{\dagger}} Y_{\smash{u}}^\dagger\)&\quad&\Im\Tr\(X_{\smash{u}}^{\smash{2}\vphantom{\dagger}}X_{\smash{d}}^{\smash{2}\vphantom{\dagger}} C_{\smash{uH}}^{\vphantom{\dagger}} Y_{\smash{u}}^\dagger\)&\quad&\Im\Tr\(X_{\smash{d}}^{\vphantom{\dagger}}X_{\smash{u}}^{\smash{2}\vphantom{\dagger}}X_{\smash{d}}^{\smash{2}\vphantom{\dagger}} C_{\smash{uH}}^{\vphantom{\dagger}} Y_{\smash{u}}^\dagger\)\\
\end{matrix} \ \right\}
\label{fullSetCuH}
\eeq
where $X_{d}^{\vphantom{\dagger}}\equiv Y_d^{\vphantom{\dagger}}Y_d^\dagger$.\footnote{The careful reader may note that the former $L_3$ in Eq.\,\eqref{diagonalSetCuH} does not appear anymore in Eq.\,\eqref{fullSetCuH}. We have removed $L_3$ because it is not independent of the first two $L$'s when two up-type quark masses are degenerate. This does not happen for any of the invariants in Eq.\,\eqref{fullSetCuH}, which are therefore preferred (see the following sections for a systematic treatment).}

Naively, the number of new flavor-invariants should match that of the new sources of CPV\@. However, observables in SMEFT are truncated at a given order in inverse powers of $\Lambda$, according to the SMEFT power counting, and it happens that not all sources of CPV contribute to physical observables at this given order as a result of non-interference. In this paper, we illustrate this fact by discussing CPV observables truncated at the leading $1/\Lambda^2$ order, to which several of the new sources of CPV at dimension-six cannot contribute.\footnote{Such CP-odd observables are at most linear in the dimension-six SMEFT coefficients, and correspond to the interference between the SM$_4$ and the leading SMEFT contributions to a given amplitude. A more thorough characterization of the observables we consider can be found in section~\ref{section:dim6NumberPhysical}.} We therefore carefully differentiate between the power counting for observables, which we truncate at order $1/\Lambda^2$, and that of SMEFT operators, which we only include up to dimension-six, {\it i.e.}~also up to order $1/\Lambda^2$. As we will explain, not all associated SMEFT coefficients can interfere with the SM$_4$ contribution to a given observable, and therefore they cannot all contribute at leading $1/\Lambda^2$ order to observables. We dub those which can {\it primary coefficients}, while we refer to the others as {\it secondary coefficients}. We perform the counting of the number of (both CP-even and CP-odd) SMEFT primary coefficients. Among those, the CP-odd fermionic ones, whose number is 699, are captured by flavor-invariants linear with respect to the dimension-six SMEFT coefficients, and an explicit and complete set of such flavor invariants is built. Consequently, we present a necessary and sufficient set of flavor-invariants, such that CP is conserved at $\cO(1/\Lambda^2)$ iff they vanish, together with $J_4$, the strong-CP phase, and the 6 CP-violating dimension-6 bosonic operators, so that they form a set of $699(+1+1+6)$ order parameters of CPV\@.

The rest of this paper is organized as follows. In section \ref{section:collective}, we review the collective nature of CP breaking in the SM$_4$ (in particular the structure of the aforementioned Jarlskog invariant), introducing concepts useful for the following, such as flavor bases and flavor invariants. We additionally emphasize that collective effects also arise beyond the SM$_4$, in SMEFT in particular. In section \ref{section:dim6Counting}, we refine the counting of dimension-six parameters of SMEFT, taking into account the interplay between the SMEFT power counting and the requirement of rephasing invariance of physical quantities. This reduces the relevant parameters to the primary ones. Focusing on new sources of CPV, we then present our strategy to capture them thanks to flavor-invariants linear in the SMEFT dimension-six coefficients. That this can be achieved is shown explicitly in section \ref{section:invariantsMinimal}, where we discuss explicit sets of CP-odd invariants for SMEFT at dimension-six. We emphasize that they need to be carefully chosen to capture all new sources of CPV for all parts of the SM$_4$ parameter space, ignoring for conciseness the case of vanishing quark masses which is subsequently treated in an appendix. We finally present conclusions and future directions in section \ref{section:conclusion}. Appendix \ref{appendix:generalCP} discusses flavor symmetries of the SM$_4$ in all relevant parts of its parameter space, which is important to determine adequate sets of CP-odd invariants as discussed in section \ref{section:invariantsMinimal}. The case of vanishing quark masses is discussed at the end of this appendix. In appendix \ref{appendix:invariantGeneralities}, we consider generic properties of flavor-invariants with three generations, in particular algebraic relations between them which are consistent with the counting of primary parameters. In appendix \ref{section:paramnf}, the counting of primary parameters, performed in section \ref{section:dim6Counting} for three generations of matter, is generalized to any number of generations. Appendices \ref{appendix:bilinears} and \ref{appendix:4Fermi} then contain a full list of linear CP-odd invariants (for operators respectively bilinear and quartic in fermion fields), which map to all independent primary Lagrangian parameters. Finally, appendix \ref{appendix:thetaQCD} includes $\theta_\text{QCD}$, which allows one to build flavor-invariants with new algebraic structures, but does not suffice to increase the number of primary coefficients.

\section{The collective nature of CP breaking in the SM(EFT)}\label{section:collective}

In order to motivate why we define CP-odd invariants, it is useful to review first one important and interesting aspect of CP breaking in SMEFT: it is collective. Indeed, it relies on the simultaneous presence of several complex parameters in the Lagrangian, which cannot all be made simultaneously real, even using the freedom to redefine fields (or equivalently, to define appropriately the CP transformation). 
In this section we review CP violation in SM$_4$, in order to establish our conventions and present several of the claims related to CP violation which will be repeatedly encountered in this paper.

\subsection{CP-violation and complex parameters}\label{section:complexParam}

The usual lore is that complex parameters in the Lagrangian violate CP\@. At the level of the fermionic Lagrangian, this claim leaves implicit crucial subtleties related to field redefinitions. The correct statement is instead that {\it the Lagrangian is CP-symmetric iff one can redefine the fields so as to make all couplings real}.\footnote{This is strictly speaking only true for models with continuous internal symmetries. When discrete symmetries are present, there exists the possibility that the couplings are pseudo-real, namely related to their complex conjugates via flavor transformations. Then one would get a CP-symmetric Lagrangian iff there exists a flavor transformation which sends all couplings to their complex conjugates at once. See Ref.~\cite{Ivanov:2015mwl} for an example, or the section 4.3 of Ref.~\cite{Trautner:2016ezn} for more details and references. In this text, we focus on the bulk of the SM(EFT) parameter space where any discrete symmetry is embedded into a continuous one (this is for instance automatic for non-degenerate spectra, see section~\ref{section:expectedRanks} and appendix~\ref{appendix:generalCP}).} In the SM$_4$, this explains why only one phase out of the six naively contained in the CKM matrix is physical and breaks CP\@. For instance, were the CKM matrix equal to the following unitary matrix
\beq
V_\CKM=
\left(
\renewcommand*{\arraystretch}{1.3}
\begin{array}{ccc}
	\frac{72-21 i}{325} & \frac{4}{13} & -\frac{12i}{13} \\
	-\frac{12}{13} & \frac{576+168 i}{1625} & \frac{49-168 i}{1625} \\
	-\frac{96-28 i}{325} & -\frac{57}{65} & -\frac{24i}{65} \\
\end{array}
\right)\ ,
 \label{wouldBeComplexCKM}
\eeq
it would not violate CP, although it is explicitly complex. Indeed, one can write
\beq
\left(
\renewcommand*{\arraystretch}{1.3}
\begin{array}{ccc}
	\frac{72-21 i}{325} & \frac{4}{13} & -\frac{12i}{13} \\
	-\frac{12}{13} & \frac{576+168 i}{1625} & \frac{49-168 i}{1625} \\
	-\frac{96-28 i}{325} & -\frac{57}{65} & -\frac{24i}{65} \\
\end{array}
\right)=
\left(
\renewcommand*{\arraystretch}{1.3}
\begin{array}{ccc}
	\frac{3-4 i}{5} & 0 & 0 \\
	0 & \frac{4-3 i}{5} & 0 \\
	0 & 0 & \frac{3-4 i}{5} \\
\end{array}
\right)
\left(
\renewcommand*{\arraystretch}{1.3}
\begin{array}{ccc}
	\frac{3}{13} & \frac{4}{13} & \frac{12}{13} \\
	-\frac{12}{13} & \frac{24}{65} & \frac{7}{65} \\
	-\frac{4}{13} & -\frac{57}{65} & \frac{24}{65} \\
\end{array}
\right)
\left(
\renewcommand*{\arraystretch}{1.3}
\begin{array}{ccc}
	\frac{4+3 i}{5} & 0 & 0 \\
	0 & \frac{3+4 i}{5} & 0 \\
	0 & 0 & \frac{4-3 i}{5} \\
\end{array}
\right)\ ,
 \eeq
and absorb all the factorized diagonal phases into the fermion fields, in order to obtain a real orthogonal CKM matrix. Such a manipulation cannot be done for the following matrix,
\beq
V_\CKM=
\left(
\renewcommand*{\arraystretch}{1.3}
\begin{array}{ccc}
	\frac{2172-5004 i}{8125} & -\frac{1784+432 i}{8125} & -\frac{2427+5196 i}{8125} \\
	-\frac{3747+3996 i}{8125} & \frac{3324+912 i}{8125} & \frac{4772-1164 i}{8125} \\
	-\frac{308+144 i}{1105} & -\frac{4389+2052 i}{5525} & \frac{1848+864 i}{5525} \\
\end{array}
\right)\ .
\eeq
However, whether it yields a CPV SM$_4$ depends on the fermion spectrum. Indeed, were two quark masses equal, the kinetic Lagrangian would have a $U(2)$ flavor symmetry, allowing for more general fermion field redefinitions. For instance, if $m_u=m_c$, we can redefine the first two flavors of up-type quarks in order to absorb the $2\times 2$ unitary matrix which appears at the top left of the first factor on the right-hand-side of the following equality:
\beq
\left(
\renewcommand*{\arraystretch}{1.3}
\begin{array}{ccc}
	\frac{2172-5004 i}{8125} & -\frac{1784+432 i}{8125} & -\frac{2427+5196 i}{8125} \\
	-\frac{3747+3996 i}{8125} & \frac{3324+912 i}{8125} & \frac{4772-1164 i}{8125} \\
	-\frac{308+144 i}{1105} & -\frac{4389+2052 i}{5525} & \frac{1848+864 i}{5525} \\
\end{array}
\right)=
\left(
\renewcommand*{\arraystretch}{1.3}
\begin{array}{ccc}
	-\frac{176+468 i}{625} & -\frac{9-12 i}{25} & 0 \\
	\frac{351-132 i}{625} & \frac{16+12 i}{25} & 0 \\
	0 & 0 & \frac{77+36 i}{85} \\
\end{array}
\right)
\left(
\renewcommand*{\arraystretch}{1.3}
\begin{array}{ccc}
	\frac{3}{13} & \frac{4}{13} & \frac{12}{13} \\
	-\frac{12}{13} & \frac{24}{65} & \frac{7}{65} \\
	-\frac{4}{13} & -\frac{57}{65} & \frac{24}{65} \\
\end{array}
\right) \ ,
\eeq
obtaining again a real orthogonal CKM matrix.

As is clear from these numerical examples, and as we will repeatedly illustrate, it is more convenient to phrase the condition for CP-violation in a way which does not require scanning over all possible field redefinitions. If the theory preserves CP, the following CP transformation of the (non-degenerate) fermionic mass eigenstates $\psi$ (together with those of bosonic fields \cite{Branco:1999fs}) leaves the Lagrangian invariant in some field basis
\beq
({\cal CP})\psi(t,\vec x)({\cal CP})^{-1}=\gamma^0C\overline{\psi}^T(t,-\vec x) \ ,
\label{CPredefinedMassEigenstates}
\eeq
where $C$ is the (antisymmetric) charge conjugation matrix such that $\gamma^\mu C=-C(\gamma^\mu)^T$. As we anticipated, this implies that the Lagrangian couplings are real (in this field basis). For instance, if we assume that there exists the following coupling in the theory,
\beq
\cL\supset c_{1212}\(\overline \psi\indices{_1}\gamma^\mu \psi\indices{_2}\)\(\overline \psi\indices{_1}\gamma_\mu \psi\indices{_2}\)+\text{h.c. }\ ,
\eeq
we learn from the invariance under the CP transformation in Eq.\,\eqref{CPredefinedMassEigenstates} that $c_{1212}$ is real. However, the opposite statement is that the theory violates CP iff the transformation in Eq.\,\eqref{CPredefinedMassEigenstates} is never a symmetry, whatever the field basis chosen. This is not equivalent to saying that $c_{1212}$ is complex in some basis, but that {\it whatever the basis chosen, there exists at least one Lagrangian parameter which is genuinely complex}\footnote{For pseudo-real couplings, the statement is rather that all complex couplings cannot be turned simultaneously into their conjugates via the same change of basis.} (which usually depends on the basis). Therefore, the condition for CPV which we look for takes the following schematic form
\begin{tcolorbox}
\centering
CPV $\iff \text{Im}\(\text{something independent of the field basis}\)\neq 0$ .
\end{tcolorbox}
\noindent
Such a basis-independent object precisely defines a CP-odd flavor invariant.  Within the framework of SMEFT, we can define flavor invariants order-by-order in the power counting.  At leading order, the condition for CPV reads:
\begin{center}
CPV at $\cO(1/\Lambda^2)\iff \text{Im}\(\text{something of $\cO(1/\Lambda^2)$ independent of the field basis}\)\neq 0$ .
\end{center}

\subsection{Flavor transformations and flavor bases}\label{section:flavorTransformations}

As we just discussed, a meaningful statement about CP violation in the SM(EFT) must account for the possibility of field redefinitions. In addition, the SM(EFT) Lagrangian is naturally written in the unbroken electroweak phase, which does not differentiate between the three fermionic generations. Therefore, it should be possible to characterize CP violation without referring to any specific flavor labeling, in particular without identifying which combinations correspond to the mass eigenstates.

When the Lagrangian is written in terms of the gauge multiplets relevant in the unbroken phase, the kinetic Lagrangian in the fermion sector (including the gauge couplings) is invariant under a $U(3)^5=U(3)_{Q_L}\times U(3)_{u_R}\times U(3)_{d_R}\times U(3)_{L_L}\times U(3)_{e_R}$ flavor group, where each factor acts on the flavor indices of the associated fermion field (we drop the chirality indices in what follows). This group is the largest under which all SMEFT coefficients can be assigned a spurious transformation so as to leave the full SM(EFT) Lagrangian unchanged.

Similarly to our conclusion in the previous section, the Lagrangian is CP-symmetric iff one can redefine the fields so as to make all couplings real. When redefining the fermion fields by means of a $U(3)^5$ flavor transformation, the precise values of all flavored couplings in SMEFT are mixed up,  with the real and imaginary parts scrambled. Consequently, any order parameter of CP breaking cannot correspond to the imaginary part of a given coefficient, but instead should map to the imaginary part of a flavor-invariant combination of coefficients.

In order to build such invariants, it is useful to notice that the flavored SMEFT couplings transform under the flavor group as spurions with transformation properties which depend on the operator they couple to. For the Yukawa couplings at dimension-four, the transformations are as listed in Table\,\ref{tab:ytrasmforma}, and each (anti)fundamental representation has a charge $(-)1$ under the associated abelian group in the decomposition $U(3)_X=SU(3)_X\times U(1)_X$, where $X=Q,u,d,L,e$.\footnote{Out of the 5 $U(1)$ factors only the gauged $U(1)_Y$ and the combinations $U(1)_{B-L}$ are conserved at the quantum level, while the other are broken by anomalies.} Performing field redefinitions which belong to the flavor group, this set of spurious charges allows one to easily compute the couplings in the redefined theory, and to easily identify objects which are independent of such redefinitions.
\begin{table}[H]
	\centering
	\begin{tabular}{c|c|c|c|c|c}
		& $SU(3)_Q$ & $SU(3)_u$ & $SU(3)_d$ & $SU(3)_L$ & $SU(3)_e$\\\hline
		$Y_u$ &$\mathbf{3}$ & $ \mathbf{\bar{3}}$ & $\mathbf{1}$ &$\mathbf{1}$ &$\mathbf{1}$  \\[0.1cm]
		$Y_d$ &$\mathbf{3}$ & $ \mathbf{1}$ & $\mathbf{\bar{3}}$ &$\mathbf{1}$ &$\mathbf{1}$  \\[0.1cm]
		$Y_e$ &$\mathbf{1}$ & $ \mathbf{1}$ & $\mathbf{1}$ &$\mathbf{3}$ &$\mathbf{\bar{3}}$
	\end{tabular}
	\caption{flavor transformation properties of the Yukawa matrices treated as spurions}
	\label{tab:ytrasmforma}
\end{table}

Using flavor transformations, one can reach flavor bases where the Yukawa matrices have a specific form, and which we will sometimes use to explicitly evaluate invariants. For instance, using the singular value decomposition, we can choose that
\beq
	Y_u=\diag(y_u,y_c,y_t) \ , \quad Y_d=V_\CKM \cdot \diag(y_d,y_s,y_b) \ , \quad Y_e=\diag(y_e,y_\mu,y_\tau) \ ,
\label{upBasis}
\eeq
where all $y$'s are real and positive and $V_\CKM$ is the Cabibbo--Kobayashi--Maskawa (CKM) matrix. We refer to this flavor basis as the {\it up basis}. Similarly, there exists a {\it down basis} where
\beq
	Y_{\smash{u}\vphantom{d}}^{\vphantom{\dagger}}=V_{\smash{\CKM}\vphantom{d}}^\dagger\cdot\diag(y_u,y_c,y_t) \ , \quad Y_d=\diag(y_d,y_s,y_b) \ , \quad Y_e=\diag(y_e,y_\mu,y_\tau) \ .
\eeq
Fixing this shape for the Yukawa couplings exhausts all flavor transformations but some diagonal ones.\footnote{Their precise form depends on the basis. In the {\it down basis}, they are of the form $\diag(e^{i\alpha_X^1},e^{i\alpha_X^2},e^{i\alpha_X^3})$, such that $\alpha_Q^i=\alpha_d^i,\alpha_L^i=\alpha_e^i$ (RH up-quark phases are unconstrained).} If, in addition, a phase choice is made on the CKM matrix (for instance imposing that all its phases are given in terms of a single one as in usual parameterizations), no flavor freedom remains but the conserved baryon and lepton number symmetries $U(1)_B\times U(1)_L$. When we make such a choice below, we use the following parametrization of the CKM matrix \cite{Chau:1984fp},
\beq
V_\CKM=\bmat
 c_{12} c_{13} & c_{13} s_{12} & s_{13} e^{-i\delta_\CKM} \\
 -c_{23} s_{12}-c_{12} s_{13} s_{23} e^{i\delta_\CKM} & c_{12} c_{23}-s_{12} s_{13} s_{23} e^{i\delta_\CKM} & c_{13} s_{23} \\
 s_{12} s_{23}-c_{12} c_{23} s_{13} e^{i\delta_\CKM} & -c_{12} s_{23}-c_{23} s_{12} s_{13} e^{i\delta_\CKM} & c_{13} c_{23}
\emat \ ,
\label{paramCKM}
\eeq
where $c_{X},\, s_{X}=\cos(\theta_{X}),\, \sin(\theta_{X})$. 

\subsection{The collective nature of CP breaking in the SM$_4$}\label{section:collectiveSM}

We now review in some detail the (well known) collective nature of CP breaking in the SM$_4$, which is useful for our purpose. 

The fact that CP breaking is collective in the SM$_4$ can be understood from the fact that it preserves CP with up to two fermionic generations \cite{Cabibbo:1963yz,Kobayashi:1973fv}, so that one needs the simultaneous presence of three generations to be sensitive to CP violation. A question is then: what is the order parameter of CP-breaking in the SM$_4$?

In order to answer this question unambiguously, one needs to mod out the impact of flavor transformations. Besides the use of invariants, a way to exhaust all flavor transformations is to work in a well-defined flavor basis, for instance in the {\it up} or {\it down basis} defined previously. Using the remaining phase rotations allowed in such bases removes all complex parameters but one, which fully specifies the flavor basis. The only leftover complex quantity can be written in a way which is independent of the phase rotations \cite{Jarlskog:1985ht,Jarlskog:1985cw,Dunietz:1985uy,Wu:1985ea,Botella:1985gb,Bernabeu:1986fc},
\beq
\cJ=\Im\(V_{\CKM,us}V_{\CKM,cb}V_{\CKM,ub}^*V_{\CKM,cs}^*\)=s_{12}^{\vphantom{2}}c_{12}^{\vphantom{2}}s_{13}^{\vphantom{2}}c_{13}^2s_{23}^{\vphantom{2}}c_{23}^{\vphantom{2}}\sin(\delta_\CKM) \ ,
\label{J4CKMpart}
\eeq
where the last equality uses Eq.\,\eqref{paramCKM}. It is straightforward to check that $\cJ=0$ for the matrix in Eq.\,\eqref{wouldBeComplexCKM}.

As we saw in section~\ref{section:complexParam}, when two masses (of same-type quarks) are degenerate, there is a larger degeneracy of {\it up} or {\it down bases}, and one can further remove the leftover complex parameter from the SM$_4$ Lagrangian. Therefore, the genuine order parameter of CP breaking in the SM, $J_4$, is proportional to\footnote{This expression depends on differences of squared masses and not, {\it e.g.}, on $m_t-m_u$. This is due to the fact that only the modulus of a fermion mass is physical: any quark mass can be made complex by an appropriate rephasing of the associated RH field without changing other SM$_4$ couplings, hence we must consider the rephasing-invariant quantity $m_\psi m_\psi^*=\abs{m_\psi}^2$ for any fermion $\psi$, which reduces to the mass squared in field bases where the mass is real.}
\beq
J_4\propto  (m_t^2-m_c^2)(m_t^2-m_u^2)(m_c^2-m_u^2)(m_b^2-m_s^2)(m_b^2-m_d^2)(m_s^2-m_d^2)\cJ  \ .
\label{J4fulltmp}
\eeq
One can show that there are no additional factors to $J_4$ \cite{Jarlskog:1985ht}.

We have constructed $J_4$ in a specific flavor basis, but it is useful to have expressions valid in all bases. In that respect, instead of looking for complex quantities invariant under mere phase rotations, one would rather consider invariants under the full flavor group. As we anticipated in the introduction, $J_4$, which goes under the name of Jarlskog invariant, then corresponds to \cite{Jarlskog:1985ht}
\beq
J_4 \equiv \mathrm{Im} \, \mathrm{Tr} \left[ Y_{\smash{u}\vphantom{d}}^{\vphantom{\dagger}} Y_{\smash{u}\vphantom{d}}^\dagger, Y_d^{\vphantom{\dagger}} Y_d^\dagger \right]^3=3 \, \mathrm{Im} \, \mathrm{Det} \left[ Y_{\smash{u}\vphantom{d}}^{\vphantom{\dagger}} Y_{\smash{u}\vphantom{d}}^\dagger, Y_d^{\vphantom{\dagger}} Y_d^\dagger \right] \, .
\label{J4def}
\eeq
Defined as above, $J_4$ is independent of the choice of flavor basis, as can be checked from the transformations in Table\,\ref{tab:ytrasmforma}. Evaluating Eq.\,\eqref{J4def} for instance in the {\it up} or {\it down basis}, the connection with Eq.\,\eqref{J4fulltmp} is made obvious,
\beq
J_4= 6 (y_t^2-y_c^2)(y_t^2-y_u^2)(y_c^2-y_u^2)(y_b^2-y_s^2)(y_b^2-y_d^2)(y_s^2-y_d^2) \cJ \ .
\label{J4full}
\eeq
It can be shown that the statement that CP is broken in the SM$_4$ is equivalent to saying that $J_4$ does not vanish \cite{Jarlskog:1985ht}, therefore it corresponds to the genuine order parameter for CP breaking in the SM$_4$\@. 

\subsection{The collective nature of CP breaking beyond the SM$_4$}\label{section:collectiveBSM}

The search for flavor-invariant order parameters for CP breaking beyond the SM$_4$ is subject to very similar discussions. As an example, let us consider the SM$_4$ with a single generation of fermions, extended by a dimension-six Yukawa coupling for the up-type quark:
\beq
\cL=\cL_{\text{SM}_4}+\frac{C_{uH}}{\Lambda^2}\abs{H}^2\overline Q_L u_R \tilde H +\text{h.c. }
\eeq
It is well known that such a coupling can generate a two-loop contribution to the electron EDM, a CPV observable, which reads at $\cO(1/\Lambda^2)$ \cite{Barr:1990vd,Brod:2013cka,Panico:2018hal}
\be
\frac{d_e}{e}=-\frac{1}{48\pi^2}\frac{vm_em_u}{m_h^2}\frac{\Im(C_{uH})}{\Lambda^2}F_1\(\frac{m_u^2}{m_h^2},0\) \ ,
\label{barrZeeElectronOneGenRealMass}
\ee
where for conciseness we only kept the dominant contribution due to photons in the loop, where $m_h$ and $v$ are the Higgs mass and vev, respectively, and where
\be
F_1\(a,0\)=\int_0^1dx\frac{\ln(\frac{a}{x(x-1)})}{a-x(x-1)}\ . 
\ee
The result in Eq.\,\eqref{barrZeeElectronOneGenRealMass} may suggest that $\Im(C_{uH})$ acts as an order parameter of CP breaking in this theory.\footnote{In the SM4, the various contributions to the electron EDM are all proportional to $\cal J$, consistently with the collective nature of CP breaking. Short-distance quark-level perturbative contributions arise at four-loop order and allow one to identify the whole structure of $J_4$ (see \cite{Pospelov:2013sca} for a parametric expression and \cite{Yamaguchi:2020dsy} for a recent appraisal identifying an additional $m_b^2$ factor), but the dominant source comes from long-distance hadronic contributions sensitive to CPV four-fermion operators \cite{Yamaguchi:2020eub}. In addition, such hadronic contributions often dominate CPV observables, e.g. in the case of paramagnetic systems (see \cite{Ema:2022yra} for the identification of a new contribution which increases the previous result by 5 order of magnitude).} However, this imaginary part could be rotated away by a chiral transformation of the up quark field (for instance by redefining $u_R\rightarrow e^{-i\arg(C_{uH})}u_R$), so one could wonder if there remains an observable electron EDM\@. The resolution to this puzzle is due to the implicit assumption that the computation is performed in a basis where the up quark has a real mass. In a generic flavor basis, the mass is complex,
\be
\cL\supset -m_u ^{\vphantom{*}}\overline u_L^{\vphantom{*}}u_R^{\vphantom{*}}- m_u^* \overline u_R^{\vphantom{*}}u_L^{\vphantom{*}}+\frac{v^2C_{uH}}{\sqrt 2 \Lambda^2}\overline u_Lu_Rh+\frac{v^2C_{uH}^*}{\sqrt 2 \Lambda^2}\overline u_Ru_L h \ ,
\ee
where we wrote the Lagrangian in the broken phase, and a careful evaluation of the two-loop diagram ({\it i.e.}~using propagators featuring complex fermion matrices) yields
\be
\frac{d_e}{e}=-\frac{1}{48\pi^2}\frac{vm_e}{m_h^2}\frac{\Im(m_u^*C^{\vphantom{*}}_{\smash{uH}})}{\Lambda^2}F_1\(\frac{\abs{m_u}^2}{m_h^2},0\) \ ,
\label{barrZeeElectronOneGenComplexMass}
\ee
which matches Eq.\,\eqref{barrZeeElectronOneGenRealMass} when $m_u$ is real, as it should. This expression allows us to identify a more satisfactory order parameter of CP-breaking, $\Im(m_u^*C^{\vphantom{*}}_{\smash{uH}})$, which does not depend on the phase convention for the up quark. Similarly to what was discussed for the SM$_4$ previously, what matters here for CP breaking is not that the Yukawa coupling has an imaginary part, but that there is an irreducible imaginary part due to the simultaneous presence of both the coupling to the Higgs of the up quark and its non-zero mass. This provides also a qualitative argument for why the result in Eq.\,\eqref{barrZeeElectronOneGenRealMass} had to be explicitly proportional to $m_u$. 

The take-away message of this section is that real or imaginary parts of coefficients are only meaningful with respect to CP breaking when the flavor basis is completely determined. In a general basis, what matters are the imaginary parts of invariant combinations of coefficients.\footnote{Examples of this also exist in the SM$_4$\@. The quantity $\epsilon$, which encodes indirect CP violation in kaon decays, is sometimes written \cite{ParticleDataGroup:2020ssz}
\beq
\epsilon\approx \frac{e^{i\pi/4}}{\sqrt 2}\frac{\Im(M_{12})}{\Delta m} \ ,
\eeq
where $M_{12}$ is associated to $K^0\leftrightarrow \overline{K^0}$ mixing and $\Delta m$ is the mass difference between kaon mass eigenstates. This formula actually assumes that $\lambda_u\equiv V_{ud}V_{us}^*$ is real. The expression which is valid independently of the phase conventions reads \cite{Branco:1999fs}
\beq
\epsilon\approx \frac{e^{i\pi/4}}{\sqrt 2}\frac{\Im(M_{12}\lambda_u^2)}{\Delta m\ \abs{\lambda_u}^2} \ .
\eeq} CP-odd invariants in SMEFT, especially with three flavors, are the subject of this paper.

\section{Characterizing CP violation at dimension six}\label{section:dim6Counting}

In this section, we discuss the number of primary parameters in SMEFT, as well as the parametrization of those which are CP-odd. 

\subsection{Primary parameters in the SMEFT}\label{section:dim6NumberPhysical}

First, we count the number of flavorful primary SMEFT parameters. We remind that they are defined to be the dimension-six SMEFT parameters which generate BSM amplitudes which can interfere with SM$_4$ amplitudes (other parameters being called secondary). Indeed, observables computed in SMEFT are subject to a power expansion, with respect to which we focus on the leading BSM order, {\it i.e. we include contributions to observables up to order $1/\Lambda^2$}. For instance, in cross-section computations, we only consider the SM$_4$ amplitude squared and the interference with the leading BSM amplitude. 
Schematically, we can express a generic amplitude as
\begin{align}
	\mathcal{A}=\mathcal{A}^{(4)}+\mathcal{A}^{(6)}+\ldots
\end{align}
where $\mathcal{A}^{(4)}$ is the leading order amplitude built with renormalizable operators, $\mathcal{A}^{(6)}$ is the next-to-leading order one, accompanied by a $1/\Lambda^2$ suppression, and the dots indicate higher order terms that we ignore. 
Then, observables such as cross sections, which are proportional to the amplitude squared, will receive contributions by 
\begin{align}
	\abs{\mathcal{A}}^2=|\mathcal{A}^{(4)}|^2+2\Re\left(\mathcal{A}^{(4)}\mathcal{A}^{(6)*}\right)+\ldots
	\label{eq:leadingorderobservables}
\end{align}
Our goal in this section is then to determine the primary parameter space that characterizes the first two terms in Eq.\,\eqref{eq:leadingorderobservables}.

This counting does not lead to a mere repetition of that of Ref.\,\cite{Alonso:2013hga}, which counts primary and secondary parameters indifferently and whose results are reviewed in the first double column of Table\,\ref{tablephysicaldim6}, due to the fact that several of the dimension-six parameters are charged under lepton numbers, unlike SM$_4$ parameters. Given that physical observables cannot be charged, such parameters can only interfere with themselves (or other charged BSM parameters) to form a neutral object, and are therefore secondary according to our classification. 

More precisely, the free fermionic Lagrangian in the broken phase of the SMEFT and in the mass basis has abelian flavor symmetries $U(1)_{u_i,d_i,L_i}$, acting on each mass eigenstate independently. By definition, those do not affect the spectrum of asymptotic states and therefore do not affect physical predictions. They simply correspond to irreducible flavor ambiguities in a basis where mass matrices are diagonal and real, which must be fixed by further specifications (for instance, phase prescriptions in the CKM matrix). A consequence is that any observable must be expressed in terms of quantities which are invariant under these $U(1)$ phase rotations,\footnote{This is a slightly different statement than the one that CP violation should be characterized in a flavor-invariant way. Although they are restricted by flavor-invariant statements, amplitudes squared with flavored external states are not flavor-invariant, but they are always invariant under phase rotations.} and therefore, any coefficient which is not invariant on its own must enter observables multiplied by another $U(1)$-charged coefficient in order to form a neutral object. This story is known to readers familiar with the notion of rephasing-invariants of the CKM matrix \cite{Greenberg:1985mr}: $V_{\CKM,ij}$ being charged under $U(1)_{d_j}-U(1)_{u_i}$, physical predictions can only depend on the moduli and quartets,\footnote{Notice that sextets and monomial with more entries of the CKM matrix can be expressed in terms of moduli and quartets using the unitarity of $V_\CKM$.}
\beq
\abs{V_{\CKM,ij}^{\vphantom{*}}}^{\smash{2}\vphantom{*}} \ , \quad V_{\CKM,ij}^{\vphantom{*}}V_{\CKM,kl}^{\vphantom{*}}V^*_{\CKM,il}V_{\CKM,kj}^*\ .
\eeq
For the SMEFT at leading order, that implies that $U(1)$-charged dimension-six coefficients must multiply $U(1)$-charged dimension-four coefficients. In the quark sector, the CKM matrix is the only such object, and there does not exist any in the lepton sector, since $U(1)_{L_i}$ is a symmetry of the SM$_4$ Lagrangian for each $i$ (remember that we work in the limit of vanishing neutrino masses). Therefore, all ``off-diagonal'' lepton coefficients 
in the first double column of Table\,\ref{tablephysicaldim6}, {\it i.e.}~those which are charged under $U(1)_{L_i}-U(1)_{L_j}$, correspond to secondary parameters. 
This requirement reduces the number of parameters to the ones in the second double column of Table\,\ref{tablephysicaldim6}.

A similar reasoning explains why adequate entries of the CKM matrix must multiply dimension-six coefficients charged under $U(1)_{u_i}$ and/or $U(1)_{d_i}$. Those coefficients must therefore contribute to observables with an additional suppression due to the smallness of the off-diagonal CKM entries. For instance, $C_{uB,13}$ (expressed in the {\it up basis}) can only enter observables as
\beq
V_{CKM,11}^*V_{CKM,31}^{\vphantom{*}}C_{uB,13}^{\vphantom{*}} \text{ or } V_{CKM,12}^*V_{CKM,32}^{\vphantom{*}}C_{uB,13}^{\vphantom{*}} \ ,
\eeq
where the unitarity of the CKM matrix allows us to not consider $V_{CKM,13}^*V_{CKM,33}^{\vphantom{*}}C_{uB,13}^{\vphantom{*}}$. Due to the gauge anomalies of the abelian symmetries $U(1)_{u_i,d_i,L_i}$, the $\theta$-angles of the different gauge factors of the SM are also charged parameters, which can interfere (non-perturbatively) with appropriate SMEFT coefficients (see appendix~\ref{appendix:thetaQCD} for a discussion of $\theta_\text{QCD}$). Nevertheless, $U(1)_{L_i}-U(1)_{L_j}$ is anomaly-free in the SM, therefore the statements made previously about SMEFT coefficients in the lepton sector also hold non-perturbatively. 

In this discussion, we ignored the specific cases of observables in the neutrino sector since we work in the limit of vanishing neutrino masses. Examples of observables which we allow include electric dipole moments (EDMs) \cite{Brod:2013cka,Panico:2018hal,Fuchs:2020uoc,Alonso-Gonzalez:2021jsa} or the CPV parameters $\epsilon_K$ and $\epsilon'$ in kaon physics~\cite{Aebischer:2018quc,Aebischer:2018csl,Aebischer:2020jto}. We also assume that the leading BSM contribution indeed corresponds to the interference at $\cO(1/\Lambda^2)$, and not to a dimension-six contribution squared (or the interference between the SM$_4$ and a dimension-eight coefficient, etc) due to some accidental suppression of the $\cO(1/\Lambda^2)$ term. Below, we will also study the SM$_4$ parameter space as a whole ({\it i.e.}, beyond values relevant for phenomenology), which includes points where several entries of the CKM matrix become unphysical and can be redefined away, turning additional dimension-six coefficients in the quark sector into secondary ones. This would for instance happen if all down-type quarks were massless: barring observables which are ill-defined when $m_d\to 0$ within our leading order observables, we find a further reduction of the relevant dimension-six coefficients (see Table\,\ref{tableRanksSMEFTMore} in Appendix~\ref{vanishingMassesAppendix}). Ref.\,\cite{Degrande:2021zpv} performs a similar counting of primary parameters (focusing on the kinematic situation where all fermions masses, but the top and bottom quark, are neglected).

\begin{table}
\centering
\resizebox{1\columnwidth}{!}{%
\renewcommand*{\arraystretch}{1.3}
\newcolumntype{a}{>{\columncolor{lightgray}}c}
\begin{tabular}{cc|c|cc|aa}
&&&&&\multicolumn{2}{c|}{\cellcolor{lightgray}\begin{minipage}{0.3\columnwidth}
	 inv. under $U(1)_{L_i}-U(1)_{L_j}$
	\end{minipage}}\\
&Type of op. & \# of ops  &\# real &\# im.&\begin{minipage}{0.14\columnwidth}
\centering	\# real 
\end{minipage}&  \begin{minipage}{0.14\columnwidth}
\centering \# im.
\end{minipage} \\ \hline
\multirow{3}{*}{\vtext{bilinears}} &Yukawa & 3 & 27 & 27 & 21 &21\\
&Dipoles& 8 & 72 & 72& 60 & 60\\
&current-current & 8 & 51 & 30&42 & 21\\ \hline
&all bilinears & 19 & 150 & 129& 123 & 102\\ \hline
\multirow{5}{*}{\vtext{4-Fermi}} &LLLL  & 5 & 171 & 126& 99 &54\\
&RRRR & 7 & 255 & 195& 186 &126\\
&LLRR & 8 & 360 & 288&246 &174\\
&LRRL & 1 & 81 & 81& 27 & 27\\
&LRLR & 4 & 324 & 324& 216 &216\\ \hline
&all 4-Fermi & 25 & 1191 & 1014&774 & 597\\ \hline
&all & &1341 &1143 &897&699
\end{tabular}%
}
\caption{Number of flavorful real and imaginary parameters in SMEFT at dimension-six (see Tables\,\ref{bilinearlist} and\,\ref{4Fermilist} for the explicit forms of the operators). The first double column counts the number of physical parameters, the second one (highlighted in gray) counts those which are also primary (see the text). 
}
\label{tablephysicaldim6}
\end{table}

\subsection{CP conservation at leading order and minimal sets of CP-odd invariants}\label{section:dim6CPconservation}

We now turn to the characterization of CP violation in SMEFT at leading order using flavor-invariants. Specifically, in the spirit of the discussion which leads to the introduction of $J_4$, we ask
\begin{tcolorbox}
Which flavor invariants vanish iff CP is conserved at leading order in SMEFT? We call such a set of CP-odd invariants of minimal cardinality a {\it minimal set}.
\end{tcolorbox}
The notion of minimal cardinality implies that there are no redundancy: the vanishing of each invariant in a minimal set provides an independent condition. The number of invariants in a minimal set must be larger or equal than the number of new primary sources of CPV\footnote{It may need to be larger: as we detail below, we count an invariant as independent if there exists at least one point in parameter space (in terms of fermion masses or CKM entries) where it cannot be expressed in terms of other invariants of the set.}\@. In the case of the SMEFT at $\cO(1/\Lambda^2)$, we find the non-trivial result that the two numbers agree for all operators (see below).

Before going further, let us discuss one subtlety associated to our definition of a minimal set of CP-odd flavor-invariants, which has to do with the parameter space considered. In the way our definition is stated, it suggests that one aims at characterizing CP-conserving points for all  possible choices of parameters, {\it i.e.}~quark masses as well as mixing angles and the CKM phase $\delta_\CKM$. However, one could also try to characterize CP-conserving points {\it within a given parameter subspace}, for instance for values of the quark masses which are non-vanishing. This is the choice we make in the main body of this paper: we build flavor invariants which vanish iff CP is conserved at leading order in SMEFT {\it under the assumption that quark masses are non-vanishing}. Our methods also allow one to identify minimal sets of flavor-invariants when one considers vanishing quark masses, but since the expression of the required invariants is more intricate than for the simpler case of non-vanishing quark masses, we leave the resolution of this question to appendix \ref{appendix:generalCP}. One could finally restrict to characterizing CP-conservation for a smaller set of parameters, {\it e.g.}~fixing the values of the quark masses or taking them non-degenerate. Our sets of invariants work in such restricted cases, but there usually exist simpler ones (which do not correspond to minimal sets on larger sets of parameters). We will encounter explicit examples in the next section.

Due to the SMEFT power counting, the conservation of CP at zeroth order first demands that $J_4=0$, so that the $J_4$ is part of any minimal set. Then, in order to build the rest of the minimal set, we look for invariants which are {\it linear with respect to the SMEFT dimension-six coefficients}, consistently with the goal of characterizing CPV in observables up to the first non-leading order, {\it i.e.}~up to the $\order{1/\Lambda^2}$ interference term in the R.H.S. of Eq.\,\eqref{eq:leadingorderobservables}.
This linearity is also valuable to check that we indeed have a necessary and sufficient condition for CP conservation, as it does not suffer from subtleties associated to non-linear invariants, found {\it e.g.}~in neutrino physics \cite{Yu:2019ihs}. Indeed, the question of whether vanishing invariants really implies vanishing CP phases is trivially answered here. We found that there does exist a minimal set of invariants linear with respect to the SMEFT dimension-six coefficients. This set is presented in subsequent sections, and represents our main result.

The linearity implies that the minimal sets are decomposed in minimal sets of invariants $\cI_a(C^{(6)})$, defined for each SMEFT dimension-six operator $\cO^{(6)}$ and its associated matrix-valued coefficient $C^{(6)}$ independently, and where the index $a$ labels the new primary sources of CPV present in $C^{(6)}$. Therefore, we can define the notion of minimal set at the level of each SMEFT operator independently. The invariants have the following form: 
\beq
\cI_a(C^{(6)})=\Im\(\text{flavor-invariant linear in }C^{(6)}\)=T^R_{ai}\(\Re C^{(6)}\)_i+T^I_{ai}\(\Im C^{(6)}\)_i\equiv\mathcal{T}_{ai}^{\vphantom{(6)}}\overrightarrow{C}^{(6)}_i \ ,
\eeq
where the last equality describes the result of evaluating the invariant in a given flavor basis. Here we define 
\beq
\overrightarrow{C}^{(6)}_i \equiv\(\(\Re C^{(6)}\)_1,\(\Re C^{(6)}\)_2,...,\(\Im C^{(6)}\)_1,...\)
\eeq
as the vector in flavor space composed by the vectors $\Re/\Im C^{(6)}$, not necessarily of same length, while the transfer matrix $\mathcal{T}_{ai}$ is defined to take the block form
\beq
\mathcal{T}=\bigg(T^R \hspace{0.3cm}T^I\bigg)\ .
\eeq
By linearity, $T^R_{ai}$ is the imaginary part of a linear combination of products of entries of dimension-four Yukawas (which are the only flavored objects at dimension-four), while $T^I_{ai}$ is the real part of a similar, albeit generically different, combination. Those matrices $T^R$ and $T^I$ depend explicitly on the operator $\cO^{(6)}$ considered.
A very convenient feature of such invariants is that they automatically project out any secondary coefficient, which cannot be arranged into invariants in a linear fashion by definition.

Showing that the set of invariants is minimal can be phrased as a condition on the matrices $T^{R/I}$:
\begin{tcolorbox}
A set of flavor invariants is a minimal set iff the rank of the transfer matrix $\mathcal{T}$ equals the number of new primary sources of CPV in $C^{(6)}$ when $J_4=0$ and never does for all sets with strictly smaller cardinality.
\end{tcolorbox}
Note that the last part of this characterization is automatic when the number of invariants, the number of primary sources of CPV and the rank are all equal. We use this condition below to check that the sets of invariants we present below are indeed minimal. We stress that the meaning of ``when $J_4=0$'' encompasses a large subset of the whole parameter space spanned by the masses and the mixing angles, as seen from the expression in Eq.\,\eqref{J4fulltmp}: it is achieved when $\theta_{ij}=0$ or $\pi/2$, or when $m_{u,i}=m_{u,j}$ or $m_{d,i}=m_{d,j}$, for any pair $i,j$. In addition, setting $J_4=0$ via one of these choices still leaves a large freedom for the remaining parameters. For instance, one may have $\theta_{ij}=0$ {\it and} $m_{u,k}=m_{u,l}$ for some $i,j,k,l=1,...,3$. A set of flavor invariants is a minimal set only if the rank of its transfer matrix corresponds to the number of new sources of CPV {\it within the whole parameter space where $J_4=0$} (up to the restriction of non-vanishing quark masses which we adopt in the main text of this paper and relax in appendix \ref{appendix:generalCP}). We will come back to this point in section~\ref{section:invariantsMinimal}.

The transfer matrix $\mathcal{T}$ acts on the flavor-space vector made out of real and imaginary entries of $C^{(6)}$ (the precise order in the labeling as well as the order between real and imaginary part is unimportant). Note that the rank does not change under the action of flavor transformations (which reshuffle real and imaginary parts, as well as the entries of $T^{R/I}$). 

\section{Minimal set of CP-odd invariants}\label{section:invariantsMinimal}

In this section, we present the minimal set of leading order CP-odd invariants in SMEFT at dimension-six, under the aforementioned assumption that all fermion masses are non-vanishing, which has an impact on how many sources of CPV are expected and which invariants correctly capture them. We treat the cases of vanishing masses in appendix~\ref{appendix:generalCP}.

\subsection{Examples}\label{section:exampleInvariants}

Let us present some parts of our minimal set of invariants for SMEFT at dimension-six. As we explained previously, the linearity with respect to the Wilson coefficients of the dimension-six CP-odd observables allows one to treat the different SMEFT operators independently. The study of all SMEFT operators proceeds along identical lines, and the full set of invariants is presented in appendices \ref{appendix:bilinears} and \ref{appendix:4Fermi}. 

We begin by considering SMEFT operators which are bilinear in fermion fields and hermitian, and therefore have the simplest non-trivial flavor structure. Invariants under unitary groups with bi-fundamental representations must feature the invariant tensor $\delta^a_b$, therefore they correspond to linear combinations of traces of products of matrices, arranged so that indices of a given fundamental representation and its conjugate are contracted in the trace, as seen for instance in Eqs.~\eqref{diagonalSetCuH}-\eqref{fullSetCuH}. In addition, there are relations between powers of $3\times 3$ matrices, and/or between their traces, derived from the Cayley--Hamilton theorem, which reduce the candidate invariants to a finite set. We explicitly present such properties in appendix \ref{appendix:invariantGeneralities}\@. For convenience we define: $X_u\equiv Y_uY_u^\dagger$ and $X_d  \equiv Y_dY_d^\dagger$. The relevant single-trace invariants linear with respect to a SMEFT coefficient $C$, for a fermion bilinear operator, take the universal form\footnote{As we will see, those structures are also the only ones needed for 4-Fermi operators.}
\beq
L_{abcd}(\tilde C)\equiv \Im\Tr(X_u^aX_d^bX_u^cX_d^d\tilde C) \ , \text{ with } a,b,c,d=0,1,2 \text{ and } a\neq c, b\neq d \ ,
\label{bilinearFormula}
\eeq
where $\tilde C=C,CY_{f=u,d,e}^\dagger$ or $Y^{\mathstrut}_fCY_{f}^\dagger$, depending on the chiral structure of the operator under study (see below for explicit formulae). We first choose $C=C_{Hu}$
for definiteness, and we find that the following property holds:
\begin{tcolorbox}
\centering
$\cL=\cL_{\text{SM}_4}+\frac{C_{Hu,ij}}{\Lambda^2}\(iH^\dagger \overleftrightarrow D_\mu H\)\overline{u_{i,R}}\gamma^\mu u_{j,R}$\\
\vspace{10pt}
preserves CP at $\cO(1/\Lambda^2$) iff
 \beq
J_4=L_{1100}\(Y_{\smash{u}}^{\vphantom{\dagger}} C_{\smash{Hu}}^{\vphantom{\dagger}}  Y_{\smash{u}}^\dagger\)=L_{2200}\(Y_{\smash{u}}^{\vphantom{\dagger}} C_{\smash{Hu}}^{\vphantom{\dagger}}  Y_{\smash{u}}^\dagger\)=L_{1122}\(Y_{\smash{u}}^{\vphantom{\dagger}} C_{\smash{Hu}}^{\vphantom{\dagger}}  Y_{\smash{u}}^\dagger\)=0
\label{minimalSetCHuExample}
\eeq
\end{tcolorbox}
Indeed, $J_4=0$ is necessary so that the leading order SM$_4$ contribution to any CP-odd observable vanishes. Once enforced, this makes the SM$_4$ Lagrangian CP-symmetric, and there remains generically three new primary sources of CPV in $C_{Hu}$. Indeed, $C_{Hu}$ is a ($3 \times 3$) hermitian matrix (which transforms as a $\mathbf{3}\otimes\bar{\mathbf{3}}$ representation of $U(3)_u$). Therefore, the minimal set for $C_{Hu}$ should at least contain three invariants. As we explained in section~\ref{section:dim6CPconservation}, in order to show that the three invariants in Eq.\,\eqref{minimalSetCHuExample} capture the three necessary conditions, it is sufficient to compute the transfer matrix $\mathcal{T}$ such that,
\beq
\bmat L_{1100}\\ L_{2200} \\ L_{1122}\emat=\bmat T^R & T^I\emat \bmat \Re C_{Hu,11}\\ \Re C_{Hu,12} \\ ... \\ \Im C_{Hu,12} \\ \Im C_{Hu,13} \\ \Im C_{Hu,23} \emat \ ,
\eeq
and show that it has rank $3$. Parametrically, and in some basis, the generic case corresponds to taking $\delta_\CKM\to 0$ in Eq.\,\eqref{paramCKM} while holding all mixing angles different from $0,\pi/2$ and all quark masses non-denegerate, which is what we assume in the current section (we treat more general cases below). Then, $T^R=0_{3\times 6}$ and the determinant of $T^I$ is found to be non-vanishing. Therefore, $L_{1100}=L_{2200}=L_{1122}=0$ implies that $\Im C_{Hu,ij}=0$ in the basis of Eq.\,\eqref{paramCKM} (or any other basis where the Yukawa matrices are real), {\it i.e.}~CP is conserved. Conversely, the conservation of CP, or equivalently $\Im C_{Hu,ij}=0$, implies that all $L$'s vanish since $T^R=0$. This proves the equivalence announced above.

One may be surprised by the fact that some simple invariants, in the sense that they feature low powers of the Yukawa matrices, are not parts of the set in Eq.\,\eqref{minimalSetCHuExample}. For instance, for $C_{Hu}$, the set formed by
\beq
L_{1100}\(Y_{\smash{u}}^{\vphantom{\dagger}} C_{\smash{Hu}}^{\vphantom{\dagger}}  Y_{\smash{u}}^\dagger\)\ , \ L_{1200}\(Y_{\smash{u}}^{\vphantom{\dagger}} C_{\smash{Hu}}^{\vphantom{\dagger}}  Y_{\smash{u}}^\dagger\) \ , \ L_{2100}\(Y_{\smash{u}}^{\vphantom{\dagger}} C_{\smash{Hu}}^{\vphantom{\dagger}}  Y_{\smash{u}}^\dagger\)
\label{wrongSetCHu}
\eeq
would pass the test performed in this section, namely the associated transfer matrix would generically have rank 3. Such invariants have been studied beyond SMEFT, and exist generally for any set of three hermitian matrices in the same adjoint representation, as explained in Ref.\,\cite{Lebedev:2002wq}. However, they would not be a valid choice of invariants, since they would not correspond to sufficient conditions whatever the values of the fermion masses and the CKM matrix. For instance, $m_t=m_c$ is another way to get $J_4=0$ other than that discussed above. In this situation, there remains three conditions necessary for CP to be conserved (see section~\ref{section:expectedRanks} for details). Therefore, imposing that a given minimal set vanishes should be equivalent to three independent conditions also when $m_t=m_c$. However (when $m_t=m_c$) we find that
\beq
(m_u^2+m_c^2)L_{1100}=L_{2100} \ ,
\eeq
where the mass-dependent factor can be expressed in terms of invariant quantities,
\beq
m_u^2+m_c^2=\frac{18\det X_u^{\vphantom{2}}-3\(\Tr X_u^{\vphantom{2}}\)^3+7\Tr X_u^{\vphantom{2}}\Tr X_u^2}{6\Tr X_u^2-2\(\Tr X_u^{\vphantom{2}}\)^2} \ .
\label{formulaSumDegenerateMasses}
\eeq
Therefore, imposing that the set in Eq.\,\eqref{wrongSetCHu} vanishes only amounts to two conditions. Instead, one can check that the vanishing of the set in Eq.\,\eqref{minimalSetCHuExample} yields three independent conditions\footnote{The interplay between degenerate cases and flavor-invariants has also been studied in Ref.~\cite{Yu:2020gre} for massive Majorana neutrinos.}, even when $m_t=m_c$. We expand on what happens however $J_4=0$ is taken in section~\ref{section:expectedRanks}, and show that the set in Eq.\,\eqref{minimalSetCHuExample} above yields a satisfactory minimal set in all cases (as long as all quark masses are non-vanishing). 

Similar reasoning applies to all SMEFT operators. Let us present the results in two more cases with slightly more complicated flavor structures, the non-hermitian bilinear operator $\cO_{uH}$ and the hermitian symmetric 4-Fermi operator $\cO_{uu}$. The Wilson coefficient $C_{uH}$ contains nine new primary sources of CPV, since it is an arbitrary ($3\times 3$) complex matrix (which transforms as a $(\mathbf{3},\bar{\mathbf{3}})$ representation of $U(3)_Q\times U(3)_u$). $C_{uu}$ contains eighteen new CPV parameters ($C_{uu,ijkl}^{\vphantom{*}}$ is ``hermitian'', {\it i.e.}~$C_{uu,ijkl}^*=C_{uu,jilk}^{\vphantom{*}}$, symmetric, $C_{uu,ijkl}^{\vphantom{*}}=C_{uu,klij}^{\vphantom{*}}$, and it transforms in the $(\mathbf{3}\otimes\bar{\mathbf{3}})^2$ of $U(3)_u$). 

For $\cO_{uH}$, we find that
\begin{tcolorbox}
\centering
$\cL=\cL_{\text{SM}_4}+\frac{C_{uH,ij}}{\Lambda^2}\overline{Q_{i,L}}\tilde Hu_{j,R}\abs{H}^2+\text{h.c. }$\\
\vspace{10pt}
preserves CP at $\cO(1/\Lambda^2$) iff
 \beq
 \bead
J_4&=L_{0000}\(C_{\smash{uH}}^{\vphantom{\dagger}} Y_{\smash{u}}^\dagger\)=L_{1000}\(C_{\smash{uH}}^{\vphantom{\dagger}} Y_{\smash{u}}^\dagger\)=L_{0100}\(C_{\smash{uH}}^{\vphantom{\dagger}} Y_{\smash{u}}^\dagger\)\\
&=L_{1100}\(C_{\smash{uH}}^{\vphantom{\dagger}} Y_{\smash{u}}^\dagger\)=L_{0110}\(C_{\smash{uH}}^{\vphantom{\dagger}} Y_{\smash{u}}^\dagger\)=L_{2200}\(C_{\smash{uH}}^{\vphantom{\dagger}} Y_{\smash{u}}^\dagger\)\\
&=L_{0220}\(C_{\smash{uH}}^{\vphantom{\dagger}} Y_{\smash{u}}^\dagger\)=L_{1220}\(C_{\smash{uH}}^{\vphantom{\dagger}} Y_{\smash{u}}^\dagger\)=L_{0122}\(C_{\smash{uH}}^{\vphantom{\dagger}} Y_{\smash{u}}^\dagger\)=0\ .
\eead
\label{minimalSetCuHExample}
\eeq
\end{tcolorbox}
We now turn to $\cO_{uu}$. For 4-Fermi operators, we generically define 
\beq
\Tr_A\(M^{(1)},M^{(2)},C\)\equiv M^{(1)}_{ji}M^{(2)}_{lk}C_{ijkl}^{\vphantom{(2)}} \ , \quad \Tr_B\(M^{(1)},M^{(2)},C\)\equiv M^{(1)}_{li}M^{(2)}_{jk}C_{ijkl}^{\vphantom{(2)}}\ ,
\label{eq:trAandtrB}
\eeq
and 
\beq
\bead
A^{abcd}_{efgh}(C)=&\Im\Tr_A\(X_u^aX_d^bX_u^cX_d^d,X_u^eX_d^{\smash{f}}X_u^gX_d^h,C\)\ , \\
B^{abcd}_{efgh}(C)=&\Im\Tr_B\(X_u^aX_d^bX_u^cX_d^d,X_u^eX_d^{\smash{f}}X_u^gX_d^h,C\) \ .
\label{eq:L4Fermi}
\eead\eeq
We further define 
\beq 
C_{\tilde u\tilde uuu,ijkl}^{\vphantom{\dagger}}\equiv \sum_{m,n}Y_{u,im}^{\vphantom{\dagger}}Y_{u,nj}^\dagger C_{uu,mnkl}^{\vphantom{\dagger}} \ ,
\label{eq:tildeCoeffs}
\eeq
and similarly for $C_{\tilde uuu\tilde u},C_{u\tilde u\tilde uu},C_{uu\tilde u\tilde u}$, $C_{\tilde u\tilde u\tilde u\tilde u}$, and for the down quark 
versions. We then find that
\begin{tcolorbox}
\centering
$\cL=\cL_{\text{SM}_4}+\frac{C_{uu,ijkl}}{\Lambda^2}\overline{u_{i,R}}\gamma^\mu u_{j,R}\,\overline{u_{k,R}}\gamma_\mu u_{l,R}$\\
\vspace{10pt}
preserves CP at $\cO(1/\Lambda^2$) iff
 \beq
\bead
J_4&=A^{0000}_{1100}\(C_{uu\tilde{u}\tilde{u}}\) =A^{1000}_{1100} \(C_{\tilde{u}\tilde{u}\tilde{u}\tilde{u}}\)= A^{0100}_{1100}\(C_{\tilde{u}\tilde{u}\tilde{u}\tilde{u}}\) \\
&=A^{0000}_{2200}\(C_{uu\tilde{u}\tilde{u}}\) = A^{1100}_{1100}\(C_{\tilde{u}\tilde{u}\tilde{u}\tilde{u}}\) = A^{0200}_{1100}\(C_{\tilde{u}\tilde{u}\tilde{u}\tilde{u}}\) \\&=
A^{0100}_{2200}\(C_{\tilde{u}\tilde{u}\tilde{u}\tilde{u}}\) = A^{0000}_{1122}\(C_{uu\tilde{u}\tilde{u}}\) = A^{1100}_{2200}\(C_{\tilde{u}\tilde{u}\tilde{u}\tilde{u}}\)\\&=
A^{1000}_{1122}\(C_{\tilde{u}\tilde{u}\tilde{u}\tilde{u}}\)= A^{0100}_{1122}\(C_{\tilde{u}\tilde{u}\tilde{u}\tilde{u}}\) = A^{1100}_{0122}\(C_{\tilde{u}\tilde{u}\tilde{u}\tilde{u}}\) \\&=
A^{1200}_{2200}\(C_{\tilde{u}\tilde{u}\tilde{u}\tilde{u}}\) = B^{0000}_{1100}\(C_{u\tilde{u}\tilde{u}u}\) = B^{0100}_{1100}\(C_{\tilde{u}\tilde{u}\tilde{u}\tilde{u}}\) \\&=
B^{0200}_{2100}\(C_{\tilde{u}\tilde{u}\tilde{u}\tilde{u}}\) = A^{1200}_{1122} \(C_{\tilde{u}\tilde{u}\tilde{u}\tilde{u}}\)= B^{1000}_{1200}\(C_{\tilde{u}\tilde{u}\tilde{u}\tilde{u}}\)=0\ . \\
\eead
\label{minimalSetCuuuuExample}
\eeq
\end{tcolorbox}
The proofs of these equivalences follow from the same logic as for $C_{Hu}$: compute $T^I$ and check that it has maximal rank, which means that the origin in invariant space is uniquely mapped to the origin in imaginary-coefficient space. Therefore, the vanishing of the invariants is equivalent to the conservation of CP (at leading order).

Finally, let us consider the leptonic case. As we argued in section~\ref{section:dim6NumberPhysical}, the fact that the SM$_4$ Lagrangian is symmetric under the lepton numbers $U(1)_{L_i}$ makes several SMEFT dimension-six coefficients in the lepton sector secondary. For the specific example of $C_{eH}$, this means that all off-diagonal entries are secondary, since they are charged under $U(1)_{L_i}-U(1)_{L_j}$ for suitable $i,j$. Therefore, although all imaginary parts of the nine entries of $C_{eH}$ violate CP when the full SMEFT expansion is considered, the only ones which can contribute to CP-odd observables at $\order{1/\Lambda^2}$ are the diagonal ones. The same reasoning applies to $C_{He}$, which therefore does not violate CP at $\order{1/\Lambda^2}$. Consequently, a minimal set for $C_{eH}$ contains three invariants, and is empty for $C_{He}$. Indeed, we find that (defining $X_e^{\vphantom{\dagger}}\equiv Y_e^{\vphantom{\dagger}}Y_e^\dagger$)
\begin{tcolorbox}
\centering
$\cL=\cL_{\text{SM}_4}+\frac{C_{He,ij}}{\Lambda^2}\(iH^\dagger \overleftrightarrow D_\mu H\)\overline{e_{i,R}}\gamma^\mu e_{j,R}+\(\frac{C_{eH,ij}}{\Lambda^2}\overline{L_{i,L}} He_{j,R}\abs{H}^2+\text{h.c. }\)$\\
\vspace{10pt}
preserves CP at $\cO(1/\Lambda^2$) iff
 \beq
J_4=\Im\Tr\(C_{\smash{eH}}^{\vphantom{\dagger}} Y_e^\dagger\)=\Im\Tr\(X_e^{\vphantom{\dagger}}C_{\smash{eH}}^{\vphantom{\dagger}} Y_e^\dagger\)=\Im\Tr\(X_e^{\smash{2}\vphantom{\dagger}}C_{\smash{eH}}^{\vphantom{\dagger}} Y_e^\dagger\)=0\ .
\label{minimalSetCeHExample}
\eeq
\end{tcolorbox}

Let us end this preview by saying that the above invariant sets are not unique: there are different sets of invariants which would equally well capture the necessary and sufficient conditions for CP-conservations at order $1/\Lambda^2$. Our construction of the above sets requires that the invariants have the lowest possible degree with respect to Yukawa matrices, and that they be as large as possible when the observed values of the fermion masses and CKM entries are plugged in.

\subsection{Expected ranks when $J_4=0$}\label{section:expectedRanks}

We now discuss the rank of the transfer matrices, related to the validity of the minimal sets presented above, in non-generic cases, namely when some fermion masses are degenerate and/or when there are texture zeros in the CKM matrix (which happens when some mixing angles take special values).

To apply consistently the definition of minimal sets from section~\ref{section:dim6CPconservation}, we need to carefully determine how many new primary sources of CPV there are when $J_4=0$, or equivalently what is the expected rank of the transfer matrix $\mathcal{T}$, irrespective of how we take $J_4\to 0$. In our generic situation of the previous section, $J_4\to 0$ meant $\delta_\CKM\to 0$ in Eq.\,\eqref{paramCKM} while holding all mixing angles different from $0,\pi/2$ and the quark masses non-degenerate. This ensures that there are no texture zeros in the CKM matrix, so that the number of CP-breaking quantities were identified with the number of imaginary parts (in the quark sector). However, that is not the only situation captured by the ambiguous ``$J_4=0$'' condition, as we anticipated in section~\ref{section:dim6CPconservation}. Indeed, mass degeneracies and/or texture zeros in the CKM matrix may lead to a conserved flavor symmetry of the SM$_4$ Lagrangian larger than $U(1)_B$, which has an impact on the number of CP-odd quantities at order $1/\Lambda^2$.

The reason is identical to that discussed in section~\ref{section:dim6NumberPhysical}: observables should be invariant under any symmetry of the spectrum of asymptotic states. Consequently, at order $1/\Lambda^2$, SMEFT coefficients must combine with SM$_4$ parameters to form invariant objects, and in particular, when the SM$_4$ Lagrangian has a flavor symmetry (which is therefore part of the symmetry group of the spectrum), only SMEFT coefficients invariant under this flavor symmetry can generate amplitudes which interfere with SM$_4$ ones.

In the generic case of a CP-preserving SM$_4$ Lagrangian ({\it i.e.}~taking $\delta_\CKM\to 0$ in Eq.\,\eqref{paramCKM} with generic values of mixing angles and quark masses, or said differently for a real CKM matrix without texture zeros), there is no flavor symmetry beyond the baryon and lepton numbers $U(1)_B\times U(1)_{L_i}$. Therefore, any additional $B$- and $L_i$-preserving SMEFT coupling in the Lagrangian is primary, and its imaginary part is a primary source of CPV\@. We leave to appendix~\ref{appendix:generalCP} the systematic discussion of all flavor symmetries of the SM$_4$ and their relation to mass degeneracies and/or texture zeros in the CKM matrix, but, for the sake of illustration, we discuss here two specific cases. 

In the first one, the CKM matrix is non-generic and has texture zeros:
\beq
V_\CKM= \left(
\begin{array}{ccc}
 * & 0 & * \\
 0 & * & 0 \\
 * & 0 & * \\
\end{array}
\right) \ .
\label{CKMs23s120}
\eeq
This texture can be achieved from Eq.\,\eqref{paramCKM} by taking $s_{23}=s_{12}=0$ (in particular, $J_4=0$ for such a texture). The flavor symmetry of the dimension-four action is now enlarged to $U(1)^2$, corresponding to two independent abelian transformations of $(Q_1,Q_3,u_1,u_3,d_1,d_3)$ and $(Q_2,u_2,d_2)$ respectively (in the {\it up} or {\it down basis}), acting as follows:
\beq
u_{i,R}\to e^{i\xi_{q_i}}u_{i,R} \ ,
\label{CPdim4s23s120}
\eeq
with $\xi_{q_1}=\xi_{q_3}$, and similarly for other quarks.

The second example is that of a degenerate fermion spectrum. We take for definiteness $m_t=m_c$ (which again implies $J_4=0$). With this degeneracy, the symmetry of the mass terms becomes non-abelian, while the phase in the CKM matrix, as well as one mixing angle, is no longer physical. Indeed, we can perform a flavor transformation (here in the {\it up basis}),
\beq
Q_L \to  \bmat
 1 & 0 & 0 \\
 0 & c_{23} & s_{23} \\
 0 & -s_{23} & c_{23} \emat Q_L \ , \quad u_R\to \bmat
 1 & 0 & 0 \\
 0 & c_{23} & s_{23} \\
 0 & -s_{23} & c_{23} \emat u_R \ ,
 \eeq
such that
\beq
V_\CKM\rightarrow \bmat c_{13} & 0 & s_{13}  \\
 0 & 1 & 0 \\
 -s_{13} & 0 & c_{13} \emat\bmat
 c_{12} & s_{12} & 0 \\
 -s_{12} & c_{12} & 0 \\
 0 & 0 & 1 \emat \ .
 \label{CKMdegenerate}
\eeq
Therefore, the values of $\theta_{23}$ and $\delta_\CKM$ have no physical impact. For generic values of $\theta_{12}$ and $\theta_{13}$, the CKM matrix in Eq.\,\eqref{CKMdegenerate} has no texture zeros, and the flavor symmetry of the SM$_4$ Lagrangian in the quark sector still corresponds to the baryon number.

All possible cases are treated similarly, and the full discussion is presented in appendix~\ref{appendix:generalCP}. The summary of this analysis (assuming no vanishing mass, see appendix~\ref{appendix:generalCP} for the more general case) is presented in Table\,\ref{tableFreeParamsCP} where, for each non-generic case of interest, we present the flavor symmetry group. The discussion in the lepton sector is similar, but features at least a $U(1)^3$ flavor freedom in the SM$_4$, since the PMNS matrix is given by the identity. When two or three charged lepton masses are degenerate, this $U(1)^3$ increases to $U(2)\times U(1)$ and $U(3)$, respectively.

\begin{table}[h!]
	\renewcommand{\arraystretch}{1.3}
	\centering
	\resizebox{1\columnwidth}{!}{%
		\begin{tabular}{ll|c}
			\multicolumn{2}{c|}{Parameter values}&\begin{minipage}{0.25\textwidth}\centering Flavor symmetries of the SM$_4$ Lagrangian
			\end{minipage}
			\\\hline
			\multirow{6}{0.25\textwidth}{$\begin{array}{l}
					m_{u}\neq m_{c}\neq m_{t}\\
					m_{d}\neq m_{s}\neq m_{b}
				\end{array}$}&$\begin{array}{l}
				\text{Generic }V_\CKM 
			\end{array}$&$U(1)_B$\\\cline{2-3}
			&$\begin{array}{ll}
				|V_{\CKM,i_0j_0}|=1\, ,&
				V_{\CKM,ij_0}=V_{\CKM,i_0j}=0\\
				i\neq i_0,\, j\neq j_0&
			\end{array}$&$U(1)^2$\\\cline{2-3}
			&$\begin{array}{l}
				|V_{\CKM,i_1j_1}|=|V_{\CKM,i_2j_2}|=|V_{\CKM,i_3j_3}|=1\quad \text{for }\begin{array}{c}
					i_1\neq i_2\neq i_3\\
					j_1\neq j_2\neq j_3
				\end{array} \\
				V_{\CKM,ij}=0 \text{ elsewhere}
			\end{array}$&$U(1)^3$\\\cline{1-3}
			\multirow{6}{0.25\textwidth}{$\begin{array}{l}
					m_{u}\neq m_{c}= m_{t}\\
					m_{d}\neq m_{s}\neq m_{b}
				\end{array}$}&$\begin{array}{l}
				\text{Generic }V_\CKM \quad (\text{see Eq.\,\eqref{CKMdegenerate}})
			\end{array}$&$U(1)_B$\\\cline{2-3}
			&$\begin{array}{ll}
				|V_{\CKM,i_0j_0}|=1\, ,&
				V_{\CKM,ij_0}=V_{\CKM,i_0j}=0\\
				i\neq i_0,\, j\neq j_0&
			\end{array}$&$U(1)^2$\\\cline{2-3}
			&$\begin{array}{l}
				|V_{\CKM,i_1j_1}|=|V_{\CKM,i_2j_2}|=|V_{\CKM,i_3j_3}|=1\quad \text{for }\begin{array}{c}
					i_1\neq i_2\neq i_3\\
					j_1\neq j_2\neq j_3
				\end{array} \\
				V_{\CKM,ij}=0 \text{ elsewhere}
			\end{array}$&$U(1)^3$\\\cline{1-3}
			\multirow{2}{0.25\textwidth}{$\begin{array}{l}
					m_{u}\neq m_{c}\neq m_{t}\\
					m_{d}= m_{s}\neq m_{b}
				\end{array}$}
			&\multirow{2}{0.5\textwidth}{$\begin{array}{l}
					\text{Same as the previous case with } V_\CKM^{\vphantom{\dagger}}\leftrightarrow  V_\CKM^\dagger
				\end{array}$}&\\
			&&\\\cline{1-3}
			\multirow{5}{0.25\textwidth}{$\begin{array}{l}
					m_{u}\neq m_{c}= m_{t}\\
					m_{d}= m_{s}\neq m_{b}
				\end{array}$}
			&$\begin{array}{l}
				\text{Generic }V_\CKM 
			\end{array}$&$U(1)^2$\\\cline{2-3}
			&$\begin{array}{l}
				|V_{\CKM,11}|=|V_{\CKM,22}|=|V_{\CKM,33}|=1\\
				V_{\CKM,ij}=0 \text{ elsewhere}
			\end{array}$&$U(1)^3$\\\cline{2-3}
			&$\begin{array}{l}
				|V_{\CKM,13}|=|V_{\CKM,22}|=|V_{\CKM,31}|=1\\
				V_{\CKM,ij}=0 \text{ elsewhere}
			\end{array}$&$U(2)\times U(1)$\\\cline{1-3}
			\multirow{3}{0.25\textwidth}{$\begin{array}{l}
					m_{u}= m_{c}= m_{t}\\
				\end{array}$}&$\begin{array}{l}
				m_{d} \neq m_{s}\neq m_{b}\end{array}$&$U(1)^3$\\\cline{2-3}
			&$\begin{array}{l}
				m_{d} = m_{s}\neq m_{b}\end{array}$&$U(2)\times U(1)$\\\cline{2-3}
			&$\begin{array}{l}
				m_{d} = m_{s} = m_{b}\end{array}$&$U(3)$\\\cline{1-3}
			\multirow{3}{0.25\textwidth}{$\begin{array}{l}
					m_{d} = m_{s} = m_{b}\,\\
				\end{array}$}&$\begin{array}{l}
				m_{u} \neq m_{c}\neq m_{t}\end{array}$&$U(1)^3$\\\cline{2-3}
			&$\begin{array}{l}
				m_{u} \neq m_{c}= m_{t}\end{array}$&$U(2)\times U(1)$\\\cline{2-3}
			&$\begin{array}{l}
				m_{u} = m_{c} = m_{t}\end{array}$&$U(3)$\\\cline{1-3}
		\end{tabular}
	}
	\caption{Flavor symmetry of the SM$_4$ Lagrangian as a function of special values for quark masses (assumed to be non-vanishing, see appendix~\ref{appendix:generalCP} for the general case) and entries of the CKM matrix. Conditions on the right are understood to be imposed on top of those on their left. Here only some of the possible combinations of mass degeneracies are treated. The other mass degeneracies lead to the same flavor symmetries provided the corresponding non-generic $V_\CKM$ are multiplied by appropriate matrices exchanging flavor labels (see footnote~\ref{foot:flavorshuffle}).}
	\label{tableFreeParamsCP}
\end{table}
This discussion allows us to work out the number of new primary sources of CPV at order $\cO(1/\Lambda^2)$, in each (non-)generic case for the CKM matrix. Let us focus again on the two previous examples, which generalize easily to all.

When the baryon or lepton numbers are the only flavor symmetries at dimension-four (in the {\it up} or {\it down basis}), all imaginary parts of the Wilson coefficients at dimension-six in the quark sector (and in the lepton sector, all imaginary parts which are not charged under the lepton numbers) can interfere with the SM$_4$\@. Instead, when the flavor symmetry increases to $U(1)^2$, several SMEFT coefficients become secondary. For instance, $C_{Hu}$ transforms as (in the {\it up} or {\it down bases})
\beq
C_{Hu,ij}\to e^{i\(\xi_{q_j}-\xi_{q_i}\)}C_{Hu,ij} \ ,
\eeq
where $\xi_{q_1}=\xi_{q_3}$ and arbitrary $\xi_{q_2}$ for the texture of Eq.\,\eqref{CKMs23s120}, and similarly for other textures. Therefore, for the texture of Eq.\,\eqref{CKMs23s120}, only $C_{Hu,13}$ is primary and $C_{Hu}$ only provides one primary source of CPV, $\Im \, C_{Hu,13}$.\footnote{One can construct non-linear invariant quantities from $C_{Hu,12/23}$, an example being $C_{Hu,12}C_{Hu,23}^*$. At leading order, however, $C_{Hu,12/23}$ cannot contribute linearly to observables.} Thus, in this case, a single invariant in the minimal set for $C_{Hu}$ is sufficient.

This exercise can be performed for all non-generic cases for the CKM matrix and for all Wilson coefficients. This results in a set of conditions for CP conservation at leading order, whose number is in one-to-one correspondence to the number of independent CP-odd invariants in a minimal set. As we just saw, those numbers depend on the flavor symmetry of the SM$_4$ Lagrangian, and are given for all SMEFT operators in Table\,\ref{tableRanksSMEFT}.
\begin{table}[h!]
\small
\centering
\renewcommand{\arraystretch}{1.3}
\resizebox{\columnwidth}{!}{%
\begin{tabular}{c|c|c|c|c|c|c|c|c|c|c|c}
&\multicolumn{4}{c|}{Bilinears}&\multicolumn{7}{c}{4-Fermi}\\\cline{1-12}
$\begin{matrix} \text{Flavour symmetries}\\\text{of the quark sector of the SM}\end{matrix}$&$\begin{matrix} C_{eH}\\C_{eW}\\C_{eB}\end{matrix}$&$\begin{matrix} C_{uH}\\C_{uG}\\C_{uW}\\C_{uB}\\C_{dH}\\C_{dG}\\C_{dW}\\C_{dB}\\C_{Hud}\end{matrix}$&$\begin{matrix} C_{HL}^{1,3}\\C_{He}\end{matrix} $&$\begin{matrix} C_{HQ}^{1,3}\\C_{Hu}\\C_{Hd}\end{matrix}$&$\begin{matrix} C_{LL}\\C_{ee}\end{matrix}$&$C_{Le}$&$\begin{matrix} C_{QQ}^{1,3}\\C_{uu}\\C_{dd}\end{matrix}$&$\begin{matrix} C_{LQ}^{1,3}\\C_{Qe}\\C_{Lu}\\C_{eu}\\C_{Ld}\\C_{ed}\end{matrix}$&$\begin{matrix} C_{ud}^{1,8}\\C_{Qu}^{1,8}\\C_{Qd}^{1,8}\end{matrix}$&$\begin{matrix} C_{LedQ}\\C_{LeQu}^{1,3}\end{matrix}$&$C_{QuQd}^{1,8}$\\
\hline
$U(1)_B$&3&9&0&3&0&3&18&9&36&27&81\\
$U(1)^2$&3&5&0&1&0&3&5&3&12&15&33\\
$U(1)^3$&3&3&0&0&0&3&0&0&3&9&15\\
$U(2)\times U(1)$&3&2&0&0&0&3&0&0&1&6&7\\
$U(3)$&3&1&0&0&0&3&0&0&0&3&2\\
\hline
Two degenerate electron-type leptons&$\times \frac{2}{3}$&$\times 1$&
&$\times 1$&
&$\times \frac{2}{3}$&$\times 1$&$\times \frac{2}{3}$&$\times 1$&$\times \frac{2}{3}$&$\times 1$\\
All electron-type leptons degenerate&$\times \frac{1}{3}$&$\times 1$&
&$\times 1$&
&$\times \frac{1}{3}$&$\times 1$&$\times \frac{1}{3}$&$\times 1$&$\times \frac{1}{3}$&$\times 1$\\
\end{tabular}
}
\caption{Numbers of new primary sources of CPV contained in each dimension-six SMEFT coefficient, for each of the possible flavor groups of the quark sector of the SM$_4$ at dimension-4 (restricting to situations where fermion masses are non-vanishing). The last two rows indicate which multiplicative coefficient should be applied to all numbers of the same column for special values of the electron-type lepton masses.}
\label{tableRanksSMEFT}
\end{table}

We can now come back to the statement that the set of invariants in Eq.\,\eqref{wrongSetCHu} is not a satisfying one for $C_{Hu}$. As seen in Table\,\ref{tableRanksSMEFT}, all its off-diagonal entries are primary when $m_t=m_c$, and all their imaginary parts violate CP (in an appropriate basis), hence we need three independent invariants to capture the conditions for CP conservation. 

Let us stress again at this point that the fact that we found a set of invariants of minimal size ({\it i.e.}~three invariants for the case of $C_{Hu}$) which captures the necessary and sufficient conditions for CP conservation in all non-generic cases listed in Table\,\ref{tableFreeParamsCP} is a non-trivial result. Nevertheless, it turns out that it can be done for all SMEFT coefficients at dimension-six, as we explicitly showed.

\section{Conclusions and future directions}\label{section:conclusion}

In this paper, we have investigated the collective properties of SMEFT at dimension-six. Their first implication which we have discussed is that only a subset of Lagrangian parameters (dubbed primary parameters) can contribute linearly to observables upon interfering with the dimension-four SM$_4$ amplitude. This is due to the existence of flavor transformations which leave the SM spectrum and thus any observable invariant, thereby demanding that covariant Lagrangian parameters combine to form invariant objects. This applies for instance to field-rephasings associated to mass-eigenstates, implying that Lagrangian coefficients must combine into rephasing-invariant objects. Associated to the SMEFT power counting, this implies that several coefficients related to dimension-six SMEFT operators cannot contribute to observables at order $1/\Lambda^2$, since they are charged under individual lepton numbers unlike all SM$_4$ parameters (in the limit of zero neutrino mass). We therefore refined the usual counting of dimension-six SMEFT parameters so as to include this effect, which resulted in the counting of primary parameters in Table\,\ref{tablephysicaldim6}.

Then, we focused on collective effects related to new sources of CP violation in the dimension-six SMEFT, both of which are captured thanks to CP-odd flavor-invariants. To respect the SMEFT power counting, we restricted to invariants linear in dimension-six coefficients, and we presented minimal sets of invariants which map in a one-to-one basis to all new primary sources of CPV\@. We proved this by showing that the points in parameter space where CP is conserved (at leading BSM order) are exactly the points where our new invariants vanish. This holds for all parts of the SM$_4$ parameter space, including degenerate cases. 
A complete list of CP-odd linear invariants can be found in Appendices \ref{appendix:bilinears} and \ref{appendix:4Fermi}\@. We remind the reader that this list is not unique.

There are several directions along which our work can be extended~\cite{followup1}. 
The first one concerns parameterizations of leading order CP-odd observables. Indeed, although our characterization of new primary sources of CPV is complete, it does not necessarily describe all the relevant parameter space of those observables. The reason is that, CP being broken at dimension-four already, the source of its breaking in the SM$_4$ can interfere with CP-even dimension-six parameters of SMEFT\@. On the other hand, remembering that physics should not depend on a specific field basis, a CPV observable can always be expressed in terms of CPV flavor-invariants. Taking all this into account suggests that one should study larger sets of CP-odd invariants than our minimal ones. In such sets, only a subset of invariants capture new primary sources of CPV, but each invariant captures a new, independent primary dimension-six quantity which corresponds to, or can interfere with, a source of CPV\@. We call such a set a {\it maximal set}. Given an invariant $\cI$ in such a set, one can always write
\beq
\cI=\sum_i c_i \cI_i + J_4 \tilde \cI \ ,
\label{maximalSetEq}
\eeq
where $\{\cI_i\}_i$ forms a minimal set in the sense described in this paper, and $\tilde\cI$ is a CP-even expression built as a ratio of polynomial flavor-invariants. Indeed, given our definition of minimal sets, all CPV observables (in particular $\cI$) must vanish at order $\cO(1/\Lambda^2)$ when $J_4=\cI_i= 0$. Nevertheless, one can check that all (CP-odd or -even) primary parameters can be captured by CP-odd invariants, thanks to a rank analysis similar to that presented in section~\ref{section:dim6CPconservation}, and consistent with the fact that all primary parameters can contribute to CPV observables, either directly or via interference with the SM$_4$ CP phase. 

Relatedly, one can numerically evaluate the invariants, which encode collective effects and the suppression they induce, using the observed values of the SM$_4$ parameters. This illustrates how accidentally small the absolute strength of each new source of CPV is, or if there are hierarchies among them, {\it etc.} This in turn explains (part of) the suppression in CPV observables \cite{Khriplovich:1993js,Pospelov:1994uf,Smith:2017dtz,Kley:2021yhn}. One can also similarly probe the effects of specific UV hypotheses on the SMEFT coefficients, such as textures derived from flavor symmetries.  A given UV model may have its own CP-odd flavor invariants, and these can be matched onto our IR invariants, inducing possible correlations amongst the IR invariants.  In other directions, it would be interesting to consider other operator bases than the Warsaw basis we use in this work, and to check which expressions our invariants map to, given that the overall number of independent sources of CPV must be conserved. One could also extend our construction beyond $\cO(1/\Lambda^2)$, {\it e.g.}~to capture squared dimension-six or interfering dimension-eight SMEFT contributions.  Finally, one could consider the RG evolution of invariants \cite{Wang:2021wdq}. We anticipate that flavor invariants, at dimension-six and beyond, are essential tools for illuminating the rich CP structure of SMEFT.
 
\acknowledgments
We thank Luca di Luzio, Brian Henning, Luca Merlo, Davide Pagani, Giuliano Panico, Pablo Quilez, Ken van Tilburg, and Neal Weiner for interesting comments and discussions. We wish to thank Jonathan Kley and Oleg Lebedev for comments on the manuscript. This work is supported by the Deutsche Forschungsgemeinschaft under Germany's Excellence Strategy  EXC 2121 ``Quantum Universe'' - 390833306. JTR is supported by the NSF CAREER grant PHY-1554858 and NSF grant PHY-19154099, and by an award from the Alexander von Humboldt Foundation.

\appendix

\section{Flavor symmetries of the SM$_4$}\label{appendix:generalCP}

In this appendix, we present the details behind Table\,\ref{tableFreeParamsCP}, and identify the possible flavor symmetries of the SM$_4$ Lagrangian in terms of textures in the CKM matrix. We remind that the flavor symmetries acts on the quark fields as follows:
\be
u_R\to U_uu_R \ ,
\label{genericFlavor}
\ee
where $U_{u}\in U(3)_u$, and similarly for all other fermionic fields $d,Q,L,e$. Which matrices $U$ lead to genuine symmetries of the SM$_4$ Lagrangian depends on the values of the masses and of the entries of the CKM matrix. In all cases of non-trivial flavor symmetries, we find that $J_4=0$, so that there exists at least one combination of CP and flavor symmetries which yield a symmetry of the SM$_4$ Lagrangian in any basis. 

In what follows, we work for definiteness in the {\it up basis} of Eq.\,\eqref{upBasis}.

\subsection{Non-vanishing quark masses}

The condition for the flavor invariance of $Y_u$ and $Y_d$ reads
\beq
Y_u^{\vphantom{\dagger}}=U_Q^\dagger Y_u^{\vphantom{\dagger}}U_u^{\vphantom{\dagger}} \ , \quad Y_d^{\vphantom{\dagger}}=U_Q^\dagger Y_d^{\vphantom{\dagger}}U_d^{\vphantom{\dagger}} \ .
\label{flavorInvariance}
\eeq
For non-vanishing quark masses, the Yukawa matrices are full rank and one can use Eq.\,\eqref{flavorInvariance} to solve for $U_{u,d}$ as a function of $U_Q$ and the Yukawa matrices:
\beq
U_u^{\vphantom{\dagger}}=Y_u^{-1\vphantom{\dagger}}U_Q^{\vphantom{\dagger}}Y_u^{\vphantom{\dagger}} \ , \quad U_d^{\vphantom{\dagger}}=Y_d^{-1\vphantom{\dagger}}U_Q^{\vphantom{\dagger}}Y_d^{\vphantom{\dagger}} \ .
\eeq
Therefore, only one matrix determines the two others, and the flavor symmetry group is at most $U(3)$. Imposing that $U_u^\dagger U_u^{\vphantom{\dagger}}=U_d^\dagger U_d^{\vphantom{\dagger}}
=\mathbb{1}$ implies (in the {\it up basis}) that
\beq
\[U_Q^{\vphantom{\dagger}},m_u^{\vphantom{\dagger}}m_u^\dagger\]=\[V_\CKM^\dagger U_Q^{\vphantom{\dagger}}V_\CKM^{\vphantom{\dagger}},m_d^{\vphantom{\dagger}}m_d^\dagger\]=0 \ ,
\eeq
where $m_{u/d}\equiv\text{diag}(m_{u/d_i})$. In the {\it up basis}, quark masses are positive and real, therefore
\beq
\[U_Q^{\vphantom{\dagger}},m_u^{\vphantom{\dagger}}\]=\[V_\CKM^\dagger U_Q^{\vphantom{\dagger}}V_\CKM^{\vphantom{\dagger}},m_d^{\vphantom{\dagger}}\]=0
\label{commutationMasses}
\eeq
and (using the explicit expression of the Yukawa matrices in the {\it up basis})
\beq
U_u^{\vphantom{\dagger}}=U_Q^{\vphantom{\dagger}} \ , \quad U_d^{\vphantom{\dagger}}=V_\CKM^\dagger U_Q^{\vphantom{\dagger}}V_\CKM^{\vphantom{\dagger}} \ .
\eeq
The commutation relations in Eq.\,\eqref{commutationMasses} are additional constraints to fulfil which depend on the spectrum, as we now explore.

\subsubsection{Non-degenerate quark masses}

If all up-type quarks are non-degenerate, the first condition in Eq.\,\eqref{commutationMasses} implies that $U_Q^{\vphantom{\dagger}}=\text{diag}\(e^{i\xi_i}\)$ and the second that $U_Q^{\vphantom{\dagger}}=V_\CKM^{\vphantom{\dagger}} \text{diag}\(e^{i\tilde \xi_i}\)V_\CKM^\dagger$, therefore
\bes
V_{\CKM,ij}^{\vphantom{\dagger}}=e^{i\(\xi_i-\tilde\xi_j\)} V_{\CKM,ij}^{\vphantom{\dagger}} \ .
\ees 
Consequently, all $\xi$'s are equal and equal to the $\tilde\xi$'s ({\it i.e.}~the flavor symmetry is given by the baryon number $U(1)_B$) unless the CKM matrix has some vanishing entries. For instance, one finds a $U(1)^2$ flavor symmetry when the CKM matrix has the following texture:
\beq
V_\CKM=\bmat * & 0 & 0 \\ 0 &* &* \\ 0 &* &*\emat  \ ,
\eeq
corresponding to the constraints $\xi_2=\xi_3=\tilde\xi_2=\tilde\xi_3$, for arbitrary $\xi_1=\tilde\xi_1$. More generally, a $U(1)^2$ flavor symmetry is obtained for any texture such that, given two integers $(i_0,j_0)$, $\abs{V_{\CKM,i_0j_0}}=1$ and $V_{\CKM,ij}=0$ for $i=i_0$ or $j=j_0$. By comparing with the explicit parametrization in Eq.\,\eqref{paramCKM}, one finds that a mixing angle has to be equal to $0$ or $\pi/2$ in all those cases, hence $J_4=0$ and there exists a basis where all SM$_4$ couplings are real.

A flavor symmetry $U(1)^3$ is obtained for all textures of $V_\CKM$ such that there is a single number of unit modulus in each row and column, such as {\it e.g.}~$V_\CKM=\mathbb{1}$ or
\beq
V_\CKM=\bmat * & 0 & 0 \\ 0 &0 &* \\ 0 &*&0 \emat \ .
\eeq

\subsubsection{Degenerate quark masses}

In  cases with quark mass degeneracies, $J_4=0$ automatically and there exists a basis where all SM$_4$ couplings are real.

\paragraph{$\mathbf{m_t=m_c}$}

Let us start with the case of two degenerate quarks of the same type, which we take to be $m_t=m_c$ for definiteness, all other masses being non-degenerate.\footnote{The equivalent case where down quark masses are degenerate is treated identically after the exchange $V_\CKM^{\vphantom{\dagger}}\leftrightarrow V_\CKM^\dagger$. We also consider the specific case $m_t=m_c$ (and $m_s=m_d$ in the case of down quarks later on), since the formulae are simpler given our parametrization of $V_\CKM$. Nevertheless, the discussion (and the parametrization) can be adapted to any other quark mass degeneracy. In particular, the remarkable textures $V_\CKM^{(ji)}$ leading to a given flavor symmetry when $m_{u_j}=m_{u_i},i<j$ are related to those when $m_t=m_c$ by
\bes
V_\CKM^{\smash{(ji)}\vphantom{\dagger}}=R_{i2}^{\vphantom{\dagger}}R_{j3}^{\vphantom{\dagger}}V_\CKM^{\smash{(tc)}\vphantom{\dagger}} \ ,
\ees
where $R_{ab}$ is the matrix which exchanges rows (or columns) $a$ and $b$. Similarly, textures obtained when $m_{d_j}=m_{d_i},i<j$ are related to those when $m_s=m_d$ by
\bes
V_\CKM^{\smash{(ji)}\vphantom{\dagger}}=V_\CKM^{\smash{(sd)}\vphantom{\dagger}}R_{i1}^{\vphantom{\dagger}}R_{j2}^{\vphantom{\dagger}} \ .
\ees
For instance, the texture which leads to a flavor symmetry $U(2)\times U(1)$ (see below) for $m_{u_j}=m_{u_i},i<j$ and $m_{d_l}=m_{d_k},k<l$ is
\bes
V_\CKM =R_{i2}R_{j3}\cdot \bmat 0 & 0 & * \\ 0 &* &0 \\ * &0 &0\emat \cdot R_{k1}R_{l2} \ .
\ees
\label{foot:flavorshuffle}}
 
The first relation in Eq.\,\eqref{commutationMasses} implies that 
\beq
U_Q=\bmat e^{i\xi_1}&0\\0&U_{u}^{(2)}\emat \text{ with } U_{u}^{(2)}\in U(2) \ ,
\eeq
while the second implies that $U_Q=V_\CKM e^{i\tilde\xi_i}V_\CKM^\dagger$, therefore
\beq
V_{\CKM}^{\vphantom{\dagger}}=\bmat e^{i\xi_1}&0\\0&U_{u}^{(2)}\emat\cdot V_{\CKM}^{\vphantom{\dagger}}\cdot\text{diag}\(e^{-i\tilde\xi_i}\) \ .
\eeq 
Upon solving this equation, one finds that, similarly to the case of non-degenerate masses, one can only obtain the flavor groups $U(1)_B$, $U(1)^2$, and $U(1)^3$. Flavor symmetries beyond baryon numbers are obtained for the textures discussed above, but the generic case for the CKM matrix when $m_t=m_c$ is given in Eq.\,\eqref{CKMdegenerate} and does not have a specific texture.

\paragraph{$\mathbf{m_t=m_c}$ and $\mathbf{m_s=m_d}$}

Let us now turn to the case where $m_t=m_c$ and $m_s=m_d$, all other masses being non-degenerate. The relations in Eq.\,\eqref{commutationMasses} imply that
 \beq
U_Q=\bmat e^{i\xi_1}&0\\0&U_{u}^{(2)}\emat=V_\CKM^{\vphantom{\dagger}}\cdot \bmat U_{d}^{(2)}&0\\0&e^{i\tilde\xi_3}\emat\cdot V_\CKM^\dagger \text{ with } U_{u}^{(2)},U_{d}^{(2)}\in U(2) \ ,
\eeq
hence
\beq
V_\CKM^{\vphantom{\dagger}}=\bmat e^{i\xi_1}&0\\0&U_{u}^{(2)}\emat\cdot V_{\CKM}^{\vphantom{\dagger}}\cdot \bmat U_{d}^{(2)}{}^\dagger&0\\0&e^{-i\tilde\xi_3}\emat \ .
\label{constraint2MassesDegen}
\eeq 
Starting from the CKM matrix in Eq.\,\eqref{CKMdegenerate}, which is generic when $m_t=m_c$, we can further rotate $\theta_{12}$ away by performing
\be
 d_R\rightarrow \bmat c_{12}&-s_{12}&0\\s_{12}&c_{12}&0\\0&0&1\emat d_R \ .
\ee
Recall that we work in the {\it up basis}, where $V_\CKM$ appears in $Y_d$ and can be affected by right-handed down-quark flavor transformations which commute with the down-quark mass matrices. We then obtain
\be
V_\CKM\rightarrow \bmat c_{13} & 0 & s_{13}  \\
 0 & 1 & 0 \\
 -s_{13} & 0 & c_{13} \emat \ , \text{ {\it i.e.}~with a texture } \bmat * & 0 & * \\ 0 &* &0 \\ * &0 &*\emat \ .
\ee
Therefore, the generic CKM matrix for the present case has a texture which allows for a flavor symmetry at least as large as $U(1)^2$. Possible larger flavor symmetries are $U(1)^3$ or $U(2)\times U(1)$, obtained for the following respective textures,\footnote{Let us present some details regarding the second case to illustrate the derivation. With such a texture, one can phase-rotate the fields so as to get
\bes
V_\CKM=\bmat 0 & 0 & 1 \\ 0 &1 &0 \\ 1&0 &0\emat \ ,
\ees
and one finds from Eq.\,\eqref{constraint2MassesDegen} that
\bes
\mathbb{1}=\bmat OU_{u}^{(2)}OU_{d}^{(2)}{}^\dagger&0\\0&e^{i\(\xi_1-\tilde\xi_3\)} \emat \ ,
\ees 
with $O\equiv\bmat0&1\\1&0\emat$. Therefore, one obtains $U_d^{\smash{(2)}\vphantom{\dagger}}=OU_{\vphantom{d}u}^{\vphantom{\dagger}\smash{(2)}}O,\tilde\xi_3=\xi_1$ and no further constraint, hence the group is $U(2)\times U(1)$.}
\beq
\bmat * & 0 & 0 \\ 0 &* &0 \\ 0 &0 &*\emat \text{ and } \bmat 0 & 0 & * \\ 0 &* &0 \\ * &0 &0\emat \ .
\eeq

\paragraph{$\mathbf{m_u=m_c=m_t}$}

When the degeneracy is maximal, the CKM matrix can be fully absorbed by a redefinition of the RH up-quarks:
\be
u_R\rightarrow V_{\CKM}u_R \ , \quad V_\CKM \rightarrow \mathbb{1} \ .
\ee
The flavor symmetry is therefore at least as large as $U(1)^3$. With such a CKM matrix, one gets that $U_Q=U_u=U_d$ and the flavor group is determined by the relations in Eq.\,\eqref{commutationMasses}: the flavor symmetry group is $U(1)^3$, $U(2)\times U(1)$, or $U(3)$, respectively when no, two, or three down-quark masses are degenerate. 

\subsection{Vanishing masses at dimension-four}\label{vanishingMassesAppendix}

When some masses vanish at dimension-four, the flavor symmetry can contain axial phases. This case complicates the power-counting, since dimension-six Yukawa couplings now yield the leading contribution to the masses, but it can be treated as suggested in section~\ref{section:dim6NumberPhysical} for the case of neutrino masses or vanishing $Y_d$.

Whenever a quark mass goes to zero, the flavor symmetry is enlarged since the LH and RH components now describe independent particles. A flavor symmetry $U(1)_B$ would then be ugraded to $U(1)_B\times U(1)_{u_{R,1}}$ when $m_u\to 0$, all other parameters being kept fixed. For non-degenerate quark masses, the flavor symmetry is always abelian, and taking one mass to zero simply adds a RH $U(1)$ factor to the flavor symmetry, as discussed just above. On the other hand, when two or three masses are degenerate, taking them to zero together adds a RH factor $U(2)$ or $U(3)$ to the flavor symmetry (even if it was abelian for non-zero masses due to some structure in the CKM matrix which distinguishes the different flavors of LH quarks). We list in Table\,\ref{tableFreeParamsCPVanishingMasses} the relevant cases, and present in Table\,\ref{tableRanksSMEFTMore} the associated numbers of new primary sources of CPV\footnote{The counting is performed as in Section~\ref{section:expectedRanks}, i.e. one counts how many complex linear invariants (under the SM$_4$ flavor symmetry) there are. Let us give some examples. In the case of a symmetry $U(1)^3\times U(3)_{u_R}$ and focussing on $C_{Hu}$, one finds that the only linear invariant reads $\delta^{ij}C_{Hu,ij}$ (summed over $i,j$). However, $C_{Hu}$ being hermitian, this combination is real in all bases and does not violate CP. In the case of a symmetry $U(1)^2\times U(1)_{u_R}$, the $U(1)^2$ factor indicates that, in some appropriate flavor basis, the SM$_4$ Lagrangian possesses two independent quark number symmetries (one of which is the usual baryon number), singling out a quark flavor. Assuming without loss of generality that this flavor is the third one and focussing on $C_{QuQd}$, linear $U(1)^2$-invariants read $C_{QuQd,ijkl}$, $C_{QuQd,33ij}$, $C_{QuQd,3ji3}$, $C_{QuQd,i33j}$, $C_{QuQd,ij33}$, $C_{QuQd,3333}$ (where $i,j,k,l=1,2$), which are $33$ complex coefficients. Then, one needs to know whether the massless quark is the one which is singled out by the quark number, i.e. $u_{R,3}$ (otherwise, without loss of generality we take the massless quark to be the first flavor). If so one must discard $C_{QuQd,33ij}$, $C_{QuQd,i33j}$ (reducing the $33$ coefficients to $25$), if not one must instead discard $C_{QuQd,i1kl}$, $C_{QuQd,31i3}$, $C_{QuQd,i133}$ (reducing to $21$ coefficients).}\@. 
\begin{table}[h!]
\centering
\begin{tabular}{l|c}
If in addition to the values in Table\,\ref{tableFreeParamsCP}:&Add to the flavor group the factor:\\
\hline
$m_u=0$& $U(1)_{u_R}$\\
$m_u=m_d=0$&$U(1)_{u_R}\times U(1)_{d_R}$\\
$m_u=m_c=0$&$U(2)_{u_R
}$\\
$m_u=m_c=m_d=0$&$U(2)_{u_R
}\times U(1)_{d_R}$\\
$m_u=m_c=m_d=m_s=0$&$U(2)_{u_R
}\times U(2)_{d_R
}$\\
$m_u=m_c=m_t=0$&$U(3)_{u_R}$\\
$m_u=m_c=m_t=m_d=0$&$U(3)_{u_R}\times U(1)_{d_R}$\\
$m_u=m_c=m_t=m_d=m_s=0$&$U(3)_{u_R}\times U(2)_{d_R
}$\\
$m_u=m_c=m_t=m_d=m_s=m_b=0$&$U(3)_{u_R}\times U(3)_{d_R}$
\end{tabular}
\caption{Additional factor to the flavor symmetry of the SM$_4$ Lagrangian when quark masses vanish.}
\label{tableFreeParamsCPVanishingMasses}
\end{table}
\begin{table}[h!]
\small
\hspace*{-1cm}
\centering
\renewcommand{\arraystretch}{1.3}
\resizebox{1.1\columnwidth}{!}{
\begin{tabular}{l|c|c|c|c|c|c|c|c|c|c|c}
\centering
&\multicolumn{4}{c|}{Bilinears}&\multicolumn{7}{c}{4-Fermi}\\\cline{1-12}
$\begin{matrix} \text{Flavour symmetries}\\\text{of the quark sector of the SM}\end{matrix}$&$\begin{matrix} C_{eH}\\C_{eW}\\C_{eB}\end{matrix}$&$\begin{matrix} C_{uH}\\C_{uG}\\C_{uW}\\C_{uB}\\C_{dH}\\C_{dG}\\C_{dW}\\C_{dB}\\C_{Hud}\end{matrix}$&$\begin{matrix} C_{HL}^{1,3}\\C_{He}\end{matrix} $&$\begin{matrix} C_{HQ}^{1,3}\\C_{Hu}\\C_{Hd}\end{matrix}$&$\begin{matrix} C_{LL}\\C_{ee}\end{matrix}$&$C_{Le}$&$\begin{matrix} C_{QQ}^{1,3}\\C_{uu}\\C_{dd}\end{matrix}$&$\begin{matrix} C_{LQ}^{1,3}\\C_{Qe}\\C_{Lu}\\C_{eu}\\C_{Ld}\\C_{ed}\end{matrix}$&$\begin{matrix} C_{ud}^{1,8}\\C_{Qu}^{1,8}\\C_{Qd}^{1,8}\end{matrix}$&$\begin{matrix} C_{LedQ}\\C_{LeQu}^{1,3}\end{matrix}$&$C_{QuQd}^{1,8}$\\
\hline
$U(1)_B$&3&9&0&3&0&3&18&9&36&27&81\\
$U(1)_B\times U(1)_{u_R}$&3&$6^4,9^4,6$&0&3,1,3&0&3&18,5,18&$9^2,3^2,9^2$&$18^2,36$&27,18&54\\
$U(1)_B\times U(1)_{u_R}\times U(1)_{d_R}$&3&$6^8,4$&0&3,$1^2$&0&3&18,$6^2$&$9^2,3^4$&$8,18^2$&$18^2$&36\\
$U(1)_B\times U(2)_{u_R}$&3&$3^4,9^4,3$&0&3,0,3&0&3&18,0,18&$9^2,0^2,9^2$&$6^2,36$&27,9&27\\
$U(1)_B\times U(2)_{u_R}\times U(1)_{d_R}$&3&$3^4,6^4,2$&0&3,0,1&0&3&18,0,6&$9^2,0^2,3^2$&$2,6,18$&18,9&18\\
\hline
$U(1)^2$&3&5&0&1&0&3&5&3&12&15&33\\
$U(1)^2\times U(1)_{u_R}$&3&(3 or 4)$^4,5^4$,3 or 4&0&1,0 or 1,1&0&3&5,0 or 5,5&3$^2$,(0 or 3)$^2$,3$^2$&(5 or 8)$^2$,12&15,9 or 12&21 or 24\\
$U(1)^2\times U(1)_{u_R}\times U(1)_{d_R}$&3&(3 or 4)$^8$,2 or 4&0&1,(0 or 1)$^2$&0&3&5,(0 or 5)$^2$&3$^2$,(0 or 3)$^4$&1 or 3 or 8,(5 or 8)$^2$&9 or 12&12 or 13 or 16\\
$U(1)^2\times U(2)_{u_R}$&3&(1 or 2)$^4$,5,1 or 2&0&1,0,1&0&3&5,0,5&$3^2,0^2,3^2$&2$^2$,12&15,3 or 6&12\\
$U(1)^2\times U(2)_{u_R}\times U(1)_{d_R}$&3&(1 or 2)$^4$,(3 or 4)$^4$,0 or 1 or 2&0&1,0,0 or 1&0&3&5,0,0 or 5&$3^2,0^2$,(0 or 3)$^2$&0 or 2,2,5 or 8&9 or 12,3&8\\
$U(1)^2\times U(2)_{u_R}\times U(2)_{d_R}$&3&(1 or 2)$^8$,0 or 1&0&1,0$^2$&0&3&5,0$^2$&$3^2,0^4$&0,2$^2$&3 or 6&4\\
\hline
$U(1)^3$&3&3&0&0&0&3&0&0&3&9&15\\
$U(1)^3\times U(1)_{u_R}$&3&$2^4,3^4,2$&0&0&0&3&0&0&$1^2,3$&9,6&10\\
$U(1)^3 \times U(1)_{u_R}\times U(1)_{d_R}$&3&$2^8$,1 or 2&0&0&0&3&0&0&0 or 1,$1^2$&6&7 or 8\\
$U(1)^3\times U(2)_{u_R}$&3&$1^4,3^4,1$&0&0&0&3&0&0&$0^2,3$&9,3&5\\
$U(1)^3\times U(2)_{u_R}\times U(1)_{d_R}$&3&$1^4,2^4$,0 or 1&0&0&0&3&0&0&$0^2,1$&6,3&3 or 4\\
$U(1)^3\times U(2)_{u_R}\times U(2)_{d_R}$&3&$1^8$,0 or 1&0&0&0&3&0&0&0&3&2\\
$U(1)^3\times U(3)_{u_R}$&3&$0^4,3^4,0$&0&0&0&3&0&0&$0^2,3$&9,0&0\\
$U(1)^3\times U(3)_{u_R}\times U(1)_{d_R}$&3&$0^4,2^4,0$&0&0&0&3&0&0&$0^2,1$&6,0&0\\
\hline
$U(2)\times U(1)$&3&2&0&0&0&3&0&0&1&6&7\\
$U(2)\times U(1)\times U(1)_{u_R}$&3&$1^4,2^4$,1&0&0&0&3&0&0&$0^2$,1&6,3&4\\
$U(2)\times U(1)\times U(1)_{u_R}\times U(1)_{d_R}$&3&1&0&0&0&3&0&0&0&3&1 or 2\\
$U(2)\times U(1)\times U(2)_{u_R}$&3&$1^4,2^4$,1&0&0&0&3&0&0&$0^2$,1&6,3&3\\
$U(2)\times U(1)\times U(2)_{u_R}\times U(1)_{d_R}$&3&$1^8,0$&0&0&0&3&0&0&0&3&2\\
$U(2)\times U(1)\times U(2)_{u_R}\times U(2)_{d_R}$&3&1&0&0&0&3&0&0&0&3&1\\
$U(2)\times U(1)\times U(3)_{u_R}$&3&$0^4,2^4,0$&0&0&0&3&0&0&$0^2,1$&6,0&0\\
$U(2)\times U(1)\times U(3)_{u_R}\times U(1)_{d_R}$&3&$0^4,1^4,0$&0&0&0&3&0&0&0&3,0&0\\
$U(2)\times U(1)\times U(3)_{u_R}\times U(2)_{d_R}$&3&$0^4,1^4,0$&0&0&0&3&0&0&0&3,0&0\\
\hline
$U(3)$&3&1&0&0&0&3&0&0&0&3&2\\
$U(3)\times U(3)_{u_R}$&3&$0^4,1^4,0$&0&0&0&3&0&0&0&3,0&0\\
$U(3)\times U(3)_{u_R}\times U(3)_{d_R}$&3&0&0&0&0&3&0&0&0&0&0\\
\hline
Two degenerate electron-type leptons&$\times \frac{2}{3}$&$\times 1$&
&$\times 1$&
&$\times \frac{2}{3}$&$\times 1$&$\times \frac{2}{3}$&$\times 1$&$\times \frac{2}{3}$&$\times 1$\\
All electron-type leptons degenerate&$\times \frac{1}{3}$&$\times 1$&
&$\times 1$&
&$\times \frac{1}{3}$&$\times 1$&$\times \frac{1}{3}$&$\times 1$&$\times \frac{1}{3}$&$\times 1$\\
One vanishing electron-type mass&$\times \frac{2}{3}$&$\times 1$&&$\times 1$&&$\times \frac{1}{3}$&$\times 1$&$\times 1$&$\times 1$&$\times \frac{2}{3}$&$\times 1$\\
Two vanishing electron-type masses&$\times \frac{1}{3}$&$\times 1$&&$\times 1$&&0&$\times 1$&$\times \frac{2}{3}$&$\times 1$&$\times \frac{1}{3}$&$\times 1$\\
All electron-type masses vanishing&$ 0$&$\times 1$&&$\times 1$&&$ 0$&$\times 1$&$ \times \frac{1}{3}$&$\times 1$&$0$&$\times 1$\\
\end{tabular}
}
\caption{Numbers of new primary sources of CPV contained in each dimension-six SMEFT coefficient. When a single number appears, it applies to all operators at the top of the concerned column. When several numbers are needed, they appear as a list, where the integer power refers to the multiplicity of a given number. An entry ``$i$ or $j$'' means that the answer depends on the details of the flavor charges. The last five rows indicate which multiplicative coefficient should be applied to all numbers of the same column for remarkable values of the electron-type lepton masses. The situation where $m_d=0$, relevant for approximations in high-energy observables, generically corresponds to the line $U(1)^3\times U(3)_{u_R}$, after a suitable replacement $u\leftrightarrow d$ (note that there is a single $\cO_{LeqQ}$ operator in the Warsaw basis, while there are two $\cO_{LeQu}^{1,3}$ operators).}
\label{tableRanksSMEFTMore}
\end{table}
Those new sources of CPV can still be captured by flavor invariants, however there are subtleties to take into account when masses vanish. Namely, the sets of invariants we present in the main text and in appendices \ref{appendix:bilinears} and \ref{appendix:4Fermi} have transfer matrices that do not maintain maximal rank in the limit of vanishing masses. Let us illustrate what we mean by this with an example. From Table~\ref{tableRanksSMEFTMore}, one learns that the three phases in $C_{HQ}^{(1,3)}$ remain primary when $m_u=m_c=0$. However, the invariants presented in the associated set in Table\,\ref{tableInvBilinears} are of the form $\text{Im}\Tr(X^{\vphantom{(1,3)}}_{\smash{u}\vphantom{HQ}}MC_{HQ}^{(1,3)})$ for some matrix $M$ built out of the Yukawas. Working in the {\it up basis} with a vanishing CKM phase, and focusing on the contribution proportional to $C_{HQ,12}^{(1,3)}$, we find
\beq
\text{Im}\Tr(X^{\vphantom{(1,3)}}_{\smash{u}\vphantom{HQ}}MC_{HQ}^{(1,3)})\supset \text{Im}\(C_{HQ,12}^{(1,3)}\)\(m_{\smash{u}\vphantom{21}}^{2\vphantom{(1,3)}}M_{21}^{\vphantom{(1,3)}}-m_{\smash{c}\vphantom{21}}^{2\vphantom{(1,3)}}M_{12}^{\vphantom{(1,3)}}\)
\eeq
which vanishes when $m_u=m_c=0$. Therefore, the set of invariants we consider does not allow us to capture the three phases in $C_{HQ}^{(1,3)}$ in such limits. Another example is that of $C_{uH}$. When $Y_d=0$, one finds only two invariants in the associated set in Table\,\ref{tableInvBilinears}, whereas three sources of CPV remain as shown in Table\,\ref{tableRanksSMEFTMore}. One could therefore conclude that one of the invariants in the set should be replaced by the missing
\beq
\text{Im}\Tr\(X_{\smash{u}}^{2\vphantom{\dagger}}C_{\smash{uH}}^{\vphantom{\dagger}}Y_{\smash{u}}^\dagger\) \ .
\eeq
However, this choice would not allow us to retain a sufficient rank for the set, as one finds 
\beq
\text{Im}\Tr\(X_{\smash{u}}^{2\vphantom{\dagger}}C_{\smash{uH}}^{\vphantom{\dagger}}Y_{\smash{u}}^\dagger\)=(m_u^2+m_t^2)\text{Im}\Tr\(X_{\smash{u}}^{\vphantom{\dagger}}C_{\smash{uH}}^{\vphantom{\dagger}}Y_{\smash{u}}^\dagger\)-m_u^2m_t^2\text{Im}\Tr\(C_{\smash{uH}}^{\vphantom{\dagger}}Y_u^\dagger\)
\eeq
when $m_u=m_c$ (one can use formulae like Eq.\,\eqref{formulaSumDegenerateMasses} to express the mass factors in terms of invariants), whereas all nine sources of CPV in $C_{uH}$ remain primary and independent in this case, as per Tables\,\ref{tableFreeParamsCP} and\,\ref{tableRanksSMEFT}. Therefore, it may seem that one needs strictly more than nine invariants to capture the nine CPV phases in $C_{uH}$, and more generally that the necessary and sufficient conditions presented in section \ref{section:exampleInvariants} are not sufficient anymore when masses can vanish. However, this is a consequence of our assumption that invariants should correspond to traces of a monomial of degree one in SMEFT coefficients, and arbitrary degree in Yukawa matrices. 
Instead, one could enlarge the set of invariants and include traces of sums over monomials of various degrees. For instance, defining instead $X_u^{\vphantom{\dagger}}\equiv 1+Y_u^{\vphantom{\dagger}}Y_u^\dagger$, and similarly for other fermions, without changing the expression of the invariants, is sufficient to ensure that the vanishing of our sets is a necessary and sufficient condition for the conservation of CP at leading order.


\vspace*{3.5ex plus 1ex minus .2ex}
\section{Generalities about invariants}\label{appendix:invariantGeneralities}

\subsection{Properties of $3\times 3$ matrices}\label{sec:propertiesof3x3}
Here we discuss some properties of generic $3\times 3$ matrices, which we use throughout the paper and will refer to later on. We will follow mostly Ref.\,\cite{Jenkins:2009dy}. The starting point is the Cayley--Hamilton theorem, which allows one to rewrite the $n$-th power of a $n\times n$ matrix $A$ in terms of the powers $<n$, and that for $n=3$ takes the form
\begin{align}
	A^3=A^2\Tr(A)-\frac{1}{2}A\left[\Tr(A)^2-\Tr(A^2)\right]+\frac{1}{6}\left[\Tr(A)^3-3\Tr(A^2)\Tr(A)+2\Tr(A^3)\right]\mathbb{1}_{3\times 3}\ .
	\label{eq:CHfor3x3}	
\end{align}
Multiplying by $A$ and taking the trace results in 
\begin{align}
	\Tr(A^4)=\frac{1}{6}\Tr(A)^4-\Tr(A^2)\Tr(A)^2+\frac{4}{3}\Tr(A^3)\Tr(A)+\frac{1}{2}\Tr(A^2)^2.
	\label{eq:reltrA4}
\end{align}
Shifting $A\to A+B+C$ in Eq.\,\eqref{eq:reltrA4}, with $B$ and $C$ some other generic $3\time 3$ matrices, and taking the terms of order $A^2BC$, one obtains
\begin{align}
	0&=\Tr(A)^2\Tr(B)\Tr(C)-\Tr(BC)\Tr(A)^2-2\Tr(AB)\Tr(A)\Tr(C)+\nonumber\\
	&-2\Tr(AC)\Tr(A)\Tr(B)+2\Tr(ABC)\Tr(A)+2\Tr(ACB)\Tr(A)+\nonumber\\
	&-\Tr(A^2)\Tr(B)\Tr(C)+2\Tr(AB)\Tr(AC)+\Tr(A^2)\Tr(BC)+\nonumber\\
	&+2\Tr(C)\Tr(A^2B)+2\Tr(B)\Tr(A^2C)-2\Tr(A^2BC)-2\Tr(A^2CB)-2\Tr(ABAC)\ .
	\label{eq:relABAC}
\end{align}
This property is useful for our purpose of building sets of invariants, as it implies that we only need to draw from a finite set. Let us focus on invariants relevant for this paper, such as those related to bi-fermion SMEFT operators, which are single-trace and linear with respect to the associated Wilson coefficient. To build such invariants, flavor-invariance imposes that we only use $X_u,X_d$ and $C$, where $X_{q=u,d}\equiv Y_qY_q^\dagger$ and $C$ is the Wilson coefficient under study (up to a specific multiplication by a Yukawa matrix for operators of LR chiral structure). 
In principle, any invariant of the form
\begin{align}
	\Tr(X_{u\vphantom{d}}^{a_1}X_d^{b_1}X_{u\vphantom{d}}^{a_2}X_d^{b_2}\ldots C)
\end{align}
is allowed. However, the formulae above imply that one cannot find third or higher powers of $X_q^{\vphantom{2}}$ in the trace, and that one can find at most one occurrence of $X_q^{\vphantom{2}}$ and another of $X_q^2$. These conditions reduce the possible single-trace invariants to a finite set (see appendix~\ref{appendix:polynomialRelations} for explicit examples).

Finally, we mention that the Cayley--Hamilton theorem also allows us to write the determinant of a $3\times 3$ matrix as
\begin{align}
	\text{Det}(A)=\frac{1}{6}\left(\Tr(A)^3-3\Tr(A)\Tr(A^2)+2\Tr(A^3)\right)\ .
	\label{eq:det3x3}
\end{align}

\subsection{Different types of invariants}
In Ref.\,\cite{Lebedev:2002wq}, the author presents a discussion of CP-violating invariants in supersymmetric models, in order to find basis independent conditions for CP violation, as done here. In that context, three types of invariants that can be built using three $3\times 3$ matrices $A$, $B$, and $C$ are proposed, namely
\begin{align}
	J_{AB}&=\Im\Tr\left(\comm{A}{B}^3\right)\ ,&K_{ABC}(p,q,r)&=\Im\Tr\left(\comm{A^p}{B^q}C^r\right)\ ,\nonumber\\
	L_C(p)&=\Im\Tr\left(C^p-\text{h.c.}\right)\ ,
\end{align}
where $A$ and $B$ are hermitian and $C$ generic. These are dubbed $J-$, $K-$ and $L-$invariants respectively. In this work we adopted a similar notation, but we only employed $L$-invariants for our set. However, we can show that this choice is general, as the remaining two types can be written in terms of the last one.\footnote{also notice that $J_{AB}$ would not suit our scopes as it is not linear in any of the two matrices in the argument. } To prove this, let us start from $J_{AB}$. First of all, using Eq.\,\eqref{eq:det3x3}, it can be shown to be equivalent to a Jarlskog-like invariant, {\it i.e.}
\begin{align}
	\Im\, \text{Det}\left(\comm{A}{B}\right)&=\frac{1}{6}\Im\left(\Tr\left(\comm{A}{B}\right)^3-3\Tr\left(\comm{A}{B}\right)\Tr\left(\comm{A}{B}^2\right)+2\Tr\left(\comm{A}{B}^3\right)\right)+\nonumber\\
	&=\frac{1}{3}\Im\Tr\left(\comm{A}{B}^3\right)\ ,
\end{align}
as the trace of a commutator vanishes. This also proves Eq.\,\eqref{eq:Jarlskoginvariant}.
Then, by expanding $\comm{A}{B}^3$ and using the cyclic property of the trace we can show
\begin{align}
	\Im\Tr\left(\comm{A}{B}^3\right)=3\Im\Tr\left(A^2B^2AB-BAB^2A^2\right)=L_{A^2B^2AB}\ .
\end{align}
Finally, we prove that any to $K$-invariant can also be expressed in terms of the $L$ ones. To do this, it is enough to prove it for $K_{ABC}(1,1,1)$ as the other cases can be obtained by redefining $A$, $B$ or $C$. Then, we split $C$ in its hermitian and anti-hermitian parts, {\it i.e.}
\begin{align}
	C_h&\equiv \frac{C+C^\dagger}{2}&C_a&\equiv \frac{C-C^\dagger}{2}\ .
\end{align}
and
\begin{align}
	K_{ABC}(1,1,1)&=\Im\Tr\left(\comm{A}{B}C\right)=\Im\Tr\left(\comm{A}{B}C_h\right)+\Im\Tr\left(\comm{A}{B}C_a\right)\ ,
\end{align}
where
\begin{align}
	\Im\Tr\left(\comm{A}{B}C_h\right)=&\Im\Tr\left(ABC_h-BAC_h\right)=\nonumber\\
	=&\frac{1}{2i}\left[\Tr\left(ABC_h\right)-\Tr\left(ABC_h\right)^*-\Tr\left(BAC_h\right)+\Tr\left(BAC_h\right)^*\right]=\nonumber\\
	=&\frac{1}{2i}\left[\Tr\left(ABC_h\right)-\Tr\left(C_hBA\right)-\Tr\left(BAC_h\right)+\Tr\left(C_hAB\right)\right]=\nonumber\\
	=&\frac{1}{i}\left[\Tr\left(ABC_h\right)-\Tr\left(C_hBA\right)\right]=2L_{ABC_h}=2L_{ABC}\ .
\end{align}
With similar steps, one can see that the piece proportional to $C_a$ vanishes, so that
\begin{align}
	K_{ABC}(1,1,1)=2L_{ABC}\ .
\end{align}

\subsection{Invariant ring and the Hilbert series}
The problem of finding a minimal set of invariants we face in this paper is reminiscent of the so called Plethystic Program \cite{Feng:2007ur, Benvenuti:2006qr, Romelsberger:2005eg, Pouliot_1999, Trautner:2020qqo} (see also Refs.\,\cite{Trautner:2018ipq,Wang:2021wdq} and references therein, and Ref. \cite{Bento:2021hyo} for other recent techniques to compute the Hilbert series). This has been the attempt to apply tools which originate from the study of polynomial rings to physics, and in particular where representation and group theory come into play. The most remarkable success of this program has perhaps been the addition of the Hilbert Series and its Plethystic Logarithm to the theoretical physicist's toolbox. In seeing how these tools apply to this case, we will adopt the logic followed in Ref.\,\cite{Trautner:2018ipq,Wang:2021wdq}. Concretely, given a set of parameters $\vec{x}$ and a symmetry group G that acts on $\vec{x}$ as some representation, {\it i.e.}~$\forall g\in G$, $\exists R(g)$ such that $\vec{x}\to R(g)\vec{x}$, one can define invariants $I(\vec{x})$ as the quantities that obey
\begin{align}
	I(\vec{x})=I(R(g)\vec{x})\ .
\end{align} 
As the sum and products of invariants still form an invariant, from an algebraic point of view we talk about a {\it ring}. Within this ring, one can find the set of invariants $\left\{I_1,\, I_2,\, \dots,\, I_m\right\}$ such that any additional invariant $I'$ can be expressed as a polynomial of those of the set,
\begin{align}
	I'=P'\left(I_1,\, I_2,\, \dots,\, I_m\right)\ .
	\label{eq:polynomialrel}
\end{align}
The invariants in $\left\{I_1,\, I_2,\, \dots,\, I_m\right\}$ are called generators, and it can be shown that, at least for all the groups relevant to physics, their number is finite. 
By construction, no relation like Eq.\,\eqref{eq:polynomialrel} can exist between the generators. Nevertheless, there could exist a polynomial $P$ such that
\begin{align}
	P\left(I_1,\, I_2,\, \dots,\, I_m\right)=0\ .
\end{align}
Relations of this kind are called {\it syzygies} in the literature, and the invariants that obey a syzygy are {\it algebraically dependent}. Taking all syzygies into account we can successively remove invariants until we get to the set of algebraically independent ones.
In this setting, the Hilbert series provides a helping hand in finding both the generators and the basic invariants \cite{Jenkins:2009dy}. It is defined as a generating function for the linearly-independent invariants:
\begin{align}
	\mathcal{H}(q)=\sum_{k=1}^{\infty} c_{k}q^k\ ,
\end{align}
where $c_0=1$. $c_k$ denotes precisely the number of linearly-independent invariants at dimension $k$, and $q$ is an arbitrary spurionic variable satisfying $\abs{q}<1$, and represents a placeholder for the building blocks of the invariants.
Let us make an example. Consider a theory with a coupling $m$ transforming under a $U(1)$ symmetry as 
\begin{align}
	m\to e^{i\phi_m}m\ .
\end{align}
Then the basic invariant is obviously $I=mm^*$, which has dimension 2, and all the invariants of this theory will have the form $I^n$. 
Hilbert series will thus have the form
\begin{align}
	\mathcal{H}(q)=1+q^2+q^4+\dots=\frac{1}{1-q^2}\ ,
\end{align}
where $q^2$ corresponds to $I$, $q^4$ to $I^2$, and so on. 
It can be shown that, in the general case of a semi-simple Lie algebra, the Hilbert series has the form
\begin{align}
	\mathcal{H}(q)=\frac{N(q)}{D(q)}\ .
\end{align}
The numerator $N(q)$ is a polynomial of degree $d_N$ with non-negative coefficients and with the property of being palindromic, {\it i.e.}
\begin{align}
	N(q)=1+c_1q+c_2q^2+\dots+c_{d_N-1}q^{d_N-1}+q^{d_N},
\end{align}
with $c_i=c_{d_N-i}$.
The denominator takes the form 
\begin{align}
	D(q)=\prod_{r=1}^p(1-q^{d_r})\ ,
\end{align}
and is thus of degree $d_D=\sum_rd_r$. The number of factors is equal to the number of parameters, {\it i.e.}~of physical observables, and coincides with the number of algebraically independent invariants. Moreover, the denominator provides information on what the algebraically independent invariants look like: a factor $(1-q^{d_r})^l$ corresponds to $l$ algebraically independent invariants of degree $d_r$.
In the previous example, only one factor is present, corresponding to a single basic (and algebraically independent) invariant $mm^*$. Indeed, we start with a complex variable $m$, and we can remove its phase using the $U(1)$, bringing the observables down to 1.
If we enlarged our toy-model to have two parameters, $m_1$ and $m_2$, transforming under the $U(1)$ symmetry as
\begin{align}
	m_1 &\to e^{i\phi_1}m_1 &  m_2 &\to e^{i\phi_2}m_2\ ,
\end{align}
we can build an example of the so called {\it multi-graded} Hilbert series by assigning different spurions to $m_1$ and $m_2$, say $q_1$ and $q_2$. The invariants in this case are built as all possible products of all possible powers of $I_{1,2}\equiv m_{1,2}m_{1,2}^*$, which means that the multi-graded Hilbert series is 
\begin{align}
	h(q_1,q_2)=(1+q^2_1+q^4_1+\dots)(1+q^2_2+q^4_2+\dots)=\frac{1}{(1-q_1^2)(1-q^2_2)}\ .
\end{align}
The multi-graded Hilbert series can give more information about the structure of the invariants, but does not have in general the properties for the numerator and denominator cited for its ungraded version. The latter can here be easily obtained by setting $q_1=q_2=q$, {\it i.e.}~$H(q)=h(q,q)$. 
In the examples we showed until now, the numerator has always taken the trivial form $N(q)=1$, and the set of generators coincided with the algebraically independent invariants. When this happens, the invariant ring is said to be {\it free}. However, this turns out not to be always the case for more complicated groups and representations. 

As one would expect, the computation we could perform straightforwardly by hand in the simple cases above quickly becomes unfeasible when larger groups are involved. Thus a general formula to compute the Hilbert series is called for. The solution is provided by the so called Molien--Weyl formula, which, for a compact, simple Lie group $G$ takes the form:
\begin{align}
	H(q)=\int \left[\dd \mu\right]_G\frac{1}{\det(\mathbb{1}-qR(g))}\ ,
\end{align}
where $\left[\dd \mu\right]_G$ denotes the Haar measure of the group $G$. If $G$ is connected, the integral can be reduced to an integral over the {\it maximal torus} of the group, {\it i.e.}~its largest abelian subgroup, which is just the direct product of $r_0$ copies of the $S^1$ unit circle, with $r_0$ the rank of the group (see {\it e.g.}~Ref.\,\cite{Henning:2017fpj}). Thus, the integral is reduced to the computation of residues inside said circles. 
The integrand $\left[\det(\mathbb{1}-qR(g))\right]^{-1}$ can be rewritten as
\begin{align}
	\frac{1}{\det(\mathbb{1}-qR(g))}=\exp(\sum_{k=1}^{\infty}\frac{q^k\chi_R(z_1^k,\dots,z_d^k)}{k})\equiv\text{PE}\left[\chi_{R}(z_1,\dots,z_d)q\right]\ ,
	\label{eq:MWformula}
\end{align}
where $\chi_R(z_1,\dots,z_d)=\sum_{j=1}^dz_j$ is the character function of $G$ in the representation $R$, $d $ is the dimension of the representation, and $z_j$ (with $j=1,\dots,d$) are the eigenvalues of $R(g)$. Eq.\,\eqref{eq:MWformula} makes use of the definition of the {\it Plethystic Exponential} (PE), which for an arbitrary function $f(x_1,\dots,x_n)$ looks like
\begin{align}
	\text{PE}\left[f(x_1,\dots,x_n)\right]=\exp(\sum_i^{\infty}\frac{f(x_1^k,\dots,x_n^k)}{k})\ .
\end{align}
The generalization to a multi-graded Hilbert series is then straightforward:
\begin{align}
	h(q_1,\dots,q_n)=\int \left[\dd \mu\right]_G\prod_{i=1}^n\text{PE}(z_1,\dots,z_d;q_i)\ .
\end{align}
An important role in this context is played by the inverse of the PE, quite fittingly called {\it plethystic logarithm} (PL) and defined so that
\begin{align}
	f(x_1,\dots,x_n)=\text{PE}\left[g(x_1,\dots,x_n)\right]\Leftrightarrow g(x_1,\dots,x_n)=\text{PL}\left[f(x_1,\dots,x_n)\right]\ .
\end{align}
It can be proved that 
\begin{align}
	\text{PL}\left[f(x_1,\dots,x_n)\right]=\sum_{k=1}^\infty\frac{\mu(k)}{k}\ln\left[f(x_1^k,\dots,x_n^k)\right]\ .
\end{align}
The value of the PL lies in the following fact: the PL of a Hilbert series is a polynomial whose leading positive terms correspond to the basic invariants, {\it i.e.}~to the generators, and whose leading negative terms correspond to the syzygies between them. Remarkably, when the invariant ring is free, this polynomial has a finite number of terms. Some complications arise when the groups and the representations that appear become increasingly non-trivial\footnote{In Ref.\,\cite{Lu:2021yej} the author argues that some of the assumptions in Ref.\,\cite{Wang:2021wdq} are imprecise, as the finiteness of the generating set of invariants is a consequence of the group being reductive, which is supposedly not the case for $U(n)$. However, as explained there and clarified by the same authors of Ref.\,\cite{Wang:2021wdq} in Ref.\,\cite{Wang:2021fgd}, the final result is nonetheless correct, as at least the ring of invariants of $U(n)$ is isomorphic to that of $GL(n,\mathbb{C})$, which is itself reductive. See further discussions in \cite{Lu:2021ada}.}\,\cite{Wang:2021wdq, Benvenuti:2006qr}. Equipped with these outstanding tools, we wish to see them applied to our case, which is the main scope of the next section.

\subsection{Finding polynomial relations between invariants}\label{appendix:polynomialRelations}
In the main body of this work, the logic we have followed to build invariants stemmed from knowing that the relative Wilson coefficient $C^{(6)}$, in a given basis, has a certain number of phases. Then we found as many independent invariants as possible, in order to obtain a transfer matrix whose rank matched the new sources of CPV in $C^{(6)}$ when $J_4=0$. 
Now, however, we could be tempted to pursue a different line of reasoning and find the relevant invariants by applying to our case the power of the Hilbert series and its Plethystic logarithm. Let us restrict to the case of quark bilinear operators. Their Wilson coefficients are generic $3\times 3$ complex matrices that we can multiply by an appropriate number of $Y_{u,d}^{(\dagger)}$ to retrofit them into a $\mathbf{3}\otimes\bar{\mathbf{3}}$ representation of $SU(3)_Q$. We refer to this combination as $C^{(6)}$ here. Then, the building blocks of our invariants are
\begin{align}
	C^{(6)}, (C^{(6)})^\dagger, X_{u,d} \in \mathbf{3}\otimes\bar{\mathbf{3}}\ .
\end{align}
and we can build the multi-graded Hilbert series 
\begin{align}
		h(c,c^\dagger, x_u,x_d)=\int \left[\dd \mu\right]_{SU(3)}\prod_{i=\left\{c,c^\dagger,x_u,x_d\right\}}\text{PE}(\vec{z};c)\ ,
\end{align}
with obvious associations between a spurion and the corresponding building block.
The resulting expression is quite long and not particularly illuminating, so we will refrain from presenting it here. However, we can look at its ungraded version
\begin{align}
	H(q)=h(q,q,q,q)=\frac{N(q)}{D(q)}\ ,
\end{align}
with
\begin{align}
	N(q)=&+q^{34}+14 q^{31}+31 q^{30}+56 q^{29}+165 q^{28}+354 q^{27}+660 q^{26}+1256 q^{25}+2097 q^{24}+\nonumber\\
	&+3184 q^{23}+4720 q^{22}+6404 q^{21}+7992 q^{20}+9536 q^{19}+10510 q^{18}+10744 q^{17}+\nonumber\\
	&+10510 q^{16}+9536 q^{15}+7992 q^{14}+6404 q^{13}+4720 q^{12}+3184 q^{11}+2097 q^{10}+\nonumber\\
	&+1256 q^9+660 q^8+354 q^7+165 q^6+56 q^5+31 q^4+14 q^3+1\ ,
\end{align}
and
\begin{align}
	D(q)=(1-q)^4(1-q^2)^{10}(1-q^3)^{10}(1-q^4)^4\ .
\end{align}
We can see that the numerator has the correct palindromic structure we expected, and more importantly the denominators contain 28 factors, correctly matching the 10 observables from the Standard Model and the 18 new (9 real + 9 imaginary) observables contained in $C^{(6)}$.
This is already quite remarkable. However, to this point we have neither an idea of how the algebraically independent invariants look like, nor a way to extract the ones that are linear in $C^{(6)}$, which is the subset we are really interested in. To gain some more insight, let us look at the Plethystic logarithm of the multi-graded Hilbert series:
\begin{align}
	PL\left[h(c,c^\dagger, x_u,x_d)\right]=&(x_u^{\vphantom{2}}+x_d^{\vphantom{2}})+(x_u^2+x_d^{\vphantom{2}}x_u^{\vphantom{2}}+x_d^2)+(x_u^3+x_u^2x_d^{\vphantom{2}}+x_u^{\vphantom{2}}x_d^2+x_d^3)+x_u^2x_d^2+\nonumber\\
	&+x_u^3x_d^3-x_u^6x_d^6+\nonumber\\
	&+\left(c+c^\dagger\right)\left[1+x_u^{\vphantom{2}}+x_d^{\vphantom{2}}+(x_u^2+x^2_d+2x_u^{\vphantom{2}}x_d^{\vphantom{2}})+(2x_u^2x_d^{\vphantom{2}}+2x_u^{\vphantom{2}}x_d^2)+\right.\nonumber\\
	&\left.+(x_u^3x_d^{\vphantom{2}}+2x_u^2x_d^2+x_d^3x_u^{\vphantom{2}})+(x_u^3x_d^2+x_u^2x_d^3)\right]+\nonumber\\
	&+\order{c^2,(c^\dagger)^2,(c+c^\dagger) x_u^3x_d^4,(c+c^\dagger) x_u^4x_d^3}\ .
	\label{eq:Plethysticlinear}
\end{align}
Since we are interested in invariants that are linear in $C^{(6)}$, we stopped the expansion at $\order{c,c^\dagger}$. The $\order{c^0(c^\dagger)^0}$ terms in this expansion correspond to the invariants one can obtained in the quark sector of the Standard Model. This case has been treated in Ref.\,\cite{Jenkins:2009dy}, and the resulting algebraically independent invariants are
\begin{align}
	\begin{array}{ll}
		I_{1,0}=\text{Tr}(X_u^{\vphantom{2}}) & I_{0,1}=\text{Tr} (X_d^{\vphantom{2}}) \\
		I_{2,0}=\text{Tr}\left(X_u^2\right) & I_{1,1}=\text{Tr} (X_u^{\vphantom{2}} X_d^{\vphantom{2}}) \\
		I_{0,2}=\text{Tr}\left(X_d^2\right) & I_{3,0}=\text{Tr}\left(X_u^3\right) \\
		I_{2,1}=\text{Tr}\left(X_u^2 X_d^{\vphantom{2}}\right) &I_{1,2}= \text{Tr}\left(X_u^{\vphantom{2}} X_d^2\right) \\
		I_{0,3}=\text{Tr}\left(X_d^3\right) & I_{2,2}=\text{Tr}\left(X_u^2 X_d^2\right) \ ,
	\end{array}
	\label{eq:SMinvariants}
\end{align}
Notice their number is 10, correctly matching the 6 masses + 3 angles + 1 phase of the Standard Model. 
The generating set contain one additional invariant, 
\begin{align}
	I_{3,3}^{(-)}=\Tr\left(X_u^2X_d^2X_u^{\vphantom{2}}X_d^{\vphantom{2}}\right)-\Tr\left(X_d^2X_u^2X_d^{\vphantom{2}}X_u^{\vphantom{2}}\right)\ ,
\end{align}
corresponding to the $x_{u\vphantom{d}}^3x_d^3$ term in Eq.\,\eqref{eq:Plethysticlinear} and which is nothing but $J_4$. In the language adopted here, this invariant does not contain any additional observable, and is just needed to capture the sign of the SM$_4$ phase $\delta$. The negative term $-x_{u\vphantom{d}}^6x_d^6$ at the end signals that there is a syzygy at degree 12, which in this case corresponds to the fact that $\left(I_{3,3}^{(-)}\right)^2$ can be expressed in terms of the remaining 10 invariants, as expected.

The part linear in $c$ and $c^\dagger$ of Eq.\,\eqref{eq:Plethysticlinear} points us at the basic invariants linear in $C^{(6)}$. We see that in this case the set of basic invariants is composed by 34 element, 17 each for $C^{(6)}$ and $(C^{(6)})^\dagger$, which is larger than the basic set. Indeed, the latter is expected to have 18 elements, corresponding to the 9 new complex observables contained in $C^{(6)}$. 
To try and build the invariants in the generating set, we will make use of the relations showed in section \ref{sec:propertiesof3x3}. Given a generic matrix $C^{(6)}\in\mathbf{3}\otimes\bar{\mathbf{3}}$ of $SU(3)_Q$, we want to contract it with as many $X_{u,d}$'s as needed to form all the possible independent invariants. Using Eq.\eqref{eq:CHfor3x3} on $X_{u,d}$, we can show that all invariants written using $X_{u,d}^n$, with $n\geq 3$, are redundant, and we only need $X_{u,d}^2$ and $X_{u,d}^{\vphantom{2}}$ as building blocks. Moreover, using Eq.\,\eqref{eq:relABAC}, one can eliminate any invariant where a matrix is repeated. Taking into account these simplifications, one can see that the set of possible invariants is finite, and is formed by these 29 objects:
\begin{align}
	\begin{array}{lll	}
		\text{Tr} (C^{(6)} ) & \text{Tr} ( X_{u\vphantom{d}}^{\vphantom{2}} C^{(6)}) & \text{Tr} ( X_d^{\vphantom{2}} C^{(6)}) \\
		\text{Tr}\left( X_{u\vphantom{d}}^2C^{(6)} \right) & \text{Tr}\left( X_d^2 C^{(6)} \right) & \text{Tr} \left(X_{u\vphantom{d}}^{\vphantom{2}} X_d^{\vphantom{2}} C^{(6)} \right) \\
		\text{Tr}\left(X_{u\vphantom{d}}^{\vphantom{2}} X_d^2 C^{(6)} \right) & \text{Tr} \left(X_d^{\vphantom{2}} X_{u\vphantom{d}} ^{\vphantom{2}}C^{(6)} \right) & \text{Tr}\left(X_d^2 X_{u\vphantom{d}}^{\vphantom{2}}C^{(6)}\right) \\
		\text{Tr}\left(X_d^{\vphantom{2}} X_{u\vphantom{d}}^2C^{(6)} \right) & \text{Tr}\left(X_{u\vphantom{d}}^2 X_d^{\vphantom{2}}C^{(6)}\right) & \text{Tr}\left(X_{u\vphantom{d}}^2 X_d^2C^{(6)}\right) \\
		\text{Tr}\left(X_d^2 X_{u\vphantom{d}}^2 C^{(6)}\right) & \text{Tr}\left(X_{u\vphantom{d}}^{\vphantom{2}} X_d^{\vphantom{2}} X_{u\vphantom{d}}^2C^{(6)} \right) & \text{Tr}\left(X_d^{\vphantom{2}} X_{u\vphantom{d}}^{\vphantom{2}} X_d^2C^{(6)} \right) \\
		\text{Tr}\left(X_{u\vphantom{d}}^2 X_d^{\vphantom{2}} X_{u\vphantom{d}}^{\vphantom{2}} C^{(6)} \right) & \text{Tr}\left(X_d^2 X_{u\vphantom{d}}^{\vphantom{2}} X_d^{\vphantom{2}}C^{(6)} \right) & \text{Tr}\left(X_{u\vphantom{d}}^{\vphantom{2}} X_d^2 X_{u\vphantom{d}}^2C^{(6)} \right) \\
		\text{Tr}\left(X_{u\vphantom{d}}^2 X_d^2 X_{u\vphantom{d}}^{\vphantom{2}}C^{(6)} \right) & \text{Tr}\left(X_d^{\vphantom{2}} X_{u\vphantom{d}}^2 X_d^2C^{(6)}\right) & \text{Tr}\left(X_d^2 X_{u\vphantom{d}}^2 X_d^{\vphantom{2}}C^{(6)} \right) \\
		\text{Tr}\left(X_{u\vphantom{d}}^{\vphantom{2}} X_d^{\vphantom{2}} X_{u\vphantom{d}}^2 X_d^2C^{(6)} \right) & \text{Tr}\left(X_{u\vphantom{d}}^{\vphantom{2}} X_d^2 X_{u\vphantom{d}}^2 X_d^{\vphantom{2}}C^{(6)} \right) & \text{Tr}\left(X_d^{\vphantom{2}} X_{u\vphantom{d}}^{\vphantom{2}} X_d^2 X_{u\vphantom{d}}^2C^{(6)} \right) \\
		\text{Tr}\left(X_d^{\vphantom{2}} X_{u\vphantom{d}}^2 X_d^2 X_{u\vphantom{d}}^{\vphantom{2}}C^{(6)} \right) & \text{Tr}\left(X_{u\vphantom{d}}^2 X_d^{\vphantom{2}} X_{u\vphantom{d}}^{\vphantom{2}} X_d^2C^{(6)}\right) & \text{Tr}\left(X_{u\vphantom{d}}^2 X_d^2 X_{u\vphantom{d}}^{\vphantom{2}} X_d^{\vphantom{2}}C^{(6)}\right) \\
		\text{Tr}\left(X_d^2 X_{u\vphantom{d}}^{\vphantom{2}} X_d^{\vphantom{2}} X_{u\vphantom{d}}^2C^{(6)} \right) & \text{Tr}\left(X_d^2 X_{u\vphantom{d}}^2 X_d^{\vphantom{2}} X_{u\vphantom{d}}^{\vphantom{2}}C^{(6)}\right)\ , & 
	\end{array}
	\label{eq:29invariants}
\end{align}
and the same for $C^{(6)}\to(C^{(6)})^\dagger$. Now, since the generating set only includes 17 elements, this means that 12 invariants of Eq.~\eqref{eq:29invariants} can be expressed as polynomials of the remaining ones and can thus be eliminated. To do this, we employ a numerical algorithm adapted from Appendix C of Ref.\,\cite{Wang:2021wdq} (see also Ref. \cite{Trautner:2018ipq}). The logic is as follows: by assigning a dummy dimension to the building blocks, {\it i.e.}~$\left[X_{u,d} \right]=1 $ and $\left[C^{(6)}\right]=1$, we can assign a dimension to all the invariants listed above. Then, we fix some given dimension $n$. Picking one of the invariants in Eq.\,\eqref{eq:29invariants}, one can then take its product with as many traces from Eq.\,\eqref{eq:SMinvariants} as needed to form a monomial $M_i$ of dimension $n$. Repeating this for all instances of Eq.\,\eqref{eq:29invariants}, we find the set $\{M_i\}$ of all the possible monomials that are dimension $n$ and linear in $C^{6}$. For example, at dimension $n=2$ one can obtain the monomials
\begin{align}
	\{M_i\}=\left\{\text{Tr}\left(C^{(6)}\right)I_{1,0},\text{Tr}\left(C^{(6)}\right)I_{0,1},\text{Tr}\left(X_{u\vphantom{d}}C^{(6)}\right),\text{Tr}\left(X_dC^{(6)}\right)\right\}\ .
\end{align}
Then we set a linear combination of these monomials to zero, {\it i.e.}
\begin{align}
	\sum_i a_iM_i=0\ ,
\end{align}
where the $a_i$'s are integer coefficients. We then plug random integer values for the entries of the matrices $X_{u,d}$ and $C^{6}$. This produces a linear equation for the $a_i$'s. Repeating this last step as many times as there are $M_i$'s, one builds a linear system for the $a_i$'s with zero constant term. The number of independent directions of the null space of the corresponding matrix matches the number of possible relations between the $M_i$'s. 
The first nontrivial result is found at dimension 5, where we get the two relations
\begin{align}
	&\text{Tr}\left(X_d^{\vphantom{2}}X_{u\vphantom{d}}^{\vphantom{2}}X_d^2C^{(6)}\right)+\text{Tr}\left(X_d^2X_{u\vphantom{d}}^{\vphantom{2}}X_d^{\vphantom{2}}C^{(6)}\right)+I_{0,1}^{\vphantom{2}} \left(\text{Tr}\left(X_d^2X_{u\vphantom{d}}^{\vphantom{2}}C^{(6)}\right)+ \text{Tr}\left(X_{u\vphantom{d}}^{\vphantom{2}}X_d^2C^{(6)}\right)\right)+\nonumber\\
	&+(-I_{0,1}^{\vphantom{2}} I_{1,0}^{\vphantom{2}}-I_{1,1}^{\vphantom{2}}) \text{Tr}\left(X_d^2C^{(6)}\right)-I_{0,1}^2 (\text{Tr}\left(X_d^{\vphantom{2}}X_{u\vphantom{d}}^{\vphantom{2}}C^{(6)}\right)+ \text{Tr}\left(X_{u\vphantom{d}}^{\vphantom{2}}X_d^{\vphantom{2}}C^{(6)}\right)+\nonumber\\
	&+\frac{1}{3}\left(2 I_{0,1}^3-3I_{0,2}^{\vphantom{2}} I_{0,1}^{\vphantom{2}}+ I_{0,3}^{\vphantom{2}}\right) \text{Tr}\left(X_{u\vphantom{d}}^{\vphantom{2}}C^{(6)}\right)+\left(I_{0,1}^2 I_{1,0}^{\vphantom{2}}-I_{1,2}^{\vphantom{2}}\right) \text{Tr}\left(X_d^{\vphantom{2}}C^{(6)}\right)+\nonumber\\
	&+\text{Tr}(C^{(6)}) \left(-\frac{2}{3}I_{0,1}^3 I_{1,0}^{\vphantom{2}}+I_{0,1}^2 I_{1,1}^{\vphantom{2}}+I_{0,2}^{\vphantom{2}} I_{0,1}^{\vphantom{2}} I_{1,0}^{\vphantom{2}}-I_{0,1}^{\vphantom{2}} I_{1,2}^{\vphantom{2}}-\frac{1}{3} I_{0,3}^{\vphantom{2}} I_{1,0}^{\vphantom{2}}\right)=0\ ,
	\label{eq:degree5relation}
\end{align}
and the same with $X_u^{\vphantom{2}}\leftrightarrow X_d^{\vphantom{2}}$, which we can use to remove $\text{Tr}\left(X_d^{\vphantom{2}}X_{u\vphantom{d}}^{\vphantom{2}}X_d^2C^{(6)}\right)$ and
\\ $\text{Tr}\left(X_{u\vphantom{d}}^{\vphantom{2}}X_d^{\vphantom{2}}X_{u\vphantom{d}}^2C^{(6)}\right)$ from the set. 
At dimension 6 we obtain two more relations, and 8 more are obtained at dimension 7. With these 12 expressions, we can reduce the set to
\begin{align}
		\begin{array}{lll	}
			\text{Tr} \left(C^{(6)} \right) & \text{Tr} \left(X_{u\vphantom{d}}^{\vphantom{2}} C^{(6)}\right) & \text{Tr} \left( X_d^{\vphantom{2}} C^{(6)}\right) \\
			\text{Tr}\left(X_{u\vphantom{d}}^2C^{(6)} \right) & \text{Tr}\left(X_d^2C^{(6)}\right) & \text{Tr} \left(X_{u\vphantom{d}}^{\vphantom{2}} X_d^{\vphantom{2}} C^{(6)} \right) \\
			\text{Tr}\left(X_{u\vphantom{d}}^{\vphantom{2}} X_d^2C^{(6)}\right) & \text{Tr} \left(X_d^{\vphantom{2}} X_{u\vphantom{d}}^{\vphantom{2}} C^{(6)}\right) & \text{Tr}\left(X_d^2 X_{u\vphantom{d}}^{\vphantom{2}}C^{(6)}\right) \\
			\text{Tr}\left(X_d^{\vphantom{2}} X_{u\vphantom{d}}^2C^{(6)} \right) & \text{Tr}\left( X_{u\vphantom{d}}^2 X_d^{\vphantom{2}}C^{(6)}\right) & \text{Tr}\left(X_{u\vphantom{d}}^2 X_d^2C^{(6)}\right) \\
			\text{Tr}\left(X_d^2 X_{u\vphantom{d}}^2C^{(6)} \right) & \text{Tr}\left(X_{u\vphantom{d}}^{\vphantom{2}} X_d^{\vphantom{2}} X_{u\vphantom{d}}^2C^{(6)} \right) & \text{Tr}\left(X_d^{\vphantom{2}} X_{u\vphantom{d}}^{\vphantom{2}} X_d^2C^{(6)} \right) \\
		   \text{Tr}\left(X_{u\vphantom{d}}^{\vphantom{2}} X_d^2 X_{u\vphantom{d}}^2C^{(6)} \right) & \text{Tr}\left(X_d^{\vphantom{2}} X_{u\vphantom{d}}^2 X_d^2C^{(6)} \right) . & \\
		\end{array}
		\label{eq:17invariants}
\end{align}
The same can be repeated for $(C^{(6)})^\dagger$. These are, quite remarkably, exactly in correspondence with the relative terms in Eq.\,\eqref{eq:Plethysticlinear}. 
However, we now wish to find the additional relations that help us express the 8 too many (complex) invariants we have in Eq~\eqref{eq:17invariants} in terms of the 9 ones we know are sufficient to express all the physical observables, {\it i.e.}~the algebraically independent ones. 
If we expand a bit further in Eq.\,\eqref{eq:Plethysticlinear}, we see that the next two terms are degree 8 and are negative, $-(c+c^\dagger)x_{u\vphantom{d}}^4x_d^3-(c+c^\dagger)x_{u\vphantom{d}}^3x_d^4$. They should then correspond to the number of syzygies at dimension 8. 
To obtain them explicitly, we just run again the described algorithm at dimension 8, obtaining indeed two syzygies of the expected degree. They include 107 terms out of the possible 808 one can build at this dimension, and allow us to remove $\text{Tr}\left(X_{u\vphantom{d}}^{\vphantom{2}} X_d^2 X_{u\vphantom{d}}^2C^{(6)} \right)$ and $ \text{Tr}\left(X_d^{\vphantom{d}} X_{u\vphantom{d}}^2 X_d^2C^{(6)}\right)$. 
Running this argument at degree 9, however, we run into a mismatch. Indeed, even though the next term in Eq.\,\eqref{eq:Plethysticlinear} would call for 4 syzygies, we only find 1, symmetric under the exchange $X_u\leftrightarrow X_d$. This is probably due to the complications that arise when the groups and representations one has to deal with start becoming less and less trivial, as in our case, and that forbid us from reading the syzygies from the negative terms directly. For a deeper discussion of this topic, see in particular Ref.\,\cite{Wang:2021wdq, Benvenuti:2006qr} and references therein.
One thing to notice, in addition, is that requiring the building blocks to be linear in $(C^{(6)})$, although justified from a physical point of view, breaks the ring structure of the invariant ring, as obviously the set we consider is no longer closed under multiplications. 

In any case, even without the Plethystic logarithm as a guide, we can just run our algorithm at increasingly higher dimensions, until no more relations are found.
Indeed, upon going up to dimension $n=13$, one manages to reduce the set down to 9 independent invariants, which we can pick to be
\begin{align}
	\begin{array}{lll	}
		\text{Tr} (C^{(6)} ) & \text{Tr} \left(X_{u\vphantom{d}}^{\vphantom{2}}C^{(6)} \right) & \text{Tr} \left(X_d^{\vphantom{2}} C^{(6)}\right) \\
		\text{Tr} \left(X_{u\vphantom{d}}^{\vphantom{2}} X_d^{\vphantom{2}}C^{(6)}  \right) &\text{Tr} \left( X_d^{\vphantom{2}} X_{u\vphantom{d}}^{\vphantom{2}} C^{(6)}\right) & \text{Tr}\left(X_d^2 X_{u\vphantom{d}}^2C^{(6)}\right)\\
		\text{Tr}\left(X_{u\vphantom{d}}^2 X_d^2C^{(6)}\right) & \text{Tr}\left(X_d^{\vphantom{2}}X_{u\vphantom{d}}^2X_d^2C^{(6)} \right) & \text{Tr}\left(X_{u\vphantom{d}}^{\vphantom{2}}X_d^2X_{u\vphantom{d}}^2C^{(6)} \right)\ ,
	\end{array}
	\label{eq:9invariants}
\end{align}
which, upon taking its imaginary parts, matches the minimal set in for a non-hermitian fermion bilinear operator Table\,\ref{tableInvBilinears}.

\section{List of dimension-6 fermionic operators and parameter counting with generic $N_f$}\label{section:paramnf}
In Tables\,\ref{bilinearlist} and\,\ref{4Fermilist} we reproduce the subset of operators from Ref.\,\cite{Grzadkowski:2010es} we are interested in in this work, namely dimension-6 operators in SMEFT containing fermions, split between bilinear and 4-Fermi operators. For each of the considered operators we list the number of real and imaginary entries and compare them with the number of (primary) real and imaginary parameters that can appear in observables at order $1/\Lambda^2$, as explained in the main text. 

In Table\,\ref{tab:countingflavor} the counting of independent primary parameters is generalized to an arbitrary number of flavors $N$.

\begin{table}
	\centering
	\renewcommand{\arraystretch}{1.5}
	\resizebox{1\columnwidth}{!}{%
	\begin{tabular}{p{0.9cm}|c|c|c|c|c|c}
		\multicolumn{7}{c}{\textbf{Bilinears}}\\\hline
		\multicolumn{2}{c|}{Label} & Operator & \begin{minipage}{0.15\columnwidth}\centering
		\# real entries \vspace*{0.1cm}
		\end{minipage}& \begin{minipage}{0.15\columnwidth}\centering
	\# imaginary entries\vspace*{0.1cm}
	\end{minipage} &\begin{minipage}{0.15\columnwidth}\centering
\cellcolor{lightgray}\# primary real entries\vspace*{0.1cm}
\end{minipage}   & \begin{minipage}{0.2\columnwidth}\centering
\cellcolor{lightgray} \# primary imaginary  entries \vspace*{0.1cm}
\end{minipage} \\\hline
	\multirow{3}{0.1\columnwidth}{
\rotatebox[origin=c]{90}{\begin{minipage}{0.1\textwidth}
\centering Modified Yukawas
	\end{minipage}}}&$Q_{eH}$       &$(H^\dag H)(\bar L_i e_j H)+\text{h.c.}$ &9&9&\cellcolor{lightgray}3&\cellcolor{lightgray}3\\
	&$Q_{uH}$          & $(H^\dag H)(\bar Q_i u_j \widetilde H )+\text{h.c.}$ &9&9&\cellcolor{lightgray}9&\cellcolor{lightgray}9\\
	&$Q_{dH}$           & $(H^\dag H)(\bar Q_i d_j H)+\text{h.c.}$ &9&9&\cellcolor{lightgray}9&\cellcolor{lightgray}9\\\hline
		\multirow{8}{0.1\columnwidth}{
		\rotatebox[origin=c]{90}{\begin{minipage}{0.1\textwidth}
				\centering Dipole
	\end{minipage}}}
	&$Q_{eW}$      & $(\bar L_i \sigma^{\mu\nu} e_j) \tau^I H W_{\mu\nu}^I+\text{h.c.}$  &9&9&\cellcolor{lightgray}3&\cellcolor{lightgray}3\\
	&$Q_{eB}$        & $(\bar L_i \sigma^{\mu\nu} e_j) H B_{\mu\nu}+\text{h.c.}$  &9&9&\cellcolor{lightgray}3&\cellcolor{lightgray}3\\
	&$Q_{uG}$        & $(\bar Q_i \sigma^{\mu\nu} T^A u_j) \widetilde H \, G_{\mu\nu}^A+\text{h.c.}$  &9&9&\cellcolor{lightgray}9&\cellcolor{lightgray}9\\
	&$Q_{uW}$        & $(\bar Q_i \sigma^{\mu\nu} u_j) \tau^I \widetilde H \, W_{\mu\nu}^I+\text{h.c.}$  &9&9&\cellcolor{lightgray}9&\cellcolor{lightgray}9\\
	&$Q_{uB}$        & $(\bar Q_i \sigma^{\mu\nu} u_j) \widetilde H \, B_{\mu\nu}+\text{h.c.}$  &9&9&\cellcolor{lightgray}9&\cellcolor{lightgray}9\\
	&$Q_{dG}$        & $(\bar Q_i \sigma^{\mu\nu} T^A d_j) H\, G_{\mu\nu}^A+\text{h.c.}$  &9&9&\cellcolor{lightgray}9&\cellcolor{lightgray}9\\
	&$Q_{dW}$         & $(\bar Q_i \sigma^{\mu\nu} d_j) \tau^I H\, W_{\mu\nu}^I+\text{h.c.}$  &9&9&\cellcolor{lightgray}9&\cellcolor{lightgray}9\\
	&$Q_{dB}$        & $(\bar Q_i \sigma^{\mu\nu} d_j) H\, B_{\mu\nu}+\text{h.c.}$ &9&9&\cellcolor{lightgray}9&\cellcolor{lightgray}9\\\hline
	\multirow{8}{0.1\columnwidth}{
		\rotatebox[origin=c]{90}{\begin{minipage}{0.1\textwidth}
				\centering Current-current
	\end{minipage}}}
	&$Q_{H L}^{(1)}$      & $(H^\dag i\overleftrightarrow{D}_\mu H)(\bar L_i \gamma^\mu L_j)$&6&3&\cellcolor{lightgray}3&\cellcolor{lightgray}0\\
	&$Q_{H L}^{(3)}$      & $(H^\dag i\overleftrightarrow{D}^I_\mu H)(\bar L_i \tau^I \gamma^\mu L_j)$&6&3&\cellcolor{lightgray}3&\cellcolor{lightgray}0\\
	&$Q_{H e}$            & $(H^\dag i\overleftrightarrow{D}_\mu H)(\bar e_i \gamma^\mu e_j)$&6&3&\cellcolor{lightgray}3&\cellcolor{lightgray}0\\
	&$Q_{H Q}^{(1)}$      & $(H^\dag i\overleftrightarrow{D}_\mu H)(\bar Q_i \gamma^\mu Q_j)$&6&3&\cellcolor{lightgray}6&\cellcolor{lightgray}3\\
	&$Q_{H Q}^{(3)}$      & $(H^\dag i\overleftrightarrow{D}^I_\mu H)(\bar Q_i \tau^I \gamma^\mu Q_j)$&6&3&\cellcolor{lightgray}6&\cellcolor{lightgray}3\\
	&$Q_{H u}$            & $(H^\dag i\overleftrightarrow{D}_\mu H)(\bar u_i \gamma^\mu u_j)$&6&3&\cellcolor{lightgray}6&\cellcolor{lightgray}3\\
	&$Q_{H d}$            & $(H^\dag i\overleftrightarrow{D}_\mu H)(\bar d_i \gamma^\mu d_j)$&6&3&\cellcolor{lightgray}6&\cellcolor{lightgray}3\\
	&$Q_{H u d}$ & $i(\widetilde H ^\dag D_\mu H)(\bar u_i \gamma^\mu d_j)+\text{h.c.}$&9&9&\cellcolor{lightgray}9&\cellcolor{lightgray}9\\
	\end{tabular}%

}

\caption{\label{bilinearlist}
		The list of dimension-6 fermionic bilinear operators of SMEFT, as given in
		Ref.\,\cite{Grzadkowski:2010es}, together with the number of real and imaginary entries they each contain, as well as the number of primary parameters (highlighted in gray, see the text for more details). When $+\hbox{h.c.}$ is specified, the hermitian conjugate of the operator must be included, too. We indicate with $i,j,k,l$ the flavor indices and with $a,b$ indices in the fundamental of $SU(2)_{L}$. $T^A$, $A=1,\ldots,8$ are the generators of the gauge $SU(3)_c$, while $\tau^I=\frac{\sigma^I}{2}$, $I=1,2,3$ are the generators of $SU(2)_L$, with $\sigma^I$ the Pauli matrices.}
\end{table}

\begin{table}
	\centering
	\renewcommand{\arraystretch}{1.5}
	\resizebox{1\columnwidth}{!}{%
		\begin{tabular}{p{0.5cm}|c|c|c|c|c|c}
			\multicolumn{7}{c}{\textbf{4-Fermi}}\\\hline
			\multicolumn{2}{c|}{Label} & Operator & \begin{minipage}{0.15\columnwidth}\centering
				\# real entries \vspace*{0.1cm}
			\end{minipage}& \begin{minipage}{0.15\columnwidth}\centering
				\# imaginary entries\vspace*{0.1cm}
			\end{minipage} &\begin{minipage}{0.15\columnwidth}\centering
				\cellcolor{lightgray}\# primary real entries\vspace*{0.1cm}
			\end{minipage}   & \begin{minipage}{0.2\columnwidth}\centering
			\cellcolor{lightgray}	\# primary imaginary  entries \vspace*{0.1cm}
			\end{minipage} \\\hline
			\multirow{5}{0.1\columnwidth}{
				\rotatebox[origin=c]{90}{\begin{minipage}{0.1\textwidth}
						\centering LLLL
			\end{minipage}}}
		&$Q_{LL}$        & $(\bar L_i \gamma_\mu L_j)(\bar L_k \gamma^\mu L_l)$ &27&18&\cellcolor{lightgray}9&\cellcolor{lightgray}0\\
		&$Q_{QQ}^{(1)}$  & $(\bar Q_i \gamma_\mu Q_j)(\bar Q_k \gamma^\mu Q_l)$ &27&18&\cellcolor{lightgray}27&\cellcolor{lightgray}18\\
		&$Q_{QQ}^{(3)}$  & $(\bar Q_i \gamma_\mu \tau^I Q_j)(\bar Q_k \gamma^\mu \tau^I Q_l)$ &27&18&\cellcolor{lightgray}27&\cellcolor{lightgray}18\\
		&$Q_{LQ}^{(1)}$                & $(\bar L_i \gamma_\mu L_j)(\bar Q_k \gamma^\mu Q_l)$ &45&36&\cellcolor{lightgray}18&\cellcolor{lightgray}9\\
		&$Q_{LQ}^{(3)}$              & $(\bar L_i \gamma_\mu \tau^I L_j)(\bar Q_k \gamma^\mu \tau^I Q_l)$&45&36&\cellcolor{lightgray}18&\cellcolor{lightgray}9\\\hline
			\multirow{7}{0.1\columnwidth}{
			\rotatebox[origin=c]{90}{\begin{minipage}{0.1\textwidth}
					\centering RRRR
		\end{minipage}}}
		&$Q_{ee}$               & $(\bar e_i \gamma_\mu e_j)(\bar e_k \gamma^\mu e_l)$ &21&15&\cellcolor{lightgray}6&\cellcolor{lightgray}0\\
		&$Q_{uu}$        & $(\bar u_i \gamma_\mu u_j)(\bar u_k \gamma^\mu u_l)$ &27&18&\cellcolor{lightgray}27&\cellcolor{lightgray}18\\
		&$Q_{dd}$        & $(\bar d_i \gamma_\mu d_j)(\bar d_k \gamma^\mu d_l)$ &27&18&\cellcolor{lightgray}27&\cellcolor{lightgray}18\\
		&$Q_{eu}$                      & $(\bar e_i \gamma_\mu e_j)(\bar u_k \gamma^\mu u_l)$& 45&36&\cellcolor{lightgray}18&\cellcolor{lightgray}9\\
		&$Q_{ed}$                      & $(\bar e_i \gamma_\mu e_j)(\bar d_k\gamma^\mu d_l)$&45&36&\cellcolor{lightgray}18&\cellcolor{lightgray}9\\
		&$Q_{ud}^{(1)}$                & $(\bar u_i \gamma_\mu u_j)(\bar d_k \gamma^\mu d_l)$ &45&36&\cellcolor{lightgray}45&\cellcolor{lightgray}36\\
		&$Q_{ud}^{(8)}$                & $(\bar u_i \gamma_\mu T^A u_j)(\bar d_k \gamma^\mu T^A d_l)$& 45&36&\cellcolor{lightgray}45&\cellcolor{lightgray}36\\\hline
			\multirow{8}{0.1\columnwidth}{
			\rotatebox[origin=c]{90}{\begin{minipage}{0.1\textwidth}
					\centering LLRR
		\end{minipage}}}
		&$Q_{Le}$               & $(\bar L_i \gamma_\mu L_j)(\bar e_k \gamma^\mu e_l)$ &45&36&\cellcolor{lightgray}12&\cellcolor{lightgray}3\\
		&$Q_{Lu}$               & $(\bar L_i \gamma_\mu L_j)(\bar u_k \gamma^\mu u_l)$ &45&36&\cellcolor{lightgray}18&\cellcolor{lightgray}9\\
		&$Q_{Ld}$               & $(\bar L_i \gamma_\mu L_j)(\bar d_k \gamma^\mu d_l)$ &45&36&\cellcolor{lightgray}18&\cellcolor{lightgray}9\\
		&$Q_{Qe}$               & $(\bar Q_i \gamma_\mu Q_j)(\bar e_k \gamma^\mu e_l)$ &45&36&\cellcolor{lightgray}18&\cellcolor{lightgray}9\\
		&$Q_{Qu}^{(1)}$         & $(\bar Q_i \gamma_\mu Q_j)(\bar u_k \gamma^\mu u_l)$ &45&36&\cellcolor{lightgray}45&\cellcolor{lightgray}36\\
		&$Q_{Qu}^{(8)}$& $(\bar Q_i \gamma_\mu T^A Q_j)(\bar u_k \gamma^\mu T^A u_l)$ &45&36&\cellcolor{lightgray}45&\cellcolor{lightgray}36\\
		&$Q_{Qd}^{(1)}$& $(\bar Q_i \gamma_\mu Q_j)(\bar d_k \gamma^\mu d_l)$ &45&36&\cellcolor{lightgray}45&\cellcolor{lightgray}36\\
		&$Q_{Qd}^{(8)}$& $(\bar Q_i \gamma_\mu T^A Q_j)(\bar d_k \gamma^\mu T^A d_l)$&45&36&\cellcolor{lightgray}45&\cellcolor{lightgray}36\\\hline
			\multirow{1}{0.1\columnwidth}{
			\rotatebox[origin=c]{90}{\begin{minipage}{0.065\textwidth}
					\centering LRRL
		\end{minipage}}}
		&\begin{minipage}{0.2\textwidth}
			\centering
			\vspace{0.5cm}
			$Q_{LedQ}$ \vspace{0.5cm}
		\end{minipage} & $(\bar L_i^a e_j)(\bar d_k Q_{la})+\text{h.c.}$
		&81&81&\cellcolor{lightgray}27&\cellcolor{lightgray}27\\\hline
			\multirow{3}{0.1\columnwidth}{
			\rotatebox[origin=c]{90}{\begin{minipage}{0.23\textwidth}
					\centering LRLR
		\end{minipage}}}
		&$Q_{QuQd}^{(1)}$& $(\bar Q_i^a u_j) \epsilon_{ab} (\bar Q_k^b d_l)+\text{h.c.}$ &81&81&\cellcolor{lightgray}81&\cellcolor{lightgray}81\\
		&$Q_{QuQd}^{(8)}$& $(\bar Q_i^a T^A u_j) \epsilon_{ab} (\bar Q_k^b T^A d_l)+\text{h.c.}$ &81&81&\cellcolor{lightgray}81&\cellcolor{lightgray}81\\
		&$Q_{LeQu}^{(1)}$& $(\bar L_i^a e_j) \epsilon_{ab} (\bar Q_k^b u_l)+\text{h.c.}$ &81&81&\cellcolor{lightgray}27&\cellcolor{lightgray}27\\
		&$Q_{LeQu}^{(3)}$& $(\bar L_i^a \sigma_{\mu\nu} e_j) \epsilon_{ab} (\bar Q_s^k \sigma^{\mu\nu} u_t)+\text{h.c.}$&81&81&\cellcolor{lightgray}27&\cellcolor{lightgray}27\\
		\end{tabular}%
	}
	\caption{\label{4Fermilist}
		The list of dimension-6 4-Fermi operators of SMEFT, as given in
		Ref.\,\cite{Grzadkowski:2010es}, together with the number of real and imaginary entries they each contain, as well as the number of primary parameters (highlighted in gray, see the text for more details). When $+\hbox{h.c.}$ is specified, the hermitian conjugate of the operator must be included, too. We indicate with $i,j,k,l$ the flavor indices and with $a,b$ indices in the fundamental of $SU(2)_{L}$. $T^A$, $A=1,\ldots,8$ are the generators of the gauge $SU(3)_c$, while $\tau^I=\frac{\sigma^I}{2}$, $I=1,2,3$ are the generators of $SU(2)_L$, with $\sigma^I$ the Pauli matrices.}
\end{table}

	\begin{table}[h!]
	\renewcommand{\arraystretch}{1.4}
	\small
	\centering
	\begin{tabular}{cc|c|ccc}
		&Type of op. & {\begin{minipage}{0.03\textwidth}
				\centering\# \\ops\vspace{0.2cm}
		\end{minipage}}  & \# real &  \# im.  &\\ \cline{1-5}
		\multirow{8}{*}{\rotatebox[origin=c]{90}{bilinears}} &	\multirow{4}{*}{Yuk.} & 	\multirow{4}{*}{3}&\multicolumn{2}{c}{\# of entries at $\order{1/\Lambda^2}$ }&\\
		&&&$3N^2$ & $3N^2$ & \\
		&&&\multicolumn{2}{c}{\cellcolor{lightgray} \# of primary parameters entering observables at $\order{1/\Lambda^2}$}& \\
		&&&\cellcolor{lightgray}$2N^2+N$&\cellcolor{lightgray}$2N^2+N$&\\\cline{2-5}
		&\multirow{2}{*}{Dipole} .& \multirow{2}{*}{8}  & $8N^2$  & $8N^2$&\\\cline{4-5} &&&\cellcolor{lightgray}$6N^2+2N$ &\cellcolor{lightgray} $6N^2+2N$&\\\cline{2-5}
		&\multirow{2}{*}{curr-curr} & \multirow{2}{*}{8} &$\frac{1}{2}N(9N+7)$& $\frac{1}{2}N(9N-7)$ &\\\cline{4-5} &&&\cellcolor{lightgray}$N(3N+5)$  & \cellcolor{lightgray}$N(3N-2)$&\\
		\cline{1-5}
		&\multirow{2}{*}{all bilinears} & \multirow{2}{*}{19}  & $\frac{1}{2} N (31 N+7)$ & $\frac{1}{2} N (31 N-7)$ &\\\cline{4-5} &&&\cellcolor{lightgray}$N (11 N+8)$ &\cellcolor{lightgray} $N (11 N+1)$&\\
		\cline{1-5}
		\multirow{10}{*}{\rotatebox[origin=c]{90}{4-Fermi}} &\multirow{2}{*}{LLLL}  & \multirow{2}{*}{5} & $\frac{1}{4} N^2 \left(7 N^2+13\right)$ & $\frac{7}{4} N^2 \left(N^2-1\right)$ &\\\cline{4-5}&&&\cellcolor{lightgray}$\frac{1}{2} N^2 \left(N^2+2 N+7\right)$ & \cellcolor{lightgray}$\frac{1}{2} N^2 \left(N^2+2 N-3\right)$ &\\\cline{2-5}
		&\multirow{2}{*}{RRRR} & \multirow{2}{*}{7} & $\frac{1}{8} N \left(21 N^3+2 N^2+31 N+2\right)$ & $\frac{1}{8} N(21N+2)(N^2-1) $&\\\cline{4-5}&&&\cellcolor{lightgray}$\frac{1}{2} N \left(3 N^3+2 N^2+8 N+1\right)$&\cellcolor{lightgray}$\frac{1}{2} N^2 \left(3 N^2+2 N-5\right)$&\\\cline{2-5}
		&\multirow{2}{*}{LLRR} & \multirow{2}{*}{8} & 4 $N^2 \left(N^2+1\right)$ & $4 N^2 \left(N^2-1\right)$&\\\cline{4-5}&&& \cellcolor{lightgray}$\frac{1}{2} N \left(4 N^3+3 N^2+9 N+2\right)$ & \cellcolor{lightgray}$\frac{1}{2} N \left(4 N^3+3 N^2-6 N-1\right)$&\\\cline{2-5}
		&\multirow{2}{*}{LRRL} &\multirow{2}{*}{1} & $N^4$ & $N^4$ &\\\cline{4-5}&&&\cellcolor{lightgray}$N^3$& \cellcolor{lightgray}$N^3$&\\\cline{2-5}
		&\multirow{2}{*}{LRLR} & \multirow{2}{*}{4} &$4N^4$ & $4N^4$&\\\cline{4-5}&&&\cellcolor{lightgray}$2 N^3 (N+1)$ & \cellcolor{lightgray}$2 N^3 (N+1)$&\\ \cline{1-5}
		&\multirow{2}{*}{all 4-Fermi} & \multirow{2}{*}{25} & $\frac{1}{8} N \left(107 N^3+2 N^2+89 N+2\right)$  & $\frac{1}{8} N \left(107 N^3+2 N^2-67 N-2\right)$ &\\\cline{4-5}&&&\cellcolor{lightgray}$\frac{1}{2} N \left(12 N^3+13 N^2+24 N+3\right)$ & \cellcolor{lightgray}$\frac{1}{2} N \left(12 N^3+13 N^2-14 N-1\right)$& \\\cline{1-5}
		&\multirow{2}{*}{all} & \multirow{2}{*}{44} &$\frac{1}{8} N \left(107 N^3+2 N^2+213 N+30\right)$ &$\frac{1}{8} N \left(107 N^3+2 N^2+57 N-30\right)  $&\\\cline{4-5}&&&\cellcolor{lightgray}$\frac{1}{2} N \left(12 N^3+13 N^2+46 N+19\right)$ &\cellcolor{lightgray}$\frac{1}{2} N \left(12 N^3+13 N^2+8 N+1\right)$& \\\cline{1-5}
	\end{tabular}
	\caption{Number of flavorful real and imaginary parameters in SMEFT at dimension-six with $N$ flavors. For each type of operator, the first line (in white) counts the number of physical parameters, while the second one (highlighted in gray) counts those which are also primary.} 
	\label{tab:countingflavor}
\end{table}

\section{Complete minimal set of invariants for 2-Fermi operators}\label{appendix:bilinears}

We list in Table\,\ref{tableInvBilinears} a valid choice of minimal sets of CP-odd flavor invariants for all dimension-six Wilson coefficients associated to operators that are bilinear in fermion fields. It can be shown that they provide independent conditions matching the numbers presented in Table\,\ref{tableRanksSMEFT}, in the generic and non-generic cases listed in Table\,\ref{tableFreeParamsCP}.
\begin{table}[h!]
\centering
\renewcommand{\arraystretch}{1.4}
\resizebox{\columnwidth}{!}{%
\begin{tabular}{l|c|c}
Wilson coefficient&Number of phases&Minimal set\\
\hline
\begin{minipage}{0.2\textwidth}
\vspace{0.2cm}
$C_e\equiv\left\{\begin{matrix}\smash{C_{eH}}\\C_{eW}\\C_{eB}\end{matrix}\right.$
\vspace{0.2cm}
\end{minipage}&3&
$\left\{ \ \begin{matrix}
L_0\(C_eY_e^\dagger\)&
L_1\(C_eY_e^\dagger\)&
L_2\(C_eY_e^\dagger\)
\end{matrix} \ \right\}$\\
\hline
\begin{minipage}{0.2\textwidth}
\vspace{0.2cm}
$C_u\equiv\left\{\begin{matrix}C_{uH}\\C_{uG}\\C_{uW}\\C_{uB}\end{matrix}\right.$
\vspace{0.2cm}
\end{minipage}
&\multirow{6}{*}{9}&
$\left\{ \ \begin{matrix}
L_{0000}\(C_{u} Y_u^\dagger\)&
L_{1000}\(C_{u} Y_u^\dagger\)&
L_{0100}\(C_{u} Y_u^\dagger\)\\
L_{1100}\(C_{u} Y_u^\dagger\)&
L_{0110}\(C_{u} Y_u^\dagger\)&
L_{2200}\(C_{u} Y_u^\dagger\)\\
L_{0220}\(C_{u} Y_u^\dagger\)&
L_{1220}\(C_{u} Y_u^\dagger\)&
L_{0122}\(C_{u} Y_u^\dagger\)
\end{matrix}\ \right\}$\\
\begin{minipage}{0.2\textwidth}
	\vspace{0.2cm}
	$C_d\equiv\left\{\begin{matrix}C_{dH}\\C_{dG}\\C_{dW}\\C_{dB}\end{matrix}\right.$
	\vspace{0.2cm}
\end{minipage}
&&
Same with $C_{u} Y_u^\dagger \rightarrow C_{d}^{\mathstrut} Y_d^\dagger$\\
$C_{Hud}$
&&
Same with $C_{u} Y_u^\dagger \rightarrow Y_u^{\mathstrut}C_{Hud}^{\mathstrut} Y_d^\dagger$\\
\hline
$C_{HL}^{(1,3)},C_{He}^{\vphantom{(1,3)}}$
&0&
$\emptyset$\\
\hline
$C_{HQ}^{(1,3)}$
&\multirow{3}{*}{3}&
$\left\{\ \begin{matrix}
L_{1100}\(C_{HQ}^{(1,3)}\)&
L_{2200}\(C_{HQ}^{(1,3)}\)&
L_{1122}\(C_{HQ}^{(1,3)}\)\\
\end{matrix}\ \right\}$\\
$C_{Hu}$
&&
Same with $C_{HQ}^{(1,3)} \to Y_u^{\mathstrut} C_{Hu}^{\mathstrut} Y_u^\dagger$\\
$C_{Hd}$
&&
Same with $C_{HQ}^{(1,3)} \to Y_d^{\mathstrut} C_{Hd}^{\mathstrut} Y_d^\dagger$\\
\end{tabular}%
}
\caption{Minimal sets of CP-odd flavor invariants for all SMEFT dimension-six Wilson coefficients associated to operators bilinear in fermion fields. We recall that $X_u\equiv Y_uY_u^\dagger$, and similarly for down quarks or electrons. We also recall the definition in Eq.\,\eqref{bilinearFormula}. We also defined for the leptons
$L_{a}(\tilde C)\equiv \Im\Tr(X_e^a\tilde C) \ , \text{ with } a=1,2  \ .$}
\label{tableInvBilinears}
\end{table}

\section{Complete minimal set of 4-Fermi invariants}\label{appendix:4Fermi}

We list in Tables\,\ref{tableInv4Fermi1},\,\ref{tableInv4Fermi2}, and\,\ref{tableInv4Fermi3} a valid choice of minimal sets of CP-odd flavor invariants for all dimension-six Wilson coefficients associated to operators quartic in fermion fields. It can be shown that they provide independent conditions matching the numbers presented in Table\,\ref{tableRanksSMEFT}, in the generic and non-generic cases listed in Table\,\ref{tableFreeParamsCP}.
\begin{table}[h!]
\centering
\renewcommand{\arraystretch}{1.4}
\resizebox{\columnwidth}{!}{%
\begin{tabular}{l|c|c}
Wilson coefficient&Number of phases&Minimal set\\
\hline
$C_{LL},C_{ee}$&0&$\emptyset$\\
\hline
$C_{Le}$&3&
\begin{minipage}{0.4\columnwidth}
	\centering
	\vspace{0.2cm}
$\left\{ \ \begin{matrix}
	B^0_0\(C_{LL\tilde e\tilde e}\)&
	B^1_0\(C_{LL\tilde e\tilde e}\)&
	B^2_0\(C_{LL\tilde e\tilde e}\) \end{matrix} \  \right\}$
\vspace{0.02cm}
\end{minipage}\\
\hline
$C_{Qe}$
&\multirow{10}{*}{9}&
\begin{minipage}{0.6\columnwidth}
	\centering
	\vspace{0.2cm}
$\left\{ \ \begin{matrix}
 A^{1100}_{0}\(C_{QQee}\)&\quad
 A^{1100}_{1}\(C_{QQee}\)&\quad
 A^{1100}_{2}\(C_{QQee}\)\\
 A^{2200}_{0}\(C_{QQee}\)&\quad
 A^{2200}_{1}\(C_{QQee}\)&\quad
 A^{2200}_{2}\(C_{QQee}\)\\
 A^{1122}_{0}\(C_{QQee}\)&\quad
 A^{1122}_{1}\(C_{QQee}\)&\quad
 A^{1122}_{2}\(C_{QQee}\)
 \end{matrix} \ \right\}$
\end{minipage}\\
\multirow{2}{0.1\textwidth}{$C_{ed}$}
&&
\multirow{2}{0.5\textwidth}{Same with $C_{QQee} \to C_{\smash{ee\tilde d\tilde d}}$ (exchanging upper with lower indices and with $Y_e^{\vphantom{\dagger}}\leftrightarrow Y_e^\dagger$)}\\
&&\\
\multirow{2}{0.1\textwidth}{$C_{eu}$}
&&
\multirow{2}{0.5\textwidth}{Same with $C_{QQee} \to C_{ee\tilde u\tilde u}$ (exchanging upper with lower indices and with $Y_e^{\vphantom{\dagger}}\leftrightarrow Y_e^\dagger$)}\\
&&\\
$C_{LQ}^{(1,3)}$
&&
$\left\{ \ \begin{matrix}
A^{0}_{1100}\(C_{LQ}^{(1,3)}\)&\quad A^1_{1100}\(C_{LQ}^{(1,3)}\)&\quad A^2_{1100}\(C_{LQ}^{(1,3)}\)\\
A^0_{2200}\(C_{LQ}^{(1,3)}\)&\quad A^1_{2200}\(C_{LQ}^{(1,3)}\)&\quad A^2_{2200}\(C_{LQ}^{(1,3)}\)\\
A^0_{1122}\(C_{LQ}^{(1,3)}\)&\quad A^1_{1122}\(C_{LQ}^{(1,3)}\)&\quad A^2_{1122}\(C_{LQ}^{(1,3)}\)
 \end{matrix} \ \right\}$\\
$C_{Ld}$
&&
Same with $C_{LQ}^{(1,3)} \to C_{LL\tilde d\tilde d}$\\
$C_{Lu}$
&&
Same with $C_{LQ}^{(1,3)} \to C_{LL\tilde u\tilde u}$\\
\hline
$C_{LeQu}^{(1,3)}$
&\multirow{3}{*}{27}&
\begin{minipage}{0.6\columnwidth}
	\centering
	\vspace{0.2cm}
$\left\{ \ \begin{matrix}
 A^0_{0000}\(C_{L\tilde eQ\tilde u}\)&\quad A^1_{0000}\(C_{L\tilde eQ\tilde u}\)&\quad A^2_{0000}\(C_{L\tilde eQ\tilde u}\)\\
 A^0_{1000}\(C_{L\tilde eQ\tilde u}\)&\quad A^1_{1000}\(C_{L\tilde eQ\tilde u}\)&\quad A^2_{1000}\(C_{L\tilde eQ\tilde u}\)\\
 A^0_{0100}\(C_{L\tilde eQ\tilde u}\)&\quad A^1_{0100}\(C_{L\tilde eQ\tilde u}\)&\quad A^2_{0100}\(C_{L\tilde eQ\tilde u}\)\\
 A^0_{1100}\(C_{L\tilde eQ\tilde u}\)&\quad A^1_{1100}\(C_{L\tilde eQ\tilde u}\)&\quad A^2_{1100}\(C_{L\tilde eQ\tilde u}\)\\
 A^0_{0110}\(C_{L\tilde eQ\tilde u}\)&\quad A^1_{0110}\(C_{L\tilde eQ\tilde u}\)&\quad A^2_{0110}\(C_{L\tilde eQ\tilde u}\)\\
  A^0_{2200}\(C_{L\tilde eQ\tilde u}\)&\quad A^1_{2200}\(C_{L\tilde eQ\tilde u}\)&\quad A^2_{2200}\(C_{L\tilde eQ\tilde u}\)\\
 A^0_{0220}\(C_{L\tilde eQ\tilde u}\)&\quad A^1_{0220}\(C_{L\tilde eQ\tilde u}\)&\quad A^2_{0220}\(C_{L\tilde eQ\tilde u}\)\\
A^0_{1220}\(C_{L\tilde eQ\tilde u}\)&\quad A^1_{1220}\(C_{L\tilde eQ\tilde u}\)&\quad A^2_{1220}\(C_{L\tilde eQ\tilde u}\)\\
A^0_{0122}\(C_{L\tilde eQ\tilde u}\)&\quad A^1_{0122}\(C_{L\tilde eQ\tilde u}\)&\quad A^2_{0122}\(C_{L\tilde eQ\tilde u}\)\\
 \end{matrix} \ \right\}$
\end{minipage}
\\
 $C_{LedQ}$
&&
Same with $C_{L\tilde eQ\tilde u} \to C_{L\tilde e\tilde dQ}$ and $A^a_{bcde}\to A^a_{edcb}$
\end{tabular}%
}
\caption{Minimal sets of CP-odd flavor invariants for all the SMEFT dimension-six Wilson coefficients associated to operators quartic in fermion fields (continued in Tables\,\ref{tableInv4Fermi2}, \ref{tableInv4Fermi3}). We recall that $X_u\equiv Y_uY_u^\dagger$, and similarly for down quarks or electrons. We use the generalized traces introduced in Eq.\,\eqref{eq:trAandtrB}, as well as the compact notations in Eqs.~\eqref{eq:L4Fermi}-\eqref{eq:tildeCoeffs}. We also defined for the leptons $A^a_b(C)\equiv \Tr_A\(X_e^a,X_e^b,C\) $, $B^a_b(C)\equiv \Tr_B\(X_e^a,X_e^b,C\) \text{ with } a,b=1,2 \ $, $A^f_{bcde}(C)\equiv \Tr_A\(X_e^f,X_{u\vphantom{d}}^bX_d^cX_{u\vphantom{d}}^dX_d^e,C\)$, $A^{abcd}_{f}(C)\equiv \Tr_A\(X_{u\vphantom{d}}^aX_d^bX_{u\vphantom{d}}^cX_d^d,\(Y_e^\dagger Y_e^{\vphantom{\dagger}}\){}^f
,C\)$ and $B^f_{bcde}(C)\equiv \Tr_B\(X_e^f,X_u^bX_d^cX_u^dX_d^e,C\)$}
\label{tableInv4Fermi1}
\end{table}

\begin{table}[h!]
	\hspace*{-1cm}
\centering
\renewcommand{\arraystretch}{1.01}
\resizebox{\columnwidth}{!}{%
\begin{tabular}{l|c|c}
Wilson coefficient&Number of phases&Minimal set\\
\hline
		$C_{QQ}^{(1,3)}$&18&
		\begin{minipage}{0.6\columnwidth}
			\centering
			\vspace{0.1cm}$\left\{ \ \begin{matrix}
				A^{0000}_{1100}\(C_{QQQQ}\) & A^{1000}_{1100}\(C_{QQQQ}\) & A^{0100}_{1100}\(C_{QQQQ}\) \\
				A^{0000}_{2200}\(C_{QQQQ}\) & A^{1100}_{1100}\(C_{QQQQ}\) & A^{1000}_{2200}\(C_{QQQQ}\) \\
				A^{0100}_{2200}\(C_{QQQQ}\) & A^{0000}_{1122}\(C_{QQQQ}\) & A^{1100}_{2200}\(C_{QQQQ}\) \\
				A^{1200}_{2100}\(C_{QQQQ}\) & A^{1000}_{1122}\(C_{QQQQ}\) & A^{0100}_{1122}\(C_{QQQQ}\) \\
				A^{1100}_{1122}\(C_{QQQQ}\) & A^{2200}_{2200}\(C_{QQQQ}\) & B^{0000}_{1100}\(C_{QQQQ}\) \\
				B^{0000}_{2200}\(C_{QQQQ}\) & B^{0000}_{1122}\(C_{QQQQ}\) & A^{2200}_{1122}\(C_{QQQQ}\) \\
			\end{matrix} \  \right\}$	\vspace{0.1cm}
		\end{minipage}\\
\hline
$C_{uu}$&18&
	\begin{minipage}{0.6\columnwidth}
	\centering
	\vspace{0.1cm}$\left\{ \
\begin{matrix}
		A^{0000}_{1100}\(C_{uu\tilde{u}\tilde{u}}\) & A^{1000}_{1100}\(C_{\tilde{u}\tilde{u}\tilde{u}\tilde{u}}\) & A^{0100}_{1100}\(C_{\tilde{u}\tilde{u}\tilde{u}\tilde{u}}\) \\
		A^{0000}_{2200}\(C_{uu\tilde{u}\tilde{u}}\) & A^{1100}_{1100}\(C_{\tilde{u}\tilde{u}\tilde{u}\tilde{u}}\) & A^{0200}_{1100}\(C_{\tilde{u}\tilde{u}\tilde{u}\tilde{u}}\) \\
		A^{0100}_{2200}\(C_{\tilde{u}\tilde{u}\tilde{u}\tilde{u}}\) & A^{0000}_{1122}\(C_{uu\tilde{u}\tilde{u}}\) & A^{1100}_{2200}\(C_{\tilde{u}\tilde{u}\tilde{u}\tilde{u}}\) \\
		A^{1000}_{1122}\(C_{\tilde{u}\tilde{u}\tilde{u}\tilde{u}}\) & A^{0100}_{1122}\(C_{\tilde{u}\tilde{u}\tilde{u}\tilde{u}}\) & A^{1100}_{0122}\(C_{\tilde{u}\tilde{u}\tilde{u}\tilde{u}}\) \\
		A^{1200}_{2200}\(C_{\tilde{u}\tilde{u}\tilde{u}\tilde{u}}\) & B^{0000}_{1100}\(C_{u\tilde{u}\tilde{u}u}\) & B^{0100}_{1100}\(C_{\tilde{u}\tilde{u}\tilde{u}\tilde{u}}\) \\
		B^{0200}_{2100}\(C_{\tilde{u}\tilde{u}\tilde{u}\tilde{u}}\) & A^{1200}_{1122}\(C_{\tilde{u}\tilde{u}\tilde{u}\tilde{u}}\) & B^{1000}_{1200}\(C_{\tilde{u}\tilde{u}\tilde{u}\tilde{u}}\) \\
\end{matrix}
 \  \right\}$\vspace{0.1cm}
\end{minipage}\\
\hline
$C_{dd}$&18&	\begin{minipage}{0.6\columnwidth}
	\centering
	\vspace{0.1cm}$\left\{ \
\begin{matrix}
	A^{0000}_{1100} \(C_{dd\tilde{d}\tilde{d}}\)& A^{1000}_{1100} \(C_{\tilde{d}\tilde{d}\tilde{d}\tilde{d}}\)& A^{0000}_{2200} \(C_{dd\tilde{d}\tilde{d}}\)\\
	A^{1100}_{2000}\(C_{\tilde{d}\tilde{d}\tilde{d}\tilde{d}}\) & A^{0100}_{1100}\(C_{\tilde{d}\tilde{d}\tilde{d}\tilde{d}}\) & A^{1100}_{1100}\(C_{\tilde{d}\tilde{d}\tilde{d}\tilde{d}}\) \\
	A^{1000}_{2200}\(C_{\tilde{d}\tilde{d}\tilde{d}\tilde{d}}\) & A^{0000}_{1122}\(C_{dd\tilde{d}\tilde{d}}\) & A^{1100}_{2200}\(C_{\tilde{d}\tilde{d}\tilde{d}\tilde{d}}\) \\
	A^{1000}_{1122} \(C_{\tilde{d}\tilde{d}\tilde{d}\tilde{d}}\)& A^{1100}_{1220} \(C_{\tilde{d}\tilde{d}\tilde{d}\tilde{d}}\)& A^{1200}_{2110}\(C_{\tilde{d}\tilde{d}\tilde{d}\tilde{d}}\) \\
	A^{2100}_{0122}\(C_{\tilde{d}\tilde{d}\tilde{d}\tilde{d}}\) & A^{2200}_{1220}\(C_{\tilde{d}\tilde{d}\tilde{d}\tilde{d}}\) & B^{0000}_{1100}\(C_{d\tilde{d}\tilde{d}d}\) \\
	B^{0100}_{2100} \(C_{\tilde{d}\tilde{d}\tilde{d}\tilde{d}}\)& B^{1000}_{1100} \(C_{\tilde{d}\tilde{d}\tilde{d}\tilde{d}}\)& B^{1200}_{2000} \(C_{\tilde{d}\tilde{d}\tilde{d}\tilde{d}}\)\\
\end{matrix}
\  \right\}$\vspace{0.1cm}
\end{minipage}\\
\hline
$C_{Qu}^{(1,8)}$&36&	\begin{minipage}{0.6\columnwidth}
	\centering
	\vspace{0.1cm}$\left\{ \
\begin{matrix}
		A^{1100}_{0000}\(C_{QQuu}\) & A^{0000}_{1100}\(C_{QQ\tilde{u}\tilde{u}}\) & A^{1000}_{1100}\(C_{QQ\tilde{u}\tilde{u}}\) \\
		A^{1100}_{0100}\(C_{QQ\tilde{u}\tilde{u}}\) & A^{1100}_{1100}\(C_{QQ\tilde{u}\tilde{u}}\) & A^{1100}_{0110}\(C_{QQ\tilde{u}\tilde{u}}\) \\
		A^{1200}_{1000}\(C_{QQ\tilde{u}\tilde{u}}\) & A^{2200}_{0000}\(C_{QQuu}\) & A^{1100}_{2200}\(C_{QQ\tilde{u}\tilde{u}}\) \\
		A^{1100}_{0220}\(C_{QQ\tilde{u}\tilde{u}}\) & A^{2200}_{0110}\(C_{QQ\tilde{u}\tilde{u}}\) & A^{1100}_{1122}\(C_{QQ\tilde{u}\tilde{u}}\) \\
		A^{1200}_{1220}\(C_{QQ\tilde{u}\tilde{u}}\) & A^{2200}_{1122}\(C_{QQ\tilde{u}\tilde{u}}\) & B^{0000}_{0100}\(C_{QQ\tilde{u}\tilde{u}}\) \\
		B^{0000}_{1000}\(C_{QQ\tilde{u}\tilde{u}}\) & B^{0000}_{0110}\(C_{QQ\tilde{u}\tilde{u}}\) & B^{0000}_{0220}\(C_{QQ\tilde{u}\tilde{u}}\) \\
		B^{0000}_{1100}\(C_{QQ\tilde{u}\tilde{u}}\) & B^{0000}_{0221}\(C_{QQ\tilde{u}\tilde{u}}\) & B^{0100}_{1000}\(C_{QQ\tilde{u}\tilde{u}}\) \\
		B^{0100}_{1100}\(C_{QQ\tilde{u}\tilde{u}}\) & B^{0100}_{2200}\(C_{QQ\tilde{u}\tilde{u}}\) & B^{0100}_{2110}\(C_{QQ\tilde{u}\tilde{u}}\) \\
		B^{0200}_{2000}\(C_{QQ\tilde{u}\tilde{u}}\) & B^{0200}_{2100}\(C_{QQ\tilde{u}\tilde{u}}\) & B^{0200}_{2110}\(C_{QQ\tilde{u}\tilde{u}}\) \\
		B^{1000}_{0110}\(C_{QQ\tilde{u}\tilde{u}}\) & B^{1000}_{0220}\(C_{QQ\tilde{u}\tilde{u}}\) & B^{1000}_{0221}\(C_{QQ\tilde{u}\tilde{u}}\) \\
		B^{1100}_{1100}\(C_{QQ\tilde{u}\tilde{u}}\) & B^{1100}_{2200}\(C_{QQ\tilde{u}\tilde{u}}\) & B^{1200}_{2100}\(C_{QQ\tilde{u}\tilde{u}}\) \\
		B^{1200}_{2210}\(C_{QQ\tilde{u}\tilde{u}}\) & B^{2100}_{1200}\(C_{QQ\tilde{u}\tilde{u}}\) & B^{0110}_{0221}\(C_{QQ\tilde{u}\tilde{u}}\) \\
\end{matrix}
\  \right\}$\vspace{0.1cm}
\end{minipage}\\
\hline
$C_{Qd}^{(1,8)}$&36&\begin{minipage}{0.6\columnwidth}
	\centering
	\vspace{0.1cm}$\left\{ \
\begin{matrix}
		A^{1100}_{0000}\(C_{QQdd}\) & A^{0000}_{1100}\(C_{QQ\smash{\tilde{d}}\smash{\tilde{d}}}\) & A^{1000}_{1100}\(C_{QQ\smash{\tilde{d}}\smash{\tilde{d}}}\) \\
		A^{1100}_{1000}\(C_{QQ\smash{\tilde{d}}\smash{\tilde{d}}}\) & A^{2200}_{0000}\(C_{QQdd}\) & A^{0100}_{1100}\(C_{QQ\smash{\tilde{d}}\smash{\tilde{d}}}\) \\
		A^{0000}_{2200}\(C_{QQ\smash{\tilde{d}}\smash{\tilde{d}}}\) & A^{1100}_{1100}\(C_{QQ\smash{\tilde{d}}\smash{\tilde{d}}}\) & A^{1100}_{2100}\(C_{QQ\smash{\tilde{d}}\smash{\tilde{d}}}\) \\
		A^{1122}_{0000}\(C_{QQdd}\) & A^{0000}_{1122}\(C_{QQ\smash{\tilde{d}}\smash{\tilde{d}}}\) & A^{1100}_{2200}\(C_{QQ\smash{\tilde{d}}\smash{\tilde{d}}}\) \\
		A^{1100}_{0220}\(C_{QQ\smash{\tilde{d}}\smash{\tilde{d}}}\) & A^{1000}_{1122}\(C_{QQ\smash{\tilde{d}}\smash{\tilde{d}}}\) & A^{1100}_{1122}\(C_{QQ\smash{\tilde{d}}\smash{\tilde{d}}}\) \\
		A^{2100}_{0122}\(C_{QQ\smash{\tilde{d}}\smash{\tilde{d}}}\) & B^{0000}_{0100}\(C_{QQ\smash{\tilde{d}}\smash{\tilde{d}}}\) & B^{0000}_{1000}\(C_{QQ\smash{\tilde{d}}\smash{\tilde{d}}}\) \\
		B^{0000}_{0110}\(C_{QQ\smash{\tilde{d}}\smash{\tilde{d}}}\) & B^{0000}_{0220}\(C_{QQ\smash{\tilde{d}}\smash{\tilde{d}}}\) & B^{0000}_{1100}\(C_{QQ\smash{\tilde{d}}\smash{\tilde{d}}}\) \\
		B^{0000}_{0221}\(C_{QQ\smash{\tilde{d}}\smash{\tilde{d}}}\) & B^{0000}_{2200}\(C_{QQ\smash{\tilde{d}}\smash{\tilde{d}}}\) & B^{0000}_{2210}\(C_{QQ\smash{\tilde{d}}\smash{\tilde{d}}}\) \\
		B^{0100}_{1000}\(C_{QQ\smash{\tilde{d}}\smash{\tilde{d}}}\) & B^{0100}_{0120}\(C_{QQ\smash{\tilde{d}}\smash{\tilde{d}}}\) & B^{0100}_{1100}\(C_{QQ\smash{\tilde{d}}\smash{\tilde{d}}}\) \\
		B^{0100}_{2210}\(C_{QQ\smash{\tilde{d}}\smash{\tilde{d}}}\) & B^{1000}_{0110}\(C_{QQ\smash{\tilde{d}}\smash{\tilde{d}}}\) & B^{1000}_{0220}\(C_{QQ\smash{\tilde{d}}\smash{\tilde{d}}}\) \\
		B^{1000}_{0221}\(C_{QQ\smash{\tilde{d}}\smash{\tilde{d}}}\) & B^{1000}_{1200}\(C_{QQ\smash{\tilde{d}}\smash{\tilde{d}}}\) & B^{1100}_{2200}\(C_{QQ\smash{\tilde{d}}\smash{\tilde{d}}}\) \\
		B^{1100}_{2210}\(C_{QQ\smash{\tilde{d}}\smash{\tilde{d}}}\) & B^{1200}_{2100}\(C_{QQ\smash{\tilde{d}}\smash{\tilde{d}}}\) & B^{2100}_{2211}\(C_{QQ\smash{\tilde{d}}\smash{\tilde{d}}}\) \\
\end{matrix}
\  \right\}$\vspace{0.1cm}
\end{minipage}\\
\end{tabular}%
}
\caption{Continuation of Table\,\ref{tableInv4Fermi1}}
\label{tableInv4Fermi2}
\end{table}

\begin{table}[h!]

\centering
\renewcommand{\arraystretch}{1.13}
\resizebox{\columnwidth}{!}{%
\begin{tabular}{l|c|c}
	Wilson coefficient&Number of phases&Minimal set\\
	\hline
$C_{ud}^{(1,8)}$&36&\begin{minipage}{0.6\columnwidth}
	\centering
	\vspace{0.1cm}
$\left\{ \
\begin{matrix}
	A^{1100}_{0000}\(C_{\tilde{u}\tilde{u}dd}\) & A^{0000}_{1100} \(C_{uu\tilde{d}\tilde{d}}\)& A^{1000}_{1100} \(C_{\tilde{u}\tilde{u}\tilde{d}\tilde{d}}\)\\
	A^{1100}_{1000}\(C_{\tilde{u}\tilde{u}\tilde{d}\tilde{d}}\) & A^{2200}_{0000}\(C_{\tilde{u}\tilde{u}dd}\) & A^{0100}_{1100}\(C_{\tilde{u}\tilde{u}\tilde{d}\tilde{d}}\) \\
	A^{0000}_{2200} \(C_{uu\tilde{d}\tilde{d}}\)& A^{1100}_{1100}\(C_{\tilde{u}\tilde{u}\tilde{d}\tilde{d}}\) & A^{1100}_{0110}\(C_{\tilde{u}\tilde{u}\tilde{d}\tilde{d}}\) \\
	A^{1000}_{2200}\(C_{\tilde{u}\tilde{u}\tilde{d}\tilde{d}}\) & A^{1100}_{2100}\(C_{\tilde{u}\tilde{u}\tilde{d}\tilde{d}}\) & A^{1122}_{0000}\(C_{\tilde{u}\tilde{u}dd}\) \\
	A^{0100}_{2200}\(C_{\tilde{u}\tilde{u}\tilde{d}\tilde{d}}\) & A^{0000}_{1122}\(C_{uu\tilde{d}\tilde{d}}\) & A^{1100}_{2200}\(C_{\tilde{u}\tilde{u}\tilde{d}\tilde{d}}\) \\
	A^{1000}_{1122} \(C_{\tilde{u}\tilde{u}\tilde{d}\tilde{d}}\)& A^{0100}_{1122}\(C_{\tilde{u}\tilde{u}\tilde{d}\tilde{d}}\)& A^{1100}_{1122}\(C_{\tilde{u}\tilde{u}\tilde{d}\tilde{d}}\) \\
	B^{0000}_{0100}\(C_{\tilde{u}\tilde{u}\tilde{d}\tilde{d}}\) & B^{0000}_{1000}\(C_{\tilde{u}\tilde{u}\tilde{d}\tilde{d}}\) & B^{0000}_{0110}\(C_{\tilde{u}\tilde{u}\tilde{d}\tilde{d}}\) \\
	B^{0000}_{1100}\(C_{\tilde{u}\tilde{u}\tilde{d}\tilde{d}}\) & B^{0000}_{0221}\(C_{\tilde{u}\tilde{u}\tilde{d}\tilde{d}}\) & B^{0000}_{2200}\(C_{\tilde{u}\tilde{u}\tilde{d}\tilde{d}}\) \\
	B^{0100}_{1000} \(C_{\tilde{u}\tilde{u}\tilde{d}\tilde{d}}\)& B^{0100}_{0110}\(C_{\tilde{u}\tilde{u}\tilde{d}\tilde{d}}\) & B^{0100}_{2110}\(C_{\tilde{u}\tilde{u}\tilde{d}\tilde{d}}\) \\
	B^{0200}_{2000} \(C_{\tilde{u}\tilde{u}\tilde{d}\tilde{d}}\)& B^{0200}_{2110}\(C_{\tilde{u}\tilde{u}\tilde{d}\tilde{d}}\) & B^{1000}_{0110} \(C_{\tilde{u}\tilde{u}\tilde{d}\tilde{d}}\)\\
	B^{1000}_{0221}\(C_{\tilde{u}\tilde{u}\tilde{d}\tilde{d}}\) & B^{1000}_{1200} \(C_{\tilde{u}\tilde{u}\tilde{d}\tilde{d}}\)& B^{1100}_{2200}\(C_{\tilde{u}\tilde{u}\tilde{d}\tilde{d}}\) \\
	B^{1100}_{2211} \(C_{\tilde{u}\tilde{u}\tilde{d}\tilde{d}}\)& B^{1200}_{2100} \(C_{\tilde{u}\tilde{u}\tilde{d}\tilde{d}}\)& B^{2100}_{1200} \(C_{\tilde{u}\tilde{u}\tilde{d}\tilde{d}}\)\\
\end{matrix}
\ \right\}$\vspace{0.1cm}
\end{minipage}\\
\hline
$C_{QuQd}^{(1,8)}$&81&\begin{minipage}{0.6\columnwidth}
	\centering
	\vspace{0.1cm}
	$\left\{ \
\begin{matrix}
		A^{0000}_{0000}\(C_{\smash{Q\tilde{u}Q\tilde{d}}}\) & A^{0000}_{1000}\(C_{\smash{Q\tilde{u}Q\tilde{d}}}\) & A^{1000}_{0000}\(C_{\smash{Q\tilde{u}Q\tilde{d}}}\) \\
		A^{1000}_{1000}\(C_{\smash{Q\tilde{u}Q\tilde{d}}}\) & A^{0000}_{0100}\(C_{\smash{Q\tilde{u}Q\tilde{d}}}\) & A^{0100}_{0000}\(C_{\smash{Q\tilde{u}Q\tilde{d}}}\) \\
		A^{0000}_{1100}\(C_{\smash{Q\tilde{u}Q\tilde{d}}}\) & A^{0000}_{0110}\(C_{\smash{Q\tilde{u}Q\tilde{d}}}\) & A^{0100}_{1000}\(C_{\smash{Q\tilde{u}Q\tilde{d}}}\) \\
		A^{1000}_{0100}\(C_{\smash{Q\tilde{u}Q\tilde{d}}}\) & A^{1100}_{0000}\(C_{\smash{Q\tilde{u}Q\tilde{d}}}\) & A^{0110}_{0000}\(C_{\smash{Q\tilde{u}Q\tilde{d}}}\) \\
		A^{1000}_{1100}\(C_{\smash{Q\tilde{u}Q\tilde{d}}}\) & A^{1000}_{0110}\(C_{\smash{Q\tilde{u}Q\tilde{d}}}\) & A^{1100}_{1000}\(C_{\smash{Q\tilde{u}Q\tilde{d}}}\) \\
		A^{0100}_{0100}\(C_{\smash{Q\tilde{u}Q\tilde{d}}}\) & A^{0100}_{1100}\(C_{\smash{Q\tilde{u}Q\tilde{d}}}\) & A^{0100}_{0110}\(C_{\smash{Q\tilde{u}Q\tilde{d}}}\) \\
		A^{0110}_{0100}\(C_{\smash{Q\tilde{u}Q\tilde{d}}}\) & A^{0000}_{2200}\(C_{\smash{Q\tilde{u}Q\tilde{d}}}\) & A^{0000}_{0220}\(C_{\smash{Q\tilde{u}Q\tilde{d}}}\) \\
		A^{0200}_{2000}\(C_{\smash{Q\tilde{u}Q\tilde{d}}}\) & A^{1100}_{1100}\(C_{\smash{Q\tilde{u}Q\tilde{d}}}\) & A^{1100}_{0110}\(C_{\smash{Q\tilde{u}Q\tilde{d}}}\) \\
		A^{2000}_{0200}\(C_{\smash{Q\tilde{u}Q\tilde{d}}}\) & A^{2100}_{0100}\(C_{\smash{Q\tilde{u}Q\tilde{d}}}\) & A^{0110}_{1100}\(C_{\smash{Q\tilde{u}Q\tilde{d}}}\) \\
		A^{0110}_{0110}\(C_{\smash{Q\tilde{u}Q\tilde{d}}}\) & A^{0210}_{1000}\(C_{\smash{Q\tilde{u}Q\tilde{d}}}\) & A^{0000}_{1220}\(C_{\smash{Q\tilde{u}Q\tilde{d}}}\) \\
		A^{1200}_{2000}\(C_{\smash{Q\tilde{u}Q\tilde{d}}}\) & A^{0000}_{0122}\(C_{\smash{Q\tilde{u}Q\tilde{d}}}\) & A^{0100}_{1220}\(C_{\smash{Q\tilde{u}Q\tilde{d}}}\) \\
		A^{1000}_{0122}\(C_{\smash{Q\tilde{u}Q\tilde{d}}}\) & A^{1100}_{2200}\(C_{\smash{Q\tilde{u}Q\tilde{d}}}\) & A^{1100}_{0220}\(C_{\smash{Q\tilde{u}Q\tilde{d}}}\) \\
		A^{1200}_{2100}\(C_{\smash{Q\tilde{u}Q\tilde{d}}}\) & A^{2100}_{1200}\(C_{\smash{Q\tilde{u}Q\tilde{d}}}\) & A^{2100}_{0210}\(C_{\smash{Q\tilde{u}Q\tilde{d}}}\) \\
		A^{2200}_{0110}\(C_{\smash{Q\tilde{u}Q\tilde{d}}}\) & A^{0110}_{2200}\(C_{\smash{Q\tilde{u}Q\tilde{d}}}\) & A^{0110}_{0220}\(C_{\smash{Q\tilde{u}Q\tilde{d}}}\) \\
		A^{0112}_{2000}\(C_{\smash{Q\tilde{u}Q\tilde{d}}}\) & A^{1100}_{1220}\(C_{\smash{Q\tilde{u}Q\tilde{d}}}\) & A^{2100}_{0112}\(C_{\smash{Q\tilde{u}Q\tilde{d}}}\) \\
		A^{1200}_{1220}\(C_{\smash{Q\tilde{u}Q\tilde{d}}}\) & A^{2200}_{2200}\(C_{\smash{Q\tilde{u}Q\tilde{d}}}\) & A^{0110}_{1122}\(C_{\smash{Q\tilde{u}Q\tilde{d}}}\) \\
		A^{0122}_{2100}\(C_{\smash{Q\tilde{u}Q\tilde{d}}}\) & A^{0220}_{0220}\(C_{\smash{Q\tilde{u}Q\tilde{d}}}\) & B^{0000}_{0000}\(C_{\smash{Q\tilde{u}Q\tilde{d}}}\) \\
		B^{0000}_{0100}\(C_{\smash{Q\tilde{u}Q\tilde{d}}}\) & B^{0000}_{1000}\(C_{\smash{Q\tilde{u}Q\tilde{d}}}\) & B^{0000}_{1100}\(C_{\smash{Q\tilde{u}Q\tilde{d}}}\) \\
		B^{0000}_{2200}\(C_{\smash{Q\tilde{u}Q\tilde{d}}}\) & B^{0000}_{0110}\(C_{\smash{Q\tilde{u}Q\tilde{d}}}\) & B^{0000}_{0122}\(C_{\smash{Q\tilde{u}Q\tilde{d}}}\) \\
		B^{0000}_{0220}\(C_{\smash{Q\tilde{u}Q\tilde{d}}}\) & B^{0100}_{0000}\(C_{\smash{Q\tilde{u}Q\tilde{d}}}\) & B^{0100}_{1000}\(C_{\smash{Q\tilde{u}Q\tilde{d}}}\) \\
		B^{0100}_{1100}\(C_{\smash{Q\tilde{u}Q\tilde{d}}}\) & B^{0100}_{2100}\(C_{\smash{Q\tilde{u}Q\tilde{d}}}\) & B^{0100}_{0120}\(C_{\smash{Q\tilde{u}Q\tilde{d}}}\) \\
		B^{0100}_{1220}\(C_{\smash{Q\tilde{u}Q\tilde{d}}}\) & B^{0200}_{1120}\(C_{\smash{Q\tilde{u}Q\tilde{d}}}\) & B^{1000}_{0000}\(C_{\smash{Q\tilde{u}Q\tilde{d}}}\) \\
		B^{1000}_{0100}\(C_{\smash{Q\tilde{u}Q\tilde{d}}}\) & B^{1000}_{1200}\(C_{\smash{Q\tilde{u}Q\tilde{d}}}\) & B^{1000}_{0110}\(C_{\smash{Q\tilde{u}Q\tilde{d}}}\) \\
		B^{1000}_{0122}\(C_{\smash{Q\tilde{u}Q\tilde{d}}}\) & B^{1000}_{0210}\(C_{\smash{Q\tilde{u}Q\tilde{d}}}\) & B^{1100}_{0000}\(C_{\smash{Q\tilde{u}Q\tilde{d}}}\) \\
		B^{1100}_{1100}\(C_{\smash{Q\tilde{u}Q\tilde{d}}}\) & B^{1100}_{2200}\(C_{\smash{Q\tilde{u}Q\tilde{d}}}\) & B^{1100}_{0110}\(C_{\smash{Q\tilde{u}Q\tilde{d}}}\) \\
		B^{1100}_{0220}\(C_{\smash{Q\tilde{u}Q\tilde{d}}}\) & B^{1100}_{1122}\(C_{\smash{Q\tilde{u}Q\tilde{d}}}\) & B^{1200}_{2100}\(C_{\smash{Q\tilde{u}Q\tilde{d}}}\) \\
		B^{2100}_{0122}\(C_{\smash{Q\tilde{u}Q\tilde{d}}}\) & B^{2200}_{0000}\(C_{\smash{Q\tilde{u}Q\tilde{d}}}\) & A^{2200}_{1122}\(C_{\smash{Q\tilde{u}Q\tilde{d}}}\) \\
\end{matrix}\  \right \}$\vspace{0.1cm}
\end{minipage}\\
\end{tabular}%
}
\caption{Continuation of Tables\,\ref{tableInv4Fermi1} and \ref{tableInv4Fermi2}}
\label{tableInv4Fermi3}
\end{table}

\section{Invariants featuring $\theta_\text{QCD}$}\label{appendix:thetaQCD}

In the main text of the paper, we focused on quantities which matter for perturbative computations in SMEFT\@. However, this left out an important contribution to CPV in the SM, the $\theta$-parameter of QCD, associated to the following topological term,
\beq
\cL_\text{QCD} \supset -\theta_{QCD}\frac{g_s^2}{16\pi^2}\Tr(G\tilde G) \ .
\eeq
$\theta_{QCD}$ has the following flavor charges (and no lepton-type charge),
\begin{center}
\renewcommand*{\arraystretch}{1.2}
\begin{tabular}{   c | c | c | c | c | c | c  }
 &  $SU(3)_{Q_L}$ & $U(1)_{Q_L}$ & $SU(3)_{u_R}$ & $U(1)_{u_R}$ & $SU(3)_{d_R}$ & $U(1)_{d_R}$    \\
 \hline 
$e^{i\theta_{QCD}}$ &   $\mathbf {1}$  & 6 &$\mathbf {1}$  & -3 & $\mathbf{1}$ & -3\\
\end{tabular}
\end{center}
These charges allow us to build the usual flavor-invariant, physical $\bar\theta$-angle, defined as follows,
\be
e^{-i\theta_{QCD}}\det Y_u \det Y_d=\abs{\det\(Y_uY_d\)} e^{-i\[\theta_{QCD}-\arg\det\(Y_uY_d\) \]}=  \abs{\det\(Y_uY_d\)} e^{-i\bar\theta} \ .
\ee
As $\theta_{QCD}$ provides a dimension-four flavor-charged quantity, one can wonder whether its presence makes new SMEFT coefficients primary, which would mean that new invariants featuring explicitly $\theta_{QCD}$ should be included in the minimal sets. The answer is however negative: the secondary sources of CPV from the dimension-six Wilson coefficients are all charged under unbroken vector-like flavor symmetries of the dimension-four Lagrangian, under which $\theta_{QCD}$ is neutral. Indeed, as the anomalous angle of a vector-like gauge theory, it only shifts under chiral transformations.

Nevertheless, some SMEFT coefficients can be arranged with $\theta_{QCD}$ to form flavor-invariants (albeit redundant in terms of primary parameter counting at dimension-six), which may yield a more natural description of some non-perturbative contributions of the strong interactions to CP-odd observables.\footnote{In the perturbative phase of QCD, the magnitude of such invariants is expected to be suppressed by an additional non-perturbative factor $e^{-8\pi^2/g_s^2}$. For low-energy observables, such as the EDMs of the neutron \cite{Pospelov:1999mv} and of the electron \cite{Choi:1990cn,Ghosh:2017uqq}, no further suppression would be needed.} Those invariants would not have the single trace structure which we used to build our sets of invariants, since $\delta_m^n$ is $U(3)^5$-invariant, while $\theta_{QCD}$ is charged under some abelian parts of the flavor group. Therefore, it will rather offset the abelian charges of determinant-like $SU(3)^5$-invariants. For instance, for the operator $C_{QuQd}$, we can form
\begin{multline*}
\text{Im}\(e^{-i\theta_{QCD}} \epsilon^{ABC}\epsilon^{abc}\epsilon^{DEF}\epsilon^{def}Y_{u,Aa}Y_{u,Bb}C_{QuQd,CcDd}Y_{d,Ee}Y_{d,Ff}\)=\nonumber\\
=\big\vert_\text{up basis}4 y_b y_s y_t y_c \text{Im}C_{QuQd,1111}+... \ .
\end{multline*}

\clearpage
\bibliographystyle{apsrev4-1_title}
\bibliography{biblio.bib}

\end{document}